%% file: main.tex
\documentclass[longauth]{aa}

\usepackage{euclid}
\usepackage{graphicx}
\usepackage{natbib}
\usepackage{scalerel}

\usepackage{longtable}

\usepackage[table]{xcolor}
\usepackage{threeparttable}

\usepackage{txfonts}
\usepackage[pdfencoding=auto,psdextra]{hyperref}
\hypersetup{
    colorlinks=true,
    linkcolor=blue,
    filecolor=magenta,      
    urlcolor=blue,
    citecolor=blue
}
\makeatletter
\renewcommand*\aa@pageof{, page \thepage{} of \pageref*{LastPage}}
\makeatother

\usepackage[utf8]{inputenc}

\newcommand*\chem[1]{\ensuremath{\mathrm{#1}}}

\usepackage[switch, modulo]{lineno}

\defcitealias{scaramella2022}{ESc22}

\begin{document}
%
%
   \title{\Euclid preparation. XXVII. A UV–NIR spectral atlas of compact planetary nebulae for wavelength calibration
       \thanks{The full spectral altas, as well as the 1D and 2D spectra, are available in electronic format on the CDS via anonymous ftp to cdsarc.cds.unistra.fr (130.79.128.5) or via \texttt{\url{https://cdsarc.cds.unistra.fr/cgi-bin/qcat?J/A+A/}}}}

\newcommand{\orcid}[1]{} 
\author{
\normalsize
Euclid Collaboration: K.~Paterson\orcid{0000-0001-8340-3486}$^{1}$\thanks{\email{paterson@mpia.de}}, M.~Schirmer\orcid{0000-0003-2568-9994}$^{1}$, Y.~Copin\orcid{0000-0002-5317-7518}$^{2}$, J.-C.~Cuillandre\orcid{0000-0002-3263-8645}$^{3}$, W.~Gillard\orcid{0000-0003-4744-9748}$^{4}$, L.~A.~Guti\'{e}rrez Soto\orcid{0000-0002-9891-8017}$^{5,6}$, L.~Guzzo\orcid{0000-0001-8264-5192}$^{7,8,9}$, H.~Hoekstra\orcid{0000-0002-0641-3231}$^{10}$, T.~Kitching\orcid{0000-0002-4061-4598}$^{11}$, S.~Paltani\orcid{0000-0002-8108-9179}$^{12}$
% ...
}

\institute{$^{1}$ Max-Planck-Institut f\"ur Astronomie, K\"onigstuhl 17, 69117 Heidelberg, Germany\\
$^{2}$ University of Lyon, Univ Claude Bernard Lyon 1, CNRS/IN2P3, IP2I Lyon, UMR 5822, 69622 Villeurbanne, France\\
$^{3}$ AIM, CEA, CNRS, Universit\'{e} Paris-Saclay, Universit\'{e} de Paris, 91191 Gif-sur-Yvette, France\\
$^{4}$ Aix-Marseille Universit\'e, CNRS/IN2P3, CPPM, Marseille, France\\
$^{5}$ Departamento de Astronomia, IAG, Universidade de S\~{a}o Paulo, Rua do Mat\~{a}o, 1226, 05509-900, S\~{a}o Paulo, Brazil\\
$^{6}$ Instituto de Astrof\'{i}sica de La Plata (CCT La Plata -- CONICET -- UNLP), B1900FWA, La Plata, Argentina\\
$^{7}$ Dipartimento di Fisica "Aldo Pontremoli", Universit\'a degli Studi di Milano, Via Celoria 16, 20133 Milano, Italy\\
$^{8}$ INAF-Osservatorio Astronomico di Brera, Via Brera 28, 20122 Milano, Italy\\
$^{9}$ INFN-Sezione di Milano, Via Celoria 16, 20133 Milano, Italy\\
$^{10}$ Leiden Observatory, Leiden University, Niels Bohrweg 2, 2333 CA Leiden, The Netherlands\\
$^{11}$ Mullard Space Science Laboratory, University College London, Holmbury St Mary, Dorking, Surrey RH5 6NT, UK\\
$^{12}$ Department of Astronomy, University of Geneva, ch. d'Ecogia 16, 1290 Versoix, Switzerland\\
$^{13}$ Centre for Astrophysics, University of Waterloo, Waterloo, Ontario N2L 3G1, Canada\\
$^{14}$ Department of Physics and Astronomy, University of Waterloo, Waterloo, Ontario N2L 3G1, Canada\\
$^{15}$ Perimeter Institute for Theoretical Physics, Waterloo, Ontario N2L 2Y5, Canada\\
$^{16}$ INAF-IASF Milano, Via Alfonso Corti 12, 20133 Milano, Italy\\
$^{17}$ NSF's NOIR, Lab 950 N. Cherry Avenue, Tucson, Arizona 85719, USA\\
$^{18}$ Caltech/IPAC, 1200 E. California Blvd., Pasadena, CA 91125\\
$^{19}$ Infrared Processing and Analysis Center, California Institute of Technology, Pasadena, CA 91125, USA\\
$^{20}$ European Space Agency/ESTEC, Keplerlaan 1, 2201 AZ Noordwijk, The Netherlands\\
$^{21}$ Institut d'Astrophysique de Paris, 98bis Boulevard Arago, 75014, Paris, France\\
$^{22}$ Institut d'Astrophysique de Paris, UMR 7095, CNRS, and Sorbonne Universit\'e, 98 bis boulevard Arago, 75014 Paris, France\\
$^{23}$ CEA Saclay, DFR/IRFU, Service d'Astrophysique, Bat. 709, 91191 Gif-sur-Yvette, France\\
$^{24}$ Universit\'e Paris-Saclay, CNRS, Institut d'astrophysique spatiale, 91405, Orsay, France\\
$^{25}$ ESAC/ESA, Camino Bajo del Castillo, s/n., Urb. Villafranca del Castillo, 28692 Villanueva de la Ca\~nada, Madrid, Spain\\
$^{26}$ Institute of Cosmology and Gravitation, University of Portsmouth, Portsmouth PO1 3FX, UK\\
$^{27}$ INAF-Osservatorio di Astrofisica e Scienza dello Spazio di Bologna, Via Piero Gobetti 93/3, 40129 Bologna, Italy\\
$^{28}$ Dipartimento di Fisica e Astronomia "Augusto Righi" - Alma Mater Studiorum Universit\`{a} di Bologna, via Piero Gobetti 93/2, 40129 Bologna, Italy\\
$^{29}$ INFN-Sezione di Bologna, Viale Berti Pichat 6/2, 40127 Bologna, Italy\\
$^{30}$ Max Planck Institute for Extraterrestrial Physics, Giessenbachstr. 1, 85748 Garching, Germany\\
$^{31}$ Universit\"ats-Sternwarte M\"unchen, Fakult\"at f\"ur Physik, Ludwig-Maximilians-Universit\"at M\"unchen, Scheinerstrasse 1, 81679 M\"unchen, Germany\\
$^{32}$ INAF-Osservatorio Astrofisico di Torino, Via Osservatorio 20, 10025 Pino Torinese (TO), Italy\\
$^{33}$ Dipartimento di Fisica, Universit\`{a} di Genova, Via Dodecaneso 33, 16146, Genova, Italy\\
$^{34}$ INFN-Sezione di Roma Tre, Via della Vasca Navale 84, 00146, Roma, Italy\\
$^{35}$ Department of Physics "E. Pancini", University Federico II, Via Cinthia 6, 80126, Napoli, Italy\\
$^{36}$ Instituto de Astrof\'isica e Ci\^encias do Espa\c{c}o, Universidade do Porto, CAUP, Rua das Estrelas, PT4150-762 Porto, Portugal\\
$^{37}$ Dipartimento di Fisica, Universit\'a degli Studi di Torino, Via P. Giuria 1, 10125 Torino, Italy\\
$^{38}$ INFN-Sezione di Torino, Via P. Giuria 1, 10125 Torino, Italy\\
$^{39}$ Institut de F\'{i}sica d'Altes Energies (IFAE), The Barcelona Institute of Science and Technology, Campus UAB, 08193 Bellaterra (Barcelona), Spain\\
$^{40}$ Port d'Informaci\'{o} Cient\'{i}fica, Campus UAB, C. Albareda s/n, 08193 Bellaterra (Barcelona), Spain\\
$^{41}$ Institute of Space Sciences (ICE, CSIC), Campus UAB, Carrer de Can Magrans, s/n, 08193 Barcelona, Spain\\
$^{42}$ Institut d'Estudis Espacials de Catalunya (IEEC), Carrer Gran Capit\'a 2-4, 08034 Barcelona, Spain\\
$^{43}$ INAF-Osservatorio Astronomico di Roma, Via Frascati 33, 00078 Monteporzio Catone, Italy\\
$^{44}$ INAF-Osservatorio Astronomico di Capodimonte, Via Moiariello 16, 80131 Napoli, Italy\\
$^{45}$ INFN section of Naples, Via Cinthia 6, 80126, Napoli, Italy\\
$^{46}$ Dipartimento di Fisica e Astronomia "Augusto Righi" - Alma Mater Studiorum Universit\'a di Bologna, Viale Berti Pichat 6/2, 40127 Bologna, Italy\\
$^{47}$ Centre National d'Etudes Spatiales -- Centre spatial de Toulouse, 18 avenue Edouard Belin, 31401 Toulouse Cedex 9, France\\
$^{48}$ Institut national de physique nucl\'eaire et de physique des particules, 3 rue Michel-Ange, 75794 Paris C\'edex 16, France\\
$^{49}$ Institute for Astronomy, University of Edinburgh, Royal Observatory, Blackford Hill, Edinburgh EH9 3HJ, UK\\
$^{50}$ Jodrell Bank Centre for Astrophysics, Department of Physics and Astronomy, University of Manchester, Oxford Road, Manchester M13 9PL, UK\\
$^{51}$ European Space Agency/ESRIN, Largo Galileo Galilei 1, 00044 Frascati, Roma, Italy\\
$^{52}$ Institute of Physics, Laboratory of Astrophysics, Ecole Polytechnique F\'{e}d\'{e}rale de Lausanne (EPFL), Observatoire de Sauverny, 1290 Versoix, Switzerland\\
$^{53}$ Departamento de F\'isica, Faculdade de Ci\^encias, Universidade de Lisboa, Edif\'icio C8, Campo Grande, PT1749-016 Lisboa, Portugal\\
$^{54}$ Instituto de Astrof\'isica e Ci\^encias do Espa\c{c}o, Faculdade de Ci\^encias, Universidade de Lisboa, Campo Grande, 1749-016 Lisboa, Portugal\\
$^{55}$ INAF-Osservatorio Astronomico di Trieste, Via G. B. Tiepolo 11, 34143 Trieste, Italy\\
$^{56}$ INAF-Osservatorio Astronomico di Padova, Via dell'Osservatorio 5, 35122 Padova, Italy\\
$^{57}$ Institute of Theoretical Astrophysics, University of Oslo, P.O. Box 1029 Blindern, 0315 Oslo, Norway\\
$^{58}$ Jet Propulsion Laboratory, California Institute of Technology, 4800 Oak Grove Drive, Pasadena, CA, 91109, USA\\
$^{59}$ Technical University of Denmark, Elektrovej 327, 2800 Kgs. Lyngby, Denmark\\
$^{60}$ Cosmic Dawn Center (DAWN), Denmark\\
$^{61}$ Universit\'e Paris-Saclay, Universit\'e Paris Cit\'e, CEA, CNRS, Astrophysique, Instrumentation et Mod\'elisation Paris-Saclay, 91191 Gif-sur-Yvette, France\\
$^{62}$ Universit\'e de Gen\`eve, D\'epartement de Physique Th\'eorique and Centre for Astroparticle Physics, 24 quai Ernest-Ansermet, CH-1211 Gen\`eve 4, Switzerland\\
$^{63}$ Department of Physics, P.O. Box 64, 00014 University of Helsinki, Finland\\
$^{64}$ Helsinki Institute of Physics, Gustaf H{\"a}llstr{\"o}min katu 2, University of Helsinki, Helsinki, Finland\\
$^{65}$ NOVA optical infrared instrumentation group at ASTRON, Oude Hoogeveensedijk 4, 7991PD, Dwingeloo, The Netherlands\\
$^{66}$ Argelander-Institut f\"ur Astronomie, Universit\"at Bonn, Auf dem H\"ugel 71, 53121 Bonn, Germany\\
$^{67}$ Department of Physics, Institute for Computational Cosmology, Durham University, South Road, DH1 3LE, UK\\
$^{68}$ Universit\'e Paris Cit\'e, CNRS, Astroparticule et Cosmologie, 75013 Paris, France\\
$^{69}$ Kapteyn Astronomical Institute, University of Groningen, PO Box 800, 9700 AV Groningen, The Netherlands\\
$^{70}$ Department of Physics and Astronomy, University of Aarhus, Ny Munkegade 120, DK-8000 Aarhus C, Denmark\\
$^{71}$ Space Science Data Center, Italian Space Agency, via del Politecnico snc, 00133 Roma, Italy\\
$^{72}$ Institute of Space Science, Bucharest, 077125, Romania\\
$^{73}$ Dipartimento di Fisica e Astronomia "G.Galilei", Universit\'a di Padova, Via Marzolo 8, 35131 Padova, Italy\\
$^{74}$ INFN-Padova, Via Marzolo 8, 35131 Padova, Italy\\
$^{75}$ Dipartimento di Fisica e Astronomia, Universit\'a di Bologna, Via Gobetti 93/2, 40129 Bologna, Italy\\
$^{76}$ Institut de Ciencies de l'Espai (IEEC-CSIC), Campus UAB, Carrer de Can Magrans, s/n Cerdanyola del Vall\'es, 08193 Barcelona, Spain\\
$^{77}$ Centre for Electronic Imaging, Open University, Walton Hall, Milton Keynes, MK7~6AA, UK\\
$^{78}$ Centro de Investigaciones Energ\'eticas, Medioambientales y Tecnol\'ogicas (CIEMAT), Avenida Complutense 40, 28040 Madrid, Spain\\
$^{79}$ Instituto de Astrof\'isica e Ci\^encias do Espa\c{c}o, Faculdade de Ci\^encias, Universidade de Lisboa, Tapada da Ajuda, 1349-018 Lisboa, Portugal\\
$^{80}$ Universidad Polit\'ecnica de Cartagena, Departamento de Electr\'onica y Tecnolog\'ia de Computadoras, 30202 Cartagena, Spain\\
$^{81}$ Institut de Recherche en Astrophysique et Plan\'etologie (IRAP), Universit\'e de Toulouse, CNRS, UPS, CNES, 14 Av. Edouard Belin, 31400 Toulouse, France\\
$^{82}$ Instituto de Astrof\'isica de Canarias, Calle V\'ia L\'actea s/n, 38204, San Crist\'obal de La Laguna, Tenerife, Spain\\
$^{83}$ INAF-Istituto di Astrofisica e Planetologia Spaziali, via del Fosso del Cavaliere, 100, 00100 Roma, Italy\\
$^{84}$ Department of Physics and Helsinki Institute of Physics, Gustaf H\"allstr\"omin katu 2, 00014 University of Helsinki, Finland\\
$^{85}$ Junia, EPA department, 41 Bd Vauban, 59800 Lille, France\\
$^{86}$ Instituto de F\'isica Te\'orica UAM-CSIC, Campus de Cantoblanco, 28049 Madrid, Spain\\
$^{87}$ CERCA/ISO, Department of Physics, Case Western Reserve University, 10900 Euclid Avenue, Cleveland, OH 44106, USA\\
$^{88}$ Laboratoire de Physique de l'\'Ecole Normale Sup\'erieure, ENS, Universit\'e PSL, CNRS, Sorbonne Universit\'e, 75005 Paris, France\\
$^{89}$ Observatoire de Paris, Universit\'e PSL, Sorbonne Universit\'e, LERMA, 750 Paris, France\\
$^{90}$ Astrophysics Group, Blackett Laboratory, Imperial College London, London SW7 2AZ, UK\\
$^{91}$ SISSA, International School for Advanced Studies, Via Bonomea 265, 34136 Trieste TS, Italy\\
$^{92}$ IFPU, Institute for Fundamental Physics of the Universe, via Beirut 2, 34151 Trieste, Italy\\
$^{93}$ INFN, Sezione di Trieste, Via Valerio 2, 34127 Trieste TS, Italy\\
$^{94}$ Dipartimento di Fisica e Scienze della Terra, Universit\'a degli Studi di Ferrara, Via Giuseppe Saragat 1, 44122 Ferrara, Italy\\
$^{95}$ Istituto Nazionale di Fisica Nucleare, Sezione di Ferrara, Via Giuseppe Saragat 1, 44122 Ferrara, Italy\\
$^{96}$ NASA Ames Research Center, Moffett Field, CA 94035, USA\\
$^{97}$ INAF, Istituto di Radioastronomia, Via Piero Gobetti 101, 40129 Bologna, Italy\\
$^{98}$ INFN-Bologna, Via Irnerio 46, 40126 Bologna, Italy\\
$^{99}$ Universit\'e C\^{o}te d'Azur, Observatoire de la C\^{o}te d'Azur, CNRS, Laboratoire Lagrange, Bd de l'Observatoire, CS 34229, 06304 Nice cedex 4, France\\
$^{100}$ Institute for Theoretical Particle Physics and Cosmology (TTK), RWTH Aachen University, 52056 Aachen, Germany\\
$^{101}$ Institute for Astronomy, University of Hawaii, 2680 Woodlawn Drive, Honolulu, HI 96822, USA\\
$^{102}$ Department of Physics \& Astronomy, University of California Irvine, Irvine CA 92697, USA\\
$^{103}$ University of Lyon, UCB Lyon 1, CNRS/IN2P3, IUF, IP2I Lyon, 4 rue Enrico Fermi, 69622 Villeurbanne, France\\
$^{104}$ INFN-Sezione di Genova, Via Dodecaneso 33, 16146, Genova, Italy\\
$^{105}$ Department of Astronomy \& Physics and Institute for Computational Astrophysics, Saint Mary's University, 923 Robie Street, Halifax, Nova Scotia, B3H 3C3, Canada\\
$^{106}$ University Observatory, Faculty of Physics, Ludwig-Maximilians-Universit{\"a}t, Scheinerstr. 1, 81679 Munich, Germany\\
$^{107}$ Ruhr University Bochum, Faculty of Physics and Astronomy, Astronomical Institute (AIRUB), German Centre for Cosmological Lensing (GCCL), 44780 Bochum, Germany\\
$^{108}$ Department of Physics, Lancaster University, Lancaster, LA1 4YB, UK\\
$^{109}$ Department of Physics and Astronomy, Vesilinnantie 5, 20014 University of Turku, Finland\\
$^{110}$ Centre de Calcul de l'IN2P3/CNRS, 21 avenue Pierre de Coubertin 69627 Villeurbanne Cedex, France\\
$^{111}$ Dipartimento di Fisica, Sapienza Universit\`a di Roma, Piazzale Aldo Moro 2, 00185 Roma, Italy\\
$^{112}$ University of Applied Sciences and Arts of Northwestern Switzerland, School of Engineering, 5210 Windisch, Switzerland\\
$^{113}$ INFN-Sezione di Roma, Piazzale Aldo Moro, 2 - c/o Dipartimento di Fisica, Edificio G. Marconi, 00185 Roma, Italy\\
$^{114}$ Aix-Marseille Universit\'e, CNRS, CNES, LAM, Marseille, France\\
$^{115}$ Centro de Astrof\'{\i}sica da Universidade do Porto, Rua das Estrelas, 4150-762 Porto, Portugal\\
$^{116}$ Dipartimento di Fisica - Sezione di Astronomia, Universit\'a di Trieste, Via Tiepolo 11, 34131 Trieste, Italy\\
$^{117}$ Institute for Computational Science, University of Zurich, Winterthurerstrasse 190, 8057 Zurich, Switzerland\\
$^{118}$ Universit\'e St Joseph; Faculty of Sciences, Beirut, Lebanon\\
$^{119}$ Institut f\"ur Theoretische Physik, University of Heidelberg, Philosophenweg 16, 69120 Heidelberg, Germany}
%
%
\abstract{The \Euclid\/ mission will conduct an extragalactic survey over 15\,000\,deg$^2$ of the extragalactic sky. The spectroscopic channel of the Near-Infrared Spectrometer and Photometer (NISP) 
has a resolution of $R\sim450$ for its blue and red grisms that collectively cover the $0.93$--$1.89\,$\micron\;range. NISP will obtain spectroscopic redshifts for $3\times10^7$ galaxies for the experiments on galaxy clustering, baryonic acoustic oscillations, and redshift space distortion. The wavelength calibration must be accurate within $5\,$\AA\ to avoid systematics in the redshifts and downstream cosmological parameters. The NISP pre-flight dispersion laws for the grisms were obtained on the ground using a Fabry-Perot etalon. Launch vibrations, zero gravity conditions, and thermal stabilisation may alter these dispersion laws, requiring an in-flight recalibration. To this end, we use the emission lines in the spectra of compact planetary nebulae (PNe), which were selected from a PN database. To ensure completeness of the PN sample, we developed a novel technique to identify compact and strong line emitters in Gaia spectroscopic data using the Gaia spectra shape coefficients. We obtained VLT/X-SHOOTER spectra from $0.3$ to $2.5$\,\micron\;for 19 PNe in excellent seeing conditions and a wide slit, mimicking \Euclid's slitless spectroscopy mode but with a ten times higher spectral resolution. Additional observations of one northern PN were obtained in the $0.80$--$1.90$\,\micron\ range with the GMOS and GNIRS instruments at the Gemini North Observatory. The collected spectra were combined into an atlas of heliocentric vacuum wavelengths with a joint statistical and systematic accuracy of 0.1\,\AA\ in the optical and 0.3\,\AA\ in the near-infrared. The wavelength atlas and the related 1D and 2D spectra are made publicly available.}
%
%
%
%
   \titlerunning{Wavelength calibration with compact PNe}
   \authorrunning{Euclid Collaboration, K. Paterson et al.}
   
   \maketitle
%
%
%
%
   
\section{\label{sc:Intro}Introduction}
The \Euclid\/ mission \citep{laureijs2011,racca2016} will employ weak gravitational lensing and galaxy clustering -- which also encompasses baryonic acoustic oscillations \citep[BAO;][]{eisenstein2005} and redshift space distortions \citep{guzzo2008} -- as cosmological probes, to determine the expansion history and growth rate of cosmic structures over the last 10 billion years \citep{blanchard2020}. These experiments address the nature and properties of dark energy, dark matter, gravitation, and the Universe's initial conditions. The accuracy of the results should be decisive for the validity of the $\Lambda$ cold dark matter ($\Lambda$CDM) concordance model and general relativity on cosmic scales 
\citep[see e.g.][]{amendola2010,wang2010,weinberg2013}.

The imaging survey of 15\,000\,deg$^2$ will be done with the visible imager 'VIS' \citep{cropper2012} in a single, wide band (0.53--0.92\,\micron) down to a $5\,\sigma$ point-source depth of 26.2\,AB\,mag. The Near-Infrared Spectrometer and Photometer \citep[NISP;][]{prieto2012,maciaszek2016} will obtain a $5\,\sigma$ point-source depth of 24.5\,AB\,mag in three wide bands covering the 0.95--2.02\,\micron\;range. A comprehensive and detailed presentation of the Euclid Wide Survey and its observational strategy is presented in \cite{scaramella2022} (hereafter \citetalias{scaramella2022}).

The spectroscopic survey will cover the same area, to a $3.5\,\sigma$ H$\alpha$ line-flux limit of $2.0\times10^{-16}$\,erg\,cm$^{-2}$\,s$^{-1}$ at a redshifted wavelength of 1.60\,\micron. The sample consists of $3\times10^7$ galaxy spectra with a resolution of $R\sim450$ for objects with a diameter of \ang{;;0.5}. Two `red grisms' in NISP cover the same $1.21$--$1.89\,$\micron\;range with opposite dispersion directions. To better decontaminate the slitless spectra in the dispersed images, these grisms are also rotated by \ang{4} in \Euclid's reference observing sequence \citepalias{scaramella2022}, yielding four different dispersion directions per survey field. Given the nonlinnearities in the dispersion law, we consider the rotated configurations as physically independent grisms. For more details see also \cite{gabarra2023}.

NISP also has a `blue grism' ($0.93$--$1.37\,$\micron) that increases the total number of grism configurations to five. The blue grism will be used solely in the observations of the \Euclid Deep Fields (50\,deg$^2$, with additional red grism coverage) and at a fixed orientation angle in NISP. However, since the deep fields will be revisited with different spacecraft roll angles, different on-sky dispersion directions of the blue grism will be realised as for the red grisms \citepalias{scaramella2022}.

\begin{figure*}[t]
\centering
\includegraphics[angle=0,width=1.0\hsize]{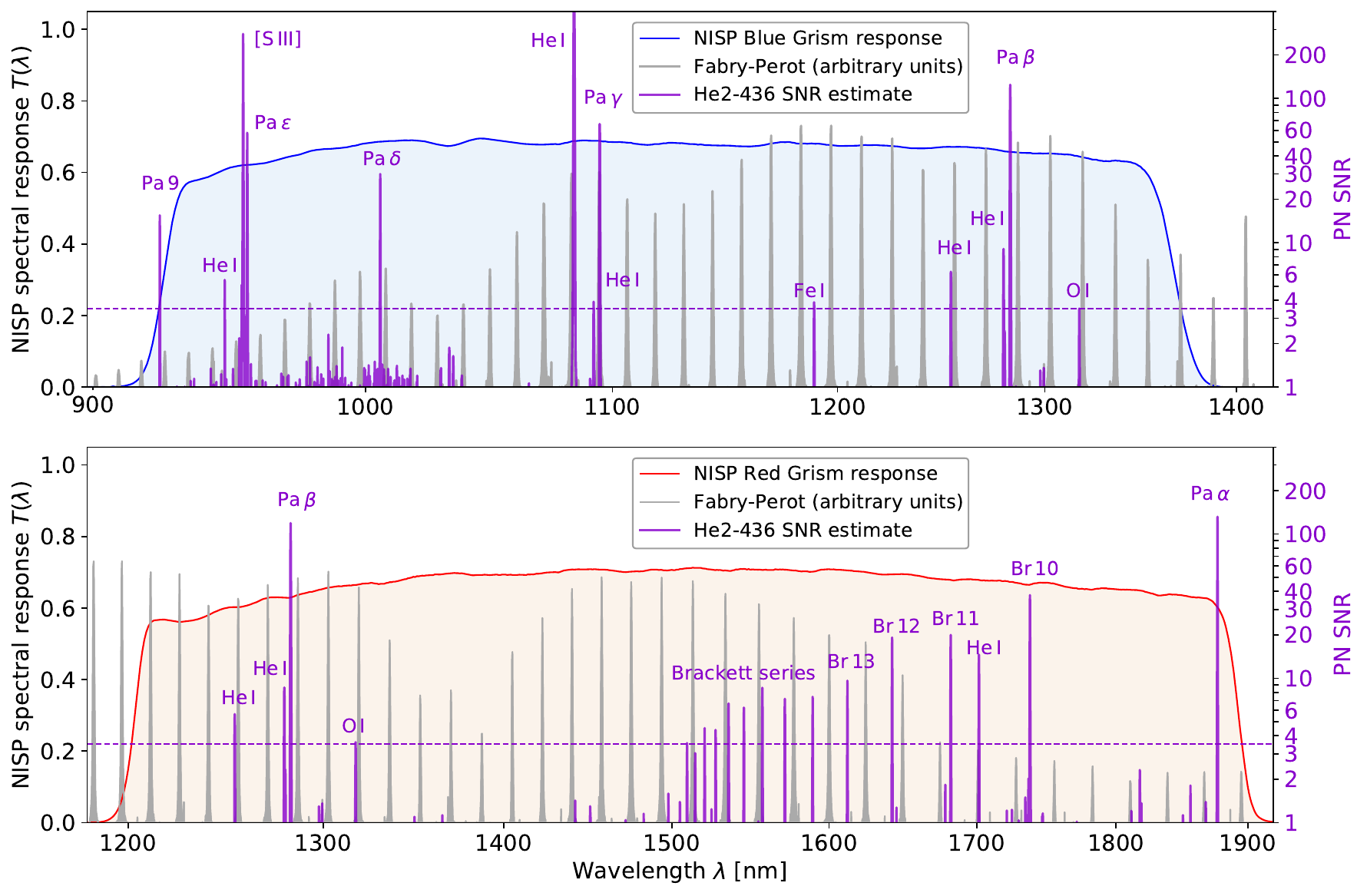}
\caption{Total system response for the NISP grisms and emission line signal-to-noise ratios (S/N). \textit{Top panel}: Total system response for the NISP blue grism (shaded area, exceeding 50\% of its peak transmission between $0.93$--$1.37\,$\micron; these data are updated with respect to those presented in \citetalias{scaramella2022}). The regularly spaced, grey spectrum shows the arbitrarily scaled Fabry-Perot emission lines used for the pre-flight calibration of the dispersion law. As an example for the in-flight calibration, the purple spectrum shows the estimated NISP S/N of He2-436, extrapolated from our X-SHOOTER observations. The horizontal dashed line displays a $3.5\,\sigma$ threshold to identify usable PN lines. \textit{Bottom panel}: Same as for the other panel, but for the NISP red grism ($1.21$--$1.89\,$\micron).}
\label{fig:fp_pn}
\end{figure*}

Accurate wavelength calibration is paramount for \Euclid's cosmological redshift measurements.  Systematic wavelength errors, in particular, if dependent on the sky position or epoch of observation, have a tremendous impact on cosmological measurements aiming to detect tiny fluctuations in the galaxy density over very large scales. These include delicate measurements, such as detecting non-Gaussianity, a signature of primordial inflation \citep[e.g.][]{castorina2019}. The NISP dispersion laws must therefore be known to be better than $5$\,{\AA} -- or 0.3 NISP pixels -- anywhere in its focal plane of 16 HAWAII-2RG detectors for the entire mission duration of six years.

On Earth, this accuracy was achieved using a Fabry-Perot emission-line spectrum (Fig.~\ref{fig:fp_pn}). Low-order deviations from this pre-flight dispersion law can occur due to acousto-mechanical vibrations during launch, zero gravity in flight, and the optics' final in-flight temperatures that are difficult to predict. After launch, during a two-month long performance verification (PV) phase, the dispersion laws -- and many other pre-flight calibration products -- will be updated. The NISP opto-mechanical design \citep{grupp2012} does not provide an on-board arc lamp for wavelength calibration. Our in-flight wavelength calibration strategy thus involves astrophysical emission-line sources for which we need to determine accurate wavelengths.  

In this paper we present ground-based observations of 20 ultra-compact planetary nebulae (PNe) comprising a spectral atlas for accurate wavelength calibration. In Sect.~\ref{sc:CalibStrategy} we motivate the wavelength calibration strategy with PN, followed by our target selection in Sect.~\ref{sec:sampleselection}. There, we also present a novel approach to identify compact line emitters in Gaia spectroscopic data. The observations and data reduction are discussed in Sect.~\ref{sec:observations}, with an emphasis on accurate wavelength calibration. Specifically, we show that no significant systematics will be introduced into \Euclid's cosmological measurements when using these PNe for wavelength calibration. In Sect.~\ref{sc:Res} we introduce the main result of this work, the spectral atlas, and some scientific results for individual PNe and emission-line ratios. We conclude in Sect.~\ref{sc:Cons}. The processed data and the spectral atlas are available online\footnote{\texttt{\url{https://euclid.esac.esa.int/msp/refdata/nisp/PN-SPECTRAL-ATLAS-V1}}}.

\section{\label{sc:CalibStrategy}Wavelength calibration with planetary nebulae}

\begin{figure}[t]
\centering
\includegraphics[angle=0,width=1.0\hsize]{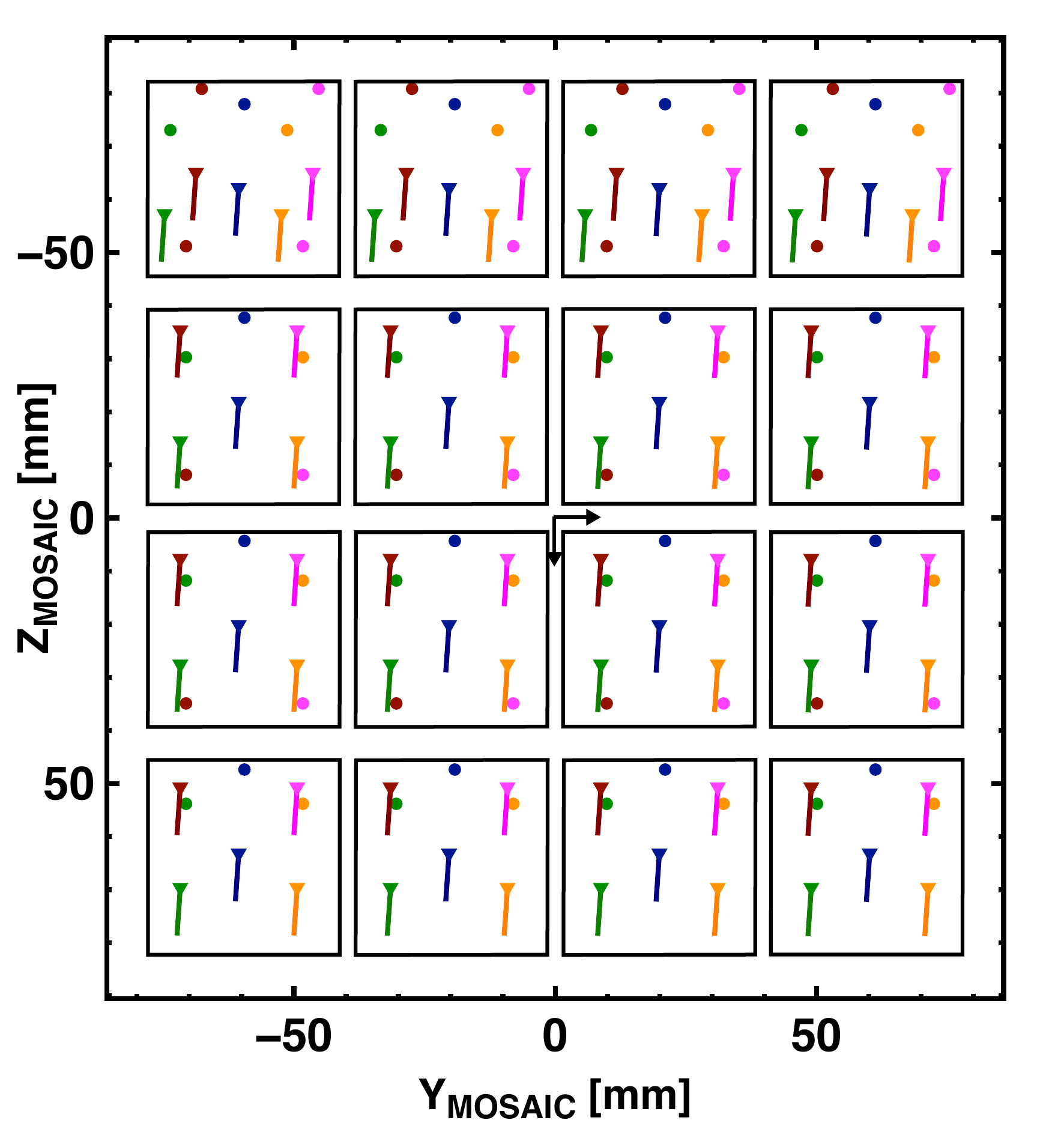}
\caption{Placement of the PN on the NISP focal plane of 16 detectors, for the \ang{4} rotated position of one of the red grisms, in the NISP R\_MOSAIC detector coordinate system. The dots show the positions of the 0th order, and the lines the location of the 1st order, colour-coded for easier association. The triangles mark the short-wavelength end of the 1st orders. The pattern is compressed for the top detector row, because we must measure both the 0th and 1st orders simultaneously to determine the dispersion laws. In the absence of the grism, the images of the PN would appear within the first order. Since the dispersion laws vary slowly, this pattern compression does not bias the result.
}
\label{fig:RGS000_grid}
\end{figure}

\subsection{\label{sec:ongroundProcedure}On-ground procedure}
Lacking the possibility of internal wavelength calibration, the NISP dispersion laws were measured on-ground with the NISP instrument itself \citep[{for details see}][]{maciaszek2022}. The characterisation was done in a vacuum and at operational temperature, using an external Fabry-Perot etalon light source with 38 and 35 emission lines in the blue and red grism transmission ranges, respectively (Fig.~\ref{fig:fp_pn}).  Additional Argon spectra were used to unambiguously identify the Fabry-Perot lines, which were then modelled with a 2D asymmetrical Gaussian profile. The full width half maximum (FWHM) of the lines ranged from 0.7\,nm at 900\,nm to 1.2\,nm at 1900\,nm. The point-like light source was placed at the nodes of a $12\times12$ grid covering the focal plane detector array (FPA). Each grism's dispersion law consists -- as a function of source position in the FPA -- of (1) the offset between the 0th order and the source position, (2) the separation between the 0th order and a reference wavelength in the 1st order, (3) the curved shape of the 1st-order spectral trace, and (4) the nonlinnear wavelength dispersion within the 1st order. We obtained the dispersion laws by fitting fourth-degree Chebyshev polynomials to each of the above. The faint 2nd order was not characterised, but its curvature and location are known from ground tests so that it can be masked if necessary.

The modelled dispersion laws predict the line positions of the Argon spectral lamp observed on ground with an RMS of 4.4\,\AA\ and a mean bias error (MBE) of 2.8\,\AA\ across the FPA. We note that these uncertainties are not purely intrinsic to the modelled dispersion laws; the low intensity of the recorded 0th order of the Argon spectra contributed as well. For reference, the dispersion laws obtained in-flight must not contribute an error larger than 5\,\AA\ (0.3 pixel) to the total wavelength error of the observed galaxy emission lines.

\subsection{\label{sc:InflightStrategy} In-flight wavelength calibration strategy}
After insertion into a L2 halo orbit \citep{howell1984}, \Euclid will enter its two-month-long PV phase prior to survey operations. To detect deviations from the pre-flight dispersion laws, an astrophysical emission-line source will be observed similarly to the on-ground calibration (Sect.~\ref{sec:ongroundProcedure}). The calibration involves four mappings: (1) from the astrometric sky \citep{gaia2016} to 0th order using astrometrically calibrated, undispersed NISP images; (2) from 0th order to a reference wavelength in the 1st order, (3) the curved shape of the 1st-order spectral trace, and (4) reconstruction of the nonlinnear dispersion in the 1st order based on the emission lines. The in-flight dispersion laws will then either replace the on-ground calibration files or complement them, depending on the actual number density and spectral sampling that can be achieved.

In the current PV plan, the emission-line source will be observed on five positions per detector (Fig.~\ref{fig:RGS000_grid}), for a total of 80 positions. This is less than the 144 positions used on-ground, since these observations are  expensive with 29\,h per grism. We expect the in-flight dispersion laws to be fairly stable over time, since \Euclid's telescope structure and mirrors are built from silicon carbide \citep[SiC;][]{bougoin2019} that features extreme stiffness and low thermal expansion; the same holds for the NISP instrument truss \citep{pamplona2016,bougoin2017}. 

Yet we know from the Gaia telescope -- also built from SiC \citep{bougoin2011} -- that focus drifts can be active at a low level even after years in space \citep{mora2016}. Therefore, immediately after observing the emission-line source and maintaining the telescope's thermo-optical state, we will observe \Euclid's self-calibration field at the North Ecliptic Pole. This field is observed monthly for monitoring purposes throughout the mission \citepalias{scaramella2022}. There, a secondary set of wavelength standards -- such as stellar absorption-line systems -- will be established, so that drifts in the dispersion laws can be caught in due time.

\subsection{Why compact planetary nebulae are the best choice}
Ideally, the astrophysical emission-line calibrators are stable over time, and spectrally and spatially unresolved by NISP. This excludes emission-line stars and many AGN. For example, massive stars with decretion disks such as luminous blue variables, and Be-type stars have variable and complicated line profiles; their emission lines may disappear, or show velocity features in excess of 300\,km\,s$^{-1}$ \citep{porter2003,groh2007}. Narrow emission-line regions around AGN are spatially extended, substructured, and kinematically broadened ($300$--$1000$\,km$\,$s$^{-1}$). Also, at low redshifts the line density in NISP spectra of AGN is insufficient, and at higher redshift the line fluxes are too low. Existing radial velocity standards \citep{soubiran2013} would deeply saturate the NISP detectors in the available spectroscopic observing modes. PNe on the other hand are well suitable. They have sufficiently many, bright near-infrared lines, and typical spectral expansion velocities of $10$--$50$ km$\,$s$^{-1}$ \citep[e.g.][]{gesicki2000,marigo2001,jacob2013,lopez2016,schoenberner2016}, at least a factor five below the NISP spectral resolution. PNe are also used to provide absolute wavelength calibration for instruments flying on other missions, such as the \textit{James Webb} Space Telescope \citep[JWST;][]{labiano2021}.

The only effective line broadening seen by the NISP slitless spectroscopy mode would come from the PN's intrinsic angular extent, which should thus not be much larger than the spatial resolution of the spectroscopy mode: the 80\% encircled energy radius, EE80, is typically \ang{;;0.48}--\ang{;;0.55} at 1.50\,\micron\ \citep[e.g.][]{grupp2019}. PN radii of up to \ang{;;0.5} should therefore be unproblematic as long as pronounced substructures are absent.

We note that morphokinematical studies of spatially extended and bright PNe are possible with slitless spectroscopy \citep[e.g.][]{steffen2009,garcia2012,clairmont2022}. However, to use these for \Euclid's wavelength calibration would require (i) complex modelling, (ii) observations with well-calibrated slitless near-infrared spectrographs from space, and (iii) the development of entirely new processing functions to analyse the \Euclid spectra of such extended sources and match them with slitless spectra from different observatories. The related effort is prohibitive. The big advantage of using compact and -- ideally unresolved -- PN is that exactly the same processing functions can be used that are in place already to extract the spectra and redshifts of galaxies at cosmological distances; consistency in the processing of science and calibration data is of utmost importance for \Euclid.

\begin{figure*}[t]
\centering
\includegraphics[angle=0,width=1.0\hsize]{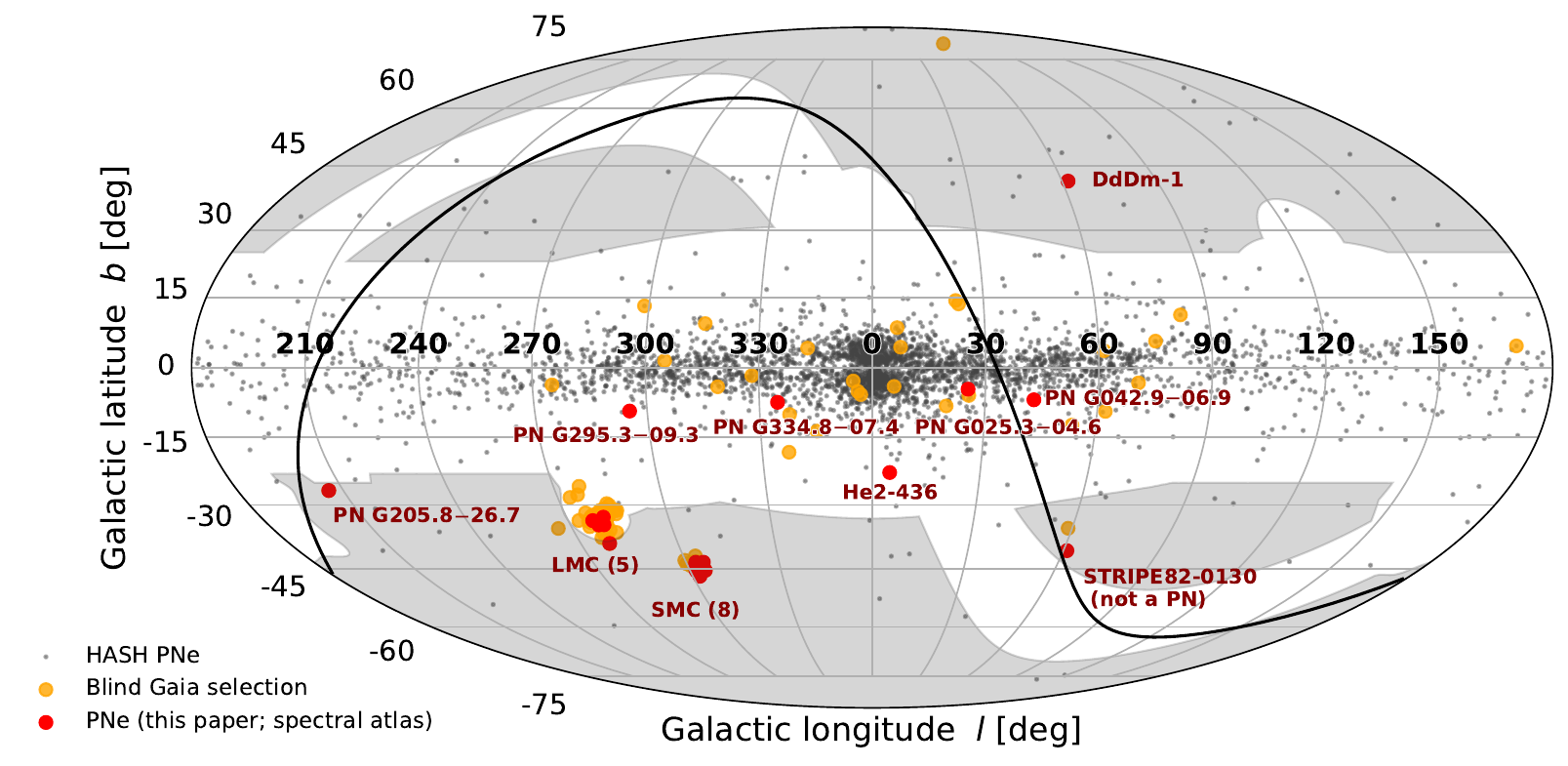}
\caption{Distribution of the 3846 Galactic PNe in the HASH database (small grey dots). Overlaid are the 20 compact PNe (red dots) that form the spectral atlas presented in this paper, and also STRIPE83-0130 (a transient included for completeness). We also show 86 blindly selected Gaia sources (orange dots; Sect.~\ref{sec:searchGaia2}) that have similar Gaia spectra as the compact PNe. Only 0.13\% of Galactic PNe in HASH have measured sub-arcsecond diameters, and only one of these (DdDm-1) is located considerably outside the Galactic plane. The black line indicates the celestial equator. The shaded areas show the Euclid Wide Survey area that avoids the ecliptic and Galactic planes \citepalias[for details see][]{scaramella2022}.}
\label{fig:PNdistribution}
\end{figure*}

In Fig.~\ref{fig:fp_pn} we compare the spectrum of one PN in our sample, He2-436, with that of the Fabry-Perot etalon. While the density of usable lines in the PN spectrum is lower, $8$--$10$ useful lines should be available with the planned integration times for both the blue and the red grism. This is sufficient to update the in-flight dispersion laws -- that vary slowly over the field -- despite the uneven distribution of lines in wavelength space.

The wavelength dispersion needs to be known with 5\,\AA\;accuracy, corresponding to 0.3 NISP pixel. Most PN lines will have a signal-to-noise ratio (S/N) considerably above $3.5\,\sigma$, hence the uncertainties of their measured line centroids are expected to be smaller than 0.3 pixel. A more accurate performance estimate is part of currently ongoing, pre-launch simulations of the PV-phase data.

\subsection{Obtaining accurate reference wavelengths from PN\label{sec:importanceObsSetup}}
To serve as an absolute wavelength calibrator for NISP, a PN must at least have strong emission lines, and 80\% of its total flux must be contained within a radius of \ang{;;0.5} or below. NISP observes slitless dispersed images, with a resolution at least a factor five too low to resolve the gas kinematics of up to 50\,km$\,$s$^{-1}$ in typical non-bipolar PNe with well-defined radii \citep{jacob2013}. Thus, the emission lines detected by NISP are line images of the PN's full spatial extent. The line-image centroid in the dispersion direction -- that is its effective wavelength -- is therefore not directly comparable to ground-based slit spectroscopy: In case of considerable substructure, slit truncation of the nebula could lead to different effective wavelengths. The PNe in our sample are compact, at most $2$--$3$ NISP pixel wide ($\ang{;;0.3}$\,pixel$^{-1}$ plate scale), lowering our sensitivity to this effect. 

Yet this is a concern, as we must use these line images to calibrate the dispersion laws to better than 0.3 pixel. The ground-based observations must therefore mimic the slitless NISP spectra as much as possible, using slits considerably wider than the PNe's spatial extents, yet narrow enough for accurate arc-lamp wavelength calibration. We also need excellent atmospheric seeing to minimise slit losses and blurring of the line image.

The ground-based spectra also need 5--10 times higher spectral resolution than NISP. Like this, systematic errors in the wavelength calibration of the ground-based spectra are reduced by a corresponding factor when propagated to the NISP data. In case of line blends, the higher spectral resolution will tell whether reliable line centroids can be determined from the lower-resolution NISP spectra.

\section{PN sample selection\label{sec:sampleselection}}

\subsection{\Euclid's visibility function}
\Euclid observes along ecliptic meridians, maintaining a Solar aspect angle of \ang{87}--\ang{110} between the target, the spacecraft, and the Sun \citepalias{scaramella2022}. At low ecliptic latitude, a target becomes visible twice a year for a few days. Visibility increases with ecliptic latitude, reaching perennial visibility within about \ang{2.5} of the ecliptic poles (continuous viewing zone). We therefore need a selection of PNe across the sky, to ensure that for any launch date we have a PN accessible in or close to PV phase; this would also accommodate unforeseen instrument anomalies that require timely recalibration.

\subsection{Notes about our selection function\label{sec:selectionoverview}}
The selection process for our compact and bright PNe was convoluted. Due to their scarcity, we had to be sure that we did not overlook suitable candidates, while  staying within the strict \Euclid mission timeline towards launch, and an increasingly constrained PV calibration plan. Ground observatory downtimes due to the pandemic were also a factor. 

The performance of the NISP spectroscopic pipeline for the PN data is not well-known at the time of writing. The pipeline is optimised for the detection of faint emission lines in low-density fields, but most PNe are found in more crowded areas. PNe have brighter lines that could still be automatically identified despite overlapping 0th, 1st and 2nd orders. This, however, depends on line flux, a telescope roll angle that is unknown because the exact time of observation is still unknown, and also the spatial distribution of field sources. Thus we did not establish quantitative crowding thresholds; but we accounted for relative crowding when deciding which of two otherwise equally valuable PN should be observed within the allocated time. As a crowding index, we compute the local number density of Gaia sources with ${\rm Gmag}<19$\,mag, based on a 13\,arcmin$^2$ circular area where sources could in principle -- depending on telescope and grism angles -- contribute to contamination of the 1st order.

 In the end, we followed paths in parallel, adjusting our criteria dynamically while observations were already ongoing. Such a tangled selection function is acceptable for the purposes of this paper, where we just needed to identify suitable PNe, but it would be problematic for systematic studies of PN populations. We adopt a simplified hindsight perspective in the rest of Sect.~\ref{sec:sampleselection}.

\subsection{Searching the HASH database}
Version 4.6 of the Hong Kong/AAO/Strasbourg H$\alpha$ (HASH) PN database \citep{parker2016} lists 3846 known PNe within the Galaxy, and many more elsewhere \citep[see e.g.][]{kwitter2022}. The total number in the Galaxy might be as high as 6000--45\,000 \citep{parker2022}, most of them highly dust-extinguished. HASH is a heterogeneous database, collecting PNe and their properties from numerous different publications. The size estimates of compact PNe depend on the image seeing unless determined with the \textit{Hubble} Space Telescope (HST), and also on the size definitions chosen by authors. Applying upper limits of \ang{;;1.5} and \ang{;;1.0} to the HASH major axis diameter results in only eleven and five PNe, respectively; we note that a considerable fraction of PNe in HASH do not have size estimates.

Figure~\ref{fig:PNdistribution} shows that 98.5\% of the Galactic HASH PNe are confined to low Galactic latitude, $|b|\leq\ang{30}$. There, the 0th and 1st orders of the slitless NISP spectra become contaminated, making PNe in the uncrowded halo much preferred. Given that only 0.13\% of HASH PNe have recorded sub-arcsecond diameters, our choices are extremely limited.

Archival HST imaging and / or slitless spectroscopy were mandatory for us to reliably select sub-arcsecond PNe from HASH. We required suitable morphologies, ideally a homogeneous or radially symmetric appearance without considerable envelopes. The only suitably compact PNe known in the Galactic halo is PN G061.9+41.3 (hereafter DdDm-1) at $b=\ang{41}$, with very low crowding. \cite{henry2008} measure a diameter of \ang{;;0.6} after deconvolving an archival HST image from 1993 that still suffered from HST's spherical aberration (see Figs.~\ref{fig:PNmorph_linear} and \ref{fig:PNmorph}). More details about this PN can also be found in \cite{otsuka2009}.

\cite{stanghellini2016} observed 51 compact PNe in the Galactic plane with HST. They define a photometric radius, $R_{\rm phot}$, containing 85\% of the flux in the HST F502N narrow-band image centred on the [\ion{O}{III}]~$\lambda5008$ line. We selected four PNe with $R_{\rm phot}<\ang{;;0.5}$, compatible with the NISP spectroscopy EE80 radius of \ang{;;0.48}--\ang{;;0.55}. Two of these (PN G025.3--04.6 and PN G042.9--06.9) show noticeable substructures in their cores in the HST images (Figs.~\ref{fig:PNmorph_linear} and \ref{fig:PNmorph}), but the NISP spectroscopic PSF is wide enough to make them usable, albeit not ideal, wavelength calibrators. Another source in \cite{stanghellini2016} is PN G205.8$-$26.7, with a ring-shaped core of \ang{;;0.8} diameter and embedded in a symmetrical fainter halo of \ang{;;2.5} diameter. While its morphology is less favourable, its crowding index is very low, and its Euclid visibility function is different to those of the other PNe, making it a valuable backup resource.

Extending the search to extragalactic PNe in the Magellanic Clouds with HST coverage \citep{shaw2001,stanghellini2002,stanghellini2003,shaw2006}, we retained 13 PNe with $R_{\rm phot}=\ang{;;0.13}$--$\ang{;;0.40}$. Their line fluxes are typically a factor five lower than for Galactic PNe, and several of them are less crowded.

HASH also contains numerous PNe in local dwarf galaxies \citep[e.g.][]{magrini2003,richer2007}, and in the Local Group \citep{pena2007,delgadoinglada2020}. However, they are mostly too faint and very crowded. We retained PN G004.8-22.7 (hereafter He2-436) in the Sagittarius dwarf elliptical galaxy, for which HST imaging and line fluxes are available from \cite{zijlstra2006}. We measured $R_{\rm phot}=\ang{;;0.21}$ in the HST narrow-band image. This is likely underestimated, as the image also contains contributions from the central star.

\begin{figure*}[t]
\centering
\includegraphics[angle=0,width=1.0\hsize]{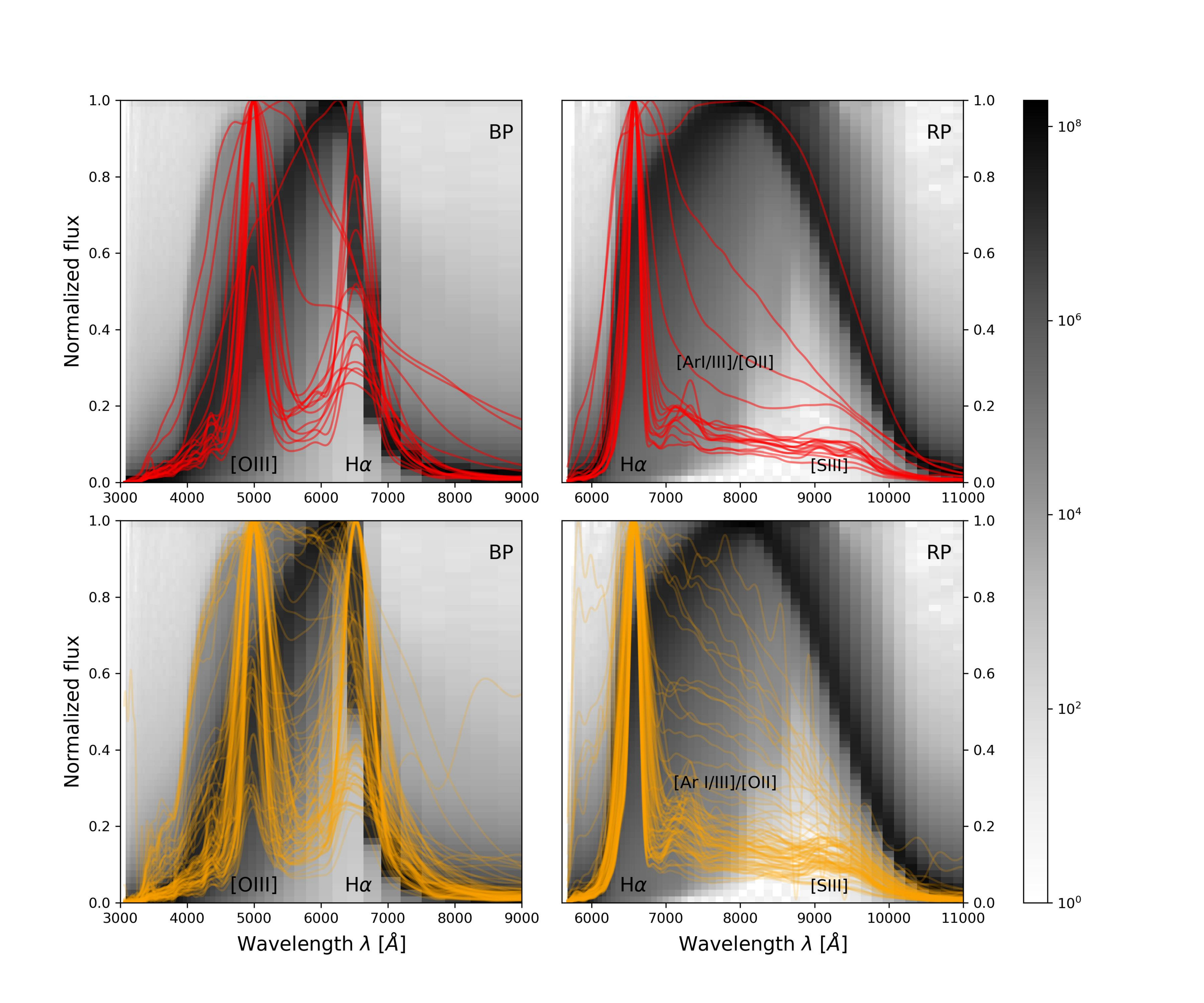}
\caption{Selected individual Gaia spectra plotted over a density map of 20 million flux-normalised Gaia BP/RP spectra. \textit{Top row}: Individual Gaia spectra of our compact PNe literature sample (red lines, see Sect.~\ref{sec:gaiaPNeSED}). The BP spectra are distinguished by strong [\ion{O}{III}]$\lambda\lambda4960,5008$ and H$\alpha$ emission. The RP spectra show H$\alpha$ emission, with secondary peaks around 7100--7300\,\AA\ from [\ion{Ar}{I/III}] and [\ion{O}{II}]; the [\ion{S}{III}]$\lambda\lambda9071,9533$ lines are also discernible. One of the PNe (G334.8$-$07.4) has a strong stellar continuum, following the bulk of the Gaia SEDs. \textit{Bottom row}: Same as above, but showing the candidates we selected blindly from all 200 million Gaia sources matching our search criteria for strong line emitters (Sect.~\ref{sec:searchGaia2}).}
\label{fig:gaia_spectra}
\end{figure*}

\subsection{Searching halo PNe in J-PLUS and S-PLUS data\label{sec:PNhalo}}
In the Galactic halo PNe are very rare, and have been the target of systematic searches before \citep[e.g.][in Sloan Digitial Sky Survey spectra]{yuan2013}. \cite{gutierrezsoto2020} searched the Javalambre and Southern Photometric Local Universe Survey data (J-PLUS and S-PLUS, respectively), using a combination of broad- and narrow-band  photometry. However, no additional useful compact PNe could be identified in an area of 1190\,deg$^2$. Together, we extended the search to yet unpublished S-PLUS data, and one potential unresolved PN candidate was identified, STRIPE82-0130.035257, albeit with comparatively weak signal in the narrow-band filters. We kept this source in our sample for spectroscopic follow-up, but the line emission turned out to be a transient event at the time of the S-PLUS observations. For completeness, the line-free spectrum of the white dwarf (WD) is included in our spectral atlas. 

\section{\label{sec:PNGaia}Selecting compact PNe in Gaia BP/RP spectra}
With the release of Gaia DR3 \citep{gaia2016,gaiaDR3} we have access to $200$ million sources over the full sky with low-resolution Gaia blue photometer (BP) and red photometer (RP) spectra. They cover the 330--670\,nm and 620--1050\,nm wavelength ranges, respectively \citep{evans2018}, facilitating a systematic search for hitherto unknown, compact PNe in the Galactic halo and elsewhere. As it turns out, a single Gaia parameter -- the ${\rm RP}_1$ shape coefficient (Sect.~\ref{sec:gaiaPNeSED}) -- is efficient to select compact H$\alpha$ emitters.

PNe generally have strong H${\alpha}$ and [\ion{O}{III}] emission, with H${\alpha}$ covered by both BP and RP spectra, and [\ion{O}{III}] by the BP spectra, only. For compact PNe, a sufficient amount of emission-line flux will enter the BP/RP extraction windows \citep{gaia2016}, distinguishing their spectra from those of normal stars.  We note that PN-related Gaia work exists; however, these are not focused on finding new PNe, but on identifying central WDs, and determining their distances and multiplicity \citep[e.g.][]{stanghellini2020,chornay2020,chornay2021a,chornay2021c,gonzalez2021}.

\subsection{\label{sec:gaiaPNeSED}Gaia SEDs of compact PNe}
To verify the detectability of strong emission lines in Gaia BP/RP spectra, and to develop suitable selection criteria, we compared the Gaia spectra of 17 of our HASH PNe against those of 20 million randomly selected Gaia sources. Three of our HASH PNe do not have BP/RP data within Gaia DR3, perhaps due to automated selection processes for the Gaia extraction windows \citep{gaia2016}, meaning a complete sample of compact PNe cannot be extracted from Gaia DR3 alone.

To better understand the Gaia data, we converted the Gaia pseudo-wavelengths -- an arbitrary unit from the photometers -- to physical wavelengths following \cite{deangeli2022} and \cite{montegriffo2022}. As can be seen in the top panels of Fig.~\ref{fig:gaia_spectra}, our compact PNe are clearly distinguished by strong [\ion{O}{III}] and H$\alpha$ emission, whereas most Gaia sources have broad spectral energy distributions (SEDs) that peak around 6150\,\AA\ in BP and 7900\,\AA\ in RP, respectively. The RP spectra of the PNe even reveal the presence of considerably weaker lines at longer wavelengths. The only exception is PN G334.8$-$07.4 with very strong continuum and thus relatively weak lines in its normalised spectrum.

Spectra in the Gaia archive are encoded by a number of shape coefficients parameterising the SED; these coefficients can be converted to the BP/RP spectra shown in Fig.~\ref{fig:gaia_spectra} with the Python {\tt Gaiaxpy} package. The 1st-order coefficient -- assuming 0 indexing -- provides a good indication of the presence of a single narrow peak above the continuum \citep[see][for more information about Gaia shape coefficients]{riello2021}. In Fig.~\ref{fig:gaia_pn} we plot the 1st order of both the normalised BP and RP coefficients (${\rm BP}_{1}$, ${\rm RP}_{1}$), for all 200 million Gaia spectra; most sources are confined to a narrow strip in the (${\rm BP}_{1}$, ${\rm RP}_{1}$) parameter space. All but one (PN G334.8$-$07.4) of our compact PNe have ${\rm RP}_{1}<-0.67$, falling well below this strip. 

The range of ${\rm BP}_1$ covered by our PNe is less well confined than 
${\rm RP}_1$. Plausibly, this is because the BP spectra are bimodal due to the simultaneous capture of the [\ion{O}{III}] and H$\alpha$ lines, and thus ${\rm BP}_1$ is insufficient to describe the essential shape of the BP spectra. We find that ${\rm BP}_3$ and ${\rm BP}_6$ are more susceptible to the bimodal nature of our BP spectra, with suitable cuts of ${\rm BP}_3 > -0.002$ and ${\rm BP}_6 < -0.1$. Thus, a multi-parametric filter could be built for more efficient selection of emission-line objects with specific SEDs. In our case, including either ${\rm BP}_3$ or ${\rm BP}_6$ would reduce the number of candidates (see Sect.~\ref{sec:searchGaia2}) by about 50\%, but did not result in new targets that were not included already using ${\rm RP}_1$ alone. Hence we did not pursue multi-parametric filters further in this paper, but we recommend to consider them for similar searches. 

\subsection{Comparison with other Gaia PNe samples\label{sec:otherGaiaPNeSamples}}
For a qualitative comparison of other PNe with ours in the (${\rm BP}_{1}$, ${\rm RP}_{1}$) space, we cross-matched all Gaia sources with those from \cite{chornay2021a} (Gaia PN central star distances). The 714 matches are shown as blue crosses in Fig.~\ref{fig:gaia_pn}, mostly following the bulk of the Gaia sources. This means that their SEDs are dominated by the central stars' continuum emission, which is expected because most PNe have spatial extents considerably larger than the Gaia extraction windows.

A small fraction of the matches extends to lower values of ${\rm RP}_{1}$, suggesting increasing relative contributions from narrow-line emission, to the point where the nebular emission dominates the stellar continuum seen by Gaia. These sources are then missed by \cite{chornay2021a}, who match HASH PNe against Gaia PN central-star candidates with a blue continuum. Thus their PNe approach our HASH sample in the (${\rm BP}_{1}$, ${\rm RP}_{1}$) parameter space, but do not infuse it.

\subsection{Searching for unknown compact PNe in Gaia\label{sec:searchGaia2}}
\begin{figure}[t]
\centering
\includegraphics[angle=0,width=1.0\hsize]{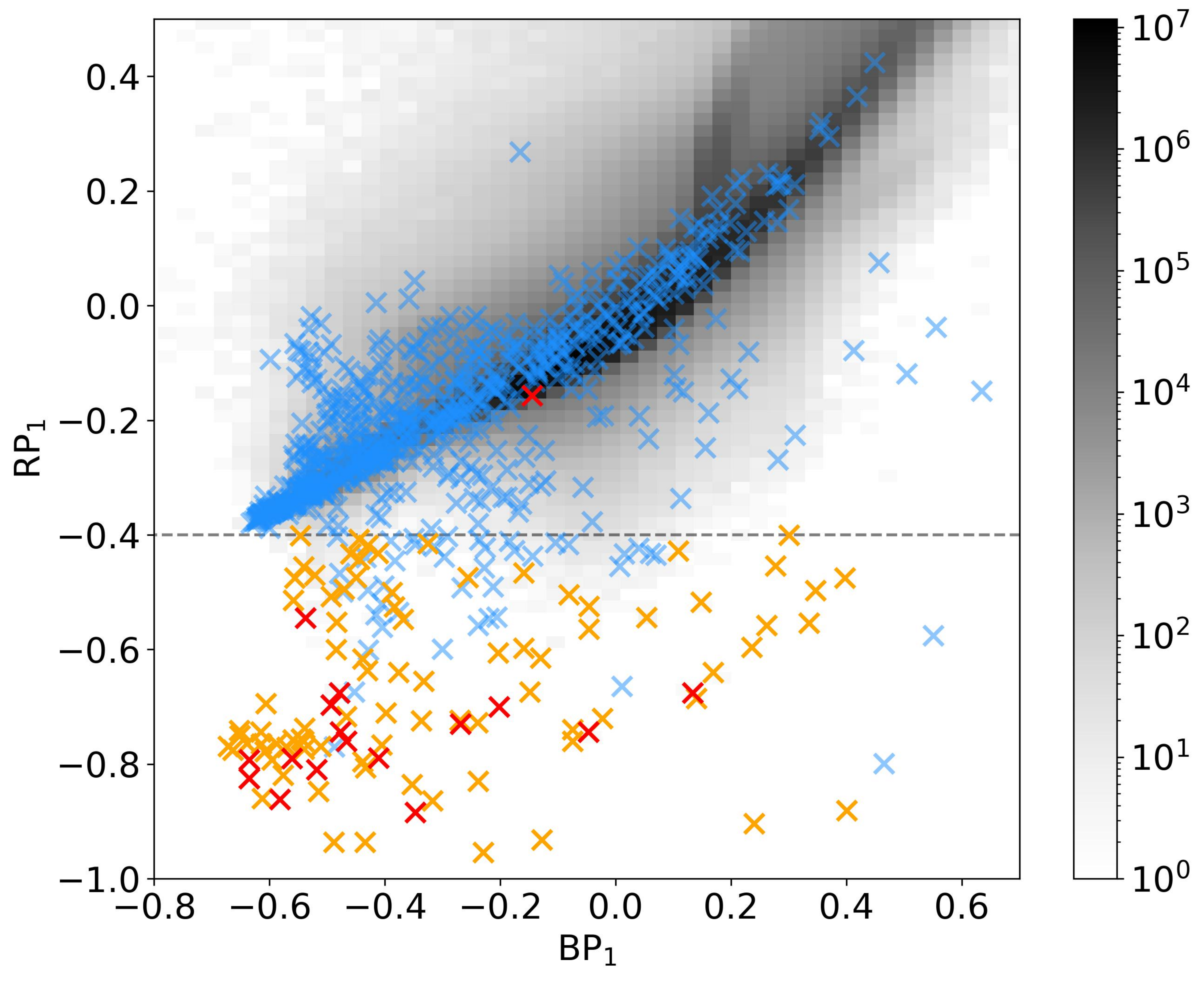}
\caption{Normalised 1st-order shape coefficients (assuming 0 indexing) of the BP and RP Gaia spectra. The logarithmic density of all 200 million Gaia sources in this space is indicated by the grey cells. Our sample of compact PNe (Table~\ref{table:observations}) is shown by the red crosses if they had Gaia DR3 BP/RP spectra, and known or candidate PNe from \cite{chornay2021a} by light blue crosses. The latter follow the bulk of the Gaia sources, suggesting that their SEDs are continuum-dominated; a small subset shows low ${\rm RP}_{1}$ values and approach our sample of compact PNe. Compact PNe candidates with ${\rm RP}_1<-0.4$ are shown in orange, but were rejected/disqualified upon further inspection  (Sect.~\ref{sec:searchGaia2}).}
\label{fig:gaia_pn}
\end{figure}

To counter any incompleteness of HASH concerning compact PNe, we conducted a blind search in the Gaia data. We focused on sources below the tip of the main strip of the Gaia population, that is ${\rm RP}_{1}<-0.4$ (dashed line in Fig.~\ref{fig:gaia_pn}). This cut selects 795 sources out of the 200 million in Gaia DR3.

Thirty-nine of these sources are recorded in HASH, in addition to those that we already selected in Sect.~\ref{sec:sampleselection}. Of those 39 PNe, 35 have size estimates in HASH and show strong emission lines in the Gaia spectra. Only one, NGC 6833, has a sub-arcsecond diameter, available HST data \citep{wright2005}, and favourable morphology. However, its field is very crowded and its Euclid visibility function is similar to those of other PNe for which we already had spectra taken. Thus it was not considered further, together with the four PNe without size estimates that have considerably higher crowding than NGC 6833. 

Besides the 17 included in our sample and the 39 Galactic HASH PNe, 739 candidates remained. To distinguish genuine sources with H${\alpha}$ and [\ion{O}{III}] emission from sources with (1) peculiar spectra, and (2) spectra with redshifted peaks (for example, AGN), we defined a flux ratio, $R$. It is evaluated for wavelengths 500/560\,nm and 656/700\,nm, corresponding to zero-redshifted [\ion{O}{III}] and H${\alpha}$ and their nearby continuum. Requiring $R>1.0$ for both lines, we identified -- and visually verified -- 86 Gaia sources with potential [\ion{O}{III}] and H${\alpha}$ emission. These sources are shown in Fig.~\ref{fig:gaia_spectra} as orange curves, and in Fig.~\ref{fig:gaia_pn} as orange crosses.

A cross-match with Vizier \citep{ochsenbein2000} revealed that 56 sources are indeed confirmed PNe, many of which are in the LMC and SMC that are already well-covered by our literature sample. Ten sources are PN-like in nature, classified as possible PN or PN candidates. Eight systems contain WDs, a number of which have experienced classical or dwarf-nova eruptions. Only two sources are non-stellar in nature, and two are classified as stellar. The remaining eight sources are not classified but appear stellar in nature. Five of these could not be cross-matched, and could therefore be genuine, previously unknown PNe in the Galaxy, albeit at high crowding levels. Their Gaia DR3 object numbers are listed in Table~\ref{table:pn}, which also contains all other sources described here. Apart from the LMC and SMC sources, only two -- an AGN and a WD candidate -- are located at higher Galactic latitudes $|b|>\ang{30;;}$. Their Gaia spectra display a prominent continuum without particularly strong emission peaks. 

\section{\label{sec:observations}Observations and reduction}
\subsection{Instrument choice}
Figure~\ref{fig:PNdistribution} shows that all of our ultra-compact PNe -- with the exception of DdDm-1 -- are accessible from the Southern Hemisphere, making X-SHOOTER \citep{vernet2011} at the Very Large Telescope (VLT) our preferred choice. X-SHOOTER offers medium-resolution spectroscopy with $R=4100$--$6500$ in the $0.3$--$2.5$\,\micron\;range, with suitable slit widths. The instrument has an atmospheric dispersion corrector and is flexure-compensated, which is highly welcome for our calibration purposes. 
In Sect.~\ref{sec:offset} we investigate the accuracy of that flexure-compensation mechanism. X-SHOOTER is offered in service mode that allows to request excellent seeing conditions.

For DdDm-1 we chose the Gemini Near-Infrared Spectrograph \citep[GNIRS;][]{elias2006b,elias2006a} at the Gemini North telescope with $R\sim2000$ in the $1.03$--$1.80$\,\micron\;range. To cover shorter NISP wavelengths down to $0.93$\,\micron, we chose the Gemini Multi-Object Spectrograph \citep[GMOS-N;][]{hook2004} with $R\sim2200$.

\subsection{\label{sec:xshooter_obs}Observations with VLT/X-SHOOTER}
All PNe but DdDm-1 were observed with VLT/X-SHOOTER from the Southern Hemisphere (Fig.~\ref{fig:PNdistribution}), through Director's Discretionary Time programme IDs 108.MQ23.001 and 110.23Q7.001 in typical seeing conditions of \ang{;;0.4}. Since the emission lines have high fluxes, the exposures were dominated by the technical overhead and could tolerate high background levels. We used the \ang{;;1.3}, \ang{;;1.2}, and \ang{;;1.2} wide long slits for the X-SHOOTER UVB ($300$--$560$\,nm), VIS\footnote{In this paper, VIS refers to the X-SHOOTER visible arm, not to the visible instrument VIS \citep{cropper2012} onboard \Euclid.} ($550$--$1020$\,nm), and NIR ($1020$--$2480$\,nm) arms, respectively, conveniently accommodating the PNe's spatial extents (Sect.~\ref{sec:importanceObsSetup}). The spectral resolutions in these arms are $R=4100$, 6500, and 4300, respectively.

The slit centroiding errors are below \ang{;;0.1} according to the X-SHOOTER user manual, contributing a systematic uncertainty of 0.1\,\AA\ in the UVB and VIS arms, and 0.3\,\AA\ in the NIR arm. The slit centroiding error cannot be directly assessed in our data, since X-SHOOTER does not save through-slit images during target acquisition. These systematic errors ultimately limit the absolute wavelength accuracy of our spectral atlas.

The individual exposure times in the three arms were guided by the available H$\beta$ line fluxes (for references, see Sect.~\ref{sec:sampleselection}). We used the {\tt NEBULAR} spectral synthesis code \citep{schirmer2016} to estimate the near-infrared hydrogen and helium line fluxes, using an electron temperature of $T_{\rm e}=10\,000$\,K, an electron density $n_{\rm e}=10\,000$\,cm$^{-3}$, and a helium abundance ratio by parts of 0.1, which are typical for PNe \citep{zhang2004}. This is simplistic given the broad range of temperatures, densities, and excitation zones present in PNe \citep[e.g.][]{martins2002}, yet sufficient to get purposeful X-SHOOTER exposure times. The near-infrared lines were our principal targets and drove the X-SHOOTER configuration, with some balancing concerning technical constraints from the UVB and VIS exposure times and readout electronics. The exposure times were sufficiently long -- as judged by the short acquisition images -- to render seeing effects isotropic, that is the measured line-image centroids are unbiased by seeing. A summary of the individual X-SHOOTER observations is given in Table~\ref{table:observations2}.

\input{PN.tex}
\input{obs_table.tex}

Data were automatically reduced with the X-SHOOTER pipeline \citep{modigliani2010,goldoni2012}, providing the rectified and wavelength-calibrated 2D spectra as well as the flux-calibrated 1D spectra on the ESO archive. We inspected the quality flags for the reduction, and ensured the cross-dispersion profile used by the pipeline for extraction looked acceptable. For X-SHOOTER, the wavelength and spatial scales of the 2D spectra are calibrated simultaneously using a mask with nine equidistant pinholes and a ThAr lamp. For our data, this provided an uncertainty (calculated by the pipeline) on the wavelength solution of the order of $10^{-4}$\,\AA\ (with a RMS on the residuals of $10^{-2}$\,\AA), along with systematic and statistical errors of $10^{-2}$ and $10^{-6}$\,\AA, respectively. We also investigated the accuracy of the arc lines used by the X-SHOOTER pipeline against the National Institute of Standards and Technology (NIST) Atomic Spectra Database\footnote{\url{https://www.nist.gov/pml/atomic-spectra-database}} v5.9 \citep{kramida2021}, and The Atomic Line List\footnote{\url{https://www.pa.uky.edu/~peter/atomic/}} v2.04. We found negligible differences of around $10^{-2}$\,\AA\ between the wavelengths used by the X-SHOOTER pipeline and those in the databases. These errors are far below our 5\,\AA\ requirement for \Euclid, and thus we accepted the pipeline reduced spectra. A detailed description of the reduction steps performed by the pipeline can be found in the X-SHOOTER user manual.

\subsection{Observations with Gemini/GNIRS}
DdDm-1 was observed through `fast turnaround' programme GN-2022A-FT-215 with an atmospheric seeing of \ang{;;0.43}. We used the \ang{;;1.0} wide long slit with the 110 lines/mm grating and the short camera. The GNIRS wavelength range of interest (1.03--1.80\,\micron) was covered with six different  central wavelengths of 1.065, 1.17, 1.22, 1.30, 1.56, and 1.68\,\micron. The first four wavelength settings were observed each with a single ABBA nodding pattern keeping the target on the slit; at each nod position, two 20\,s exposures were taken and coadded, totalling 160\,s integration time. For the last two wavelength settings the ABBA pattern was executed twice, yielding 320\,s integration time. Observations were taken at the parallactic angle. Flats and arcs were taken before each science observation. A summary of the GNIRS observations is given in Table~\ref{table:observations2}.

Due to the selection of a wrong order-blocking filter in the non-standard 1.17\,\micron\;setting, the 1.11--1.22\,\micron\;interval is not usable in our data. Our X-SHOOTER PN spectra show that this range is devoid of suitably bright emission lines for \Euclid calibration, and thus these observations were not repeated. The spectral dispersion for each setup is 0.93, 1.12, 1.40, and 1.39\,\AA\,pixel$^{-1}$, for central wavelengths of 1.065, 1.22, 1.30, 1.56, and 1.68\,\micron, respectively.

We note that with the GNIRS cross-dispersed mode (six simultaneously mapped spectral orders) one could cover the $0.80$--$2.5$\,\micron\;range, albeit only with four separate central wavelength settings in case of the high-resolution 110 lines/mm grating. This would divide the wavelength range into 24 individual chunks, each carrying a comparatively low number of arc lines for wavelength calibration. We decided against this mode in favour of conceptually simpler, long-slit observations. They cover a larger, contiguous wavelength range, and thus more arc lines are available for better wavelength calibration.

Data were reduced with the {\tt Gemini/GNIRS IRAF} package\footnote{\url{https://www.gemini.edu/instrumentation/gnirs/data-reduction}}. Even with the chosen setup, the spectra still cover a comparatively small wavelength coverage -- and thus have a lower number of arc lines --  so the uncertainty of the wavelength calibration is considerably larger than for X-SHOOTER. To estimate the error on the wavelength solution, we compared the wavelength of the lines identified in the arc spectra to The Atomic Line List, finding a median RMS of around 1\,\AA\ across all used GNIRS setups. We also compared the reference arc wavelengths used by IRAF against The Atomic Line List. Differences are small, typically 0.04\,\AA, suggesting that the lack of lines is the dominant contributor to the error of the wavelength solution.

Although residuals of 1\,\AA\ are below our 5\,\AA\ requirement, we reverted to OH sky lines from \cite{rousselot2000} for wavelength calibration. About three times as many sky lines than arc lines could be used, resulting in a 12\% lower RMS of the wavelength residuals; we adopted this improved wavelength solution.

To flux-calibrate the GNIRS spectra, we fitted a black-body SED with a temperature of 10\,000\,K to the telluric standard star, HIP\,81126 \citep{gaiaDR3}. The black-body spectrum was then renormalised to match the telluric's 2MASS \citep{cutri2003} $H$-band magnitude that was converted to the AB mag system following \cite{blanton2007}.

\subsection{Observations with Gemini/GMOS}
Our observations of DdDm-1 with GMOS were also a part of the same programme as GNIRS. Data were taken in dark and clear conditions with an atmospheric seeing of \ang{;;0.41}. We used the R831 grating with the RG610 order-sorting filter, a central wavelength setting of 925\,nm, and a \ang{;;1.0} wide slit. This setup provides a spectral resolution of about $R=2200$, and covers the 805--1043\,nm range. Observations were taken at the parallactic angle. 

We used the GMOS `Nod\&Shuffle' mode \citep{glazebrook2001}, alternatingly exposing and storing two spectra of the same source on the detector every 30\,s. In this way, a very accurate subtraction of airglow lines could be achieved in software later-on. The total integration time was 480\,s, and the data were reduced with the {\tt Gemini/GMOS IRAF} package. We compared the wavelength of the lines identified in the arc spectra to NIST, finding a RMS of 0.7\,\AA. This is higher than for X-SHOOTER ($10^{-2}$\,\AA), but still well below our requirement. Thus we accepted the wavelengths used by {\tt Gemini/GMOS IRAF} and did not attempt to feed it a custom-made alternative list of NIST reference arc wavelengths.

To flux-calibrate the GMOS spectra, we fitted a black-body SED with a temperature of 81\,300\,K \citep{Latour2015} to the GMOS telluric standard star, BD+28 4211. The black-body spectrum was then renormalised to match the telluric's Pan-STARRS \citep{chambers2016} $z$-band magnitude. After telluric correction, we rescaled the science spectrum of DdDm-1 to match the flux density of the GNIRS spectrum in the wavelength interval common to both data sets. 

\subsection{\label{sec:telluriccorrection}Telluric correction}
Telluric standard stars were observed during standard nightly operations for both the VLT and Gemini programmes. For X-SHOOTER this usually occurred within four hours and 0.3 airmass of the targets; we selected the telluric spectrum closest in airmass to the science spectrum, avoiding underexposure and saturation that would adversely affect the correction. For GNIRS a standard star was observed close in airmass and hour angle directly after the PN, for each configuration. For GMOS, the telluric was taken a few days earlier, compliant with the Gemini baseline calibration programme.

For GNIRS, we computed the telluric correction using a black-body model with a temperature of 10\,000\,K \citep{gaiaDR3} to describe the continuum of the telluric. We used an error of 180\,K \citep{fouesneau2022} for the model, and propagated all errors to the science spectra. Likewise, for 
GMOS we computed the telluric correction using a black-body model with a temperature of $\num{81300}\pm\num{1219}$\,K for the hot subdwarf \citep{Latour2015}. Again, all errors were propagated to the science spectrum.

\subsubsection{Correcting atmospheric absorption with {\tt MOLECFIT}}
For X-SHOOTER, the large wavelength coverage allows a more accurate computation -- compared to GNIRS and GMOS -- of the telluric correction with {\tt MOLECFIT} \citep{smette2015,kausch2015}. This procedure includes the following steps: (1) normalising the flux of the telluric standard; (2) determining the molecular column densities by fitting absorption on small, unsaturated wavelength ranges free of intrinsic stellar features, and where those molecules dominate; (3) computing the correction over the full wavelength range for the included molecules and water vapour content; (4) applying the correction to the science spectrum accounting for airmass differences. 

Telluric corrections are computed for VIS and NIR, but not for UVB that does not contain considerable atmospheric absorption lines. For VIS, we fitted for \chem{H_2O} and \chem{O_2} in the small wavelength ranges of 0.91--0.92, 0.69--0.70 and 0.93--0.94\,\micron; this corresponds to step (2) in the previous paragraph. These molecules were then also used to compute the telluric correction for the full VIS wavelength range (step 3). For NIR, we fitted for \chem{H_2O}, \chem{CO_2}, \chem{CH_4} and \chem{O_2} in the small wavelength ranges of 1.12--1.13, 1.26--1.27, 1.47--1.48, 1.8--1.81, 2.06--2.07, and 2.35--2.36\,\micron. As no wavelength range is available to fit for CO without contamination from other molecular species, we fixed the value to a relative column density of 1.0. This is usually close to the true value based on the standard MIPAS profile \citep[Michelson Interferometer for Passive Atmospheric Sounding, see e.g.][]{gessner2001,wang2004}. To compute the telluric correction for the full NIR arm, we included both the fitted and fixed molecular species.

\subsubsection{Estimating {\tt MOLECFIT} uncertainties}
{\tt MOLECFIT} computes and applies a telluric correction using the best-fit column densities of the included molecules, but it does not provide an error estimate of the correction. This is problematic, because the deep atmospheric absorption bands in the $0.9$--$2.0$\,\micron\;range fully overlap with the wide Euclid spectral bands (Fig.~\ref{fig:fp_pn}); \Euclid is not affected by atmospheric absorption, and thus we need reliable line calibrators also in these absorption bands, such as heavily absorbed Pa$\alpha$ at 18\,756\,\AA. The subject is complicated by the fact that inside the absorption bands the correction factors may become very large; they also vary sharply as a function of wavelength, giving rise to spurious features that might be wrongly interpreted as emission lines.

The best-fit column densities from {\tt MOLECFIT} have an associated error based on the fit within the specified wavelength ranges. We exploited this to estimate the error of the telluric correction, by computing 1000 telluric corrections for each X-SHOOTER science spectrum, using random draws from the best-fit parameter distributions. The standard deviation of these 1000 models at each wavelength then estimates the correction error, so that we could reliably select true, significant emission lines also in absorption bands (see e.g.\ Fig.~\ref{fig:PN_xshooter_spectrum}).

\subsubsection{\label{sec:molecfitwave}Additional wavelength correction with {\tt MOLECFIT}}
During the fitting, {\tt MOLECFIT} rebins the wavelength of the model spectrum -- that accounts for all the fitted molecules -- to the telluric spectrum. The difference between the rebinned model wavelengths and the input wavelengths of the telluric spectrum offers an additional wavelength correction. Its reliability depends on the quality of the fit, and on the interpolation outside of the fitting wavelength ranges. Thus, {\tt MOLECFIT} does not automatically apply this correction to the science spectra.

We investigated this correction and found linear trends, when using a 1st-order polynomial, with maximum corrections of 0.2\,\AA\ and 0.5\,\AA, for the VIS and NIR arms, respectively. Higher-order corrections were poorly constrained at the ends of the spectra, and thus not pursued further. Although the corrections suggested by {\tt MOLECFIT} are much smaller than our requirement of 5\,\AA, we checked whether they could reduce much smaller, yet significant offsets visible in the data (Fig.~\ref{fig:PN_wave_offset}). We find that for some PNe the correction in either VIS and/or NIR decreased the observed offsets considerably. For others however, the correction provided little improvement, and could even degrade the offsets further. We could not find a relation with airmass, seeing, and time of observation passed between the telluric and the PN observations. Possibly, we see uncorrected flexure and temperature effects in X-SHOOTER. Based on Fig.~\ref{fig:PN_wave_offset}, we applied the correction only if the offsets were reduced, independently for the VIS and NIR arms. For details, see Sect.~\ref{sec:offset}.

\subsection{\label{sc:air2vac}Air-to-vacuum wavelength conversion}
Our next step was to convert air wavelengths to vacuum wavelengths for the X-SHOOTER and GMOS spectra; the GNIRS pipeline readily uses vacuum wavelengths. The X-SHOOTER pipeline uses NIST air wavelengths for the ThAr arc lamp for a temperature of $20^\circ$\,C and an absolute pressure of 101.325\,kPa, whereas in reality X-SHOOTER operates partially in vacuum and at lower ambient pressure and temperature. Based on measurements of several thousand arc lines, the pipeline-provided air wavelengths are indeed as if obtained under NIST standard ambient conditions, with minimal systematics.

For the air-to-vacuum conversion we used the modified Edl\'en equation \citep{birch1993,birch1994} for the refractive index $n$ of air\footnote{See also \url{https://emtoolbox.nist.gov/Wavelength/Documentation.asp}}, with improved numerical precision as given by Nikolai Piskunov\footnote{\url{https://www.astro.uu.se/valdwiki/Air-to-vacuum\%20conversion}},
\begin{alignat}{2}
    \lambda_{\rm vac} & = n\,(\lambda_{\rm air})\,\lambda_{\rm air}\\
    n(\lambda_{\rm air}) & = 1+ a_1 + \frac{a_2}{a_3-s^2} + \frac{a_4}{a_5-s^2}\\
    s & = 10\,000\,\text{\r{A}} / \lambda_{\rm air}\\
    a_1 & = 8.336624212083\times10^{-5}\\
    a_2 & = 2.408926869968\times10^{-2}\\
    a_3 & = 130.1065924522\\
    a_4 & = 1.599740894897\times10^{-4}\\
    a_5 & = 38.92568793293\;.
\end{alignat}
We emphasise that this expression is strictly valid only for NIST ambient conditions, and note that it is also used in the Vienna Atomic Line Database \citep[VALD;][]{piskunov1995,ryabchikova2015}. Expressions including temperature and pressure dependencies are given in \cite{birch1993,birch1994}, which are entirely negligible for our purposes.

\subsection{\label{sc:heliocentric}Heliocentric wavelength correction}
The observed wavelengths are still modulated by up to $\pm1.5$\,\AA\ for a PN in the ecliptic plane and at 1.5\,\micron\;wavelength, due to Earth's revolution around the Sun. To transform the wavelengths to a heliocentric system, we used the {\tt radial\_velocity\_correction} function from the {\tt astropy.coordinates} package \citep{astropy2013,astropy2018}, providing a precision of 3\,m\,s$^{-1}$ or $1.5\times10^{-4}$\,\AA. This correction also accounts for Earth's rotation.

\subsection{\label{sec:specjoining}Joining the wavelength ranges into a single spectrum}
Finally, we combined the different spectra from the X-SHOOTER arms (and in case of DdDm-1, the GMOS and GNIRS settings) into a single spectrum. The overlapping areas between spectra were joined by simple cuts in wavelength: a spectrum was truncated once its uncertainty consistently exceeded that of its adjacent spectrum. We found that this occurs fairly consistently across targets for X-SHOOTER; the UVB spectrum extends until 5562\,\AA, and the VIS spectrum until 10\,203\,\AA. In the NIR, the spectra are truncated at $24\,000$\,\AA, above which the errors exceed the flux. For GMOS, the wavelengths extend up to 10\,440\,\AA.

It is important to note that we did not resample the wavelength axis in the joint 1D spectra, to avoid the introduction of uncertainties from resampling. We retained the native dispersion plate scale of 0.2\,\AA\ pixel$^{-1}$ for UVB and VIS, and 
0.6\,\AA\ pixel$^{-1}$ for NIR. An example X-SHOOTER spectrum is shown in Fig.~\ref{fig:PN_xshooter_spectrum}. For GMOS the spectral dispersion is 0.38\,\AA\,pixel$^{-1}$; while for GNIRS it varies between 0.93 and 1.4\,\AA\,pixel$^{-1}$, depending on the spectral order. 

\subsection{\label{sc:Lines}Continuum subtraction and line detection}
We estimated the continuum using a two-step process. First, we computed the continuum by finding the median value every 500 data points (100\,\AA\ for UVB and VIS, 300\,\AA\ for NIR), and then interpolated between these points over the full wavelength range using cubic splines. After automatic line detection and masking (see next paragraph) on the continuum-subtracted spectrum, the continuum was recomputed before repeating the line detection. For GMOS and GNIRS, the step size for the median computation was 50 data points (20\,\AA\ for GMOS and 50\,\AA\ for GNIRS).

For line detection, a first automatic pass was done using the {\tt find\_lines\_threshold} function using a noise factor of three from the {\tt Astropy} {\tt specutils.fitting} package. For each automatically detected line, we fitted a 1D Gaussian (including a constant additive term to account for any local residuals in the continuum subtraction) to the spectrum using the 100 pixels centred on the line. To reject false positives from detector effects and other broader features, we required the fitted width of the line to be within 0.2--3\,\AA. To remove spurious fits, we also required the uncertainty of the central line wavelength to be less than the standard deviation of the fitted Gaussian. We incorporated the intrinsic wavelength errors by adding in quadrature the statistical solution errors to the error on the central wavelength of each line when available.

As the last step for line detection, we used an interactive plot of the continuum-subtracted spectrum to visually verify the automatically detected lines. For features, such as cosmic rays or those caused by poor sky subtraction, incorrectly detected as a line, or lines with poorly determined centres, we deleted the automatically determined positions. For missed emission lines or refitting lines, we selected the position of the line on the plot, allowing us to manually enter the wavelength range on which to fit a Gaussian. We also inspected the 2D spectrum, (1) as a sanity check for identified lines, (2) to help with the identification of low-S/N lines, and (3) to identify any additional line structures not evident in the 1D spectrum (see Fig.~\ref{fig:PNlinemap}). 

\subsection{Line-flux computation\label{sec:linefluxcomputation}}
To compute line fluxes, we integrated over the best-fit Gaussian within the $\pm3\sigma$ interval, with $\sigma$ being the standard deviation of the individual line's best-fit Gaussian. The total line fluxes, FWHM, and effective heliocentric vacuum wavelengths are available online. Close line blends are not resolved since we fitted a single Gaussian, only. Flux estimates for line blends will have lower accuracy. We also note that some line fluxes could be underestimated for PN with spatial extends larger than the 1D extraction aperture.
    
\subsection{Line identification in NIST and The Atomic Line List\label{sec:lineidentification}}
We primarily identified lines in NIST, and in the few cases where no obvious transition was found, we searched The Atomic Line List. In particular for faint, allowed lines, the identification can be ambiguous. Due to similarities in their electronic configurations, the transitions for some atomic and ionic species can also be clustered in wavelength, such as for Fe, He and O, all common nebular lines. In these cases, we selected the element that is more commonly represented elsewhere in the spectrum, for example PN G295.3$-$09.3 is particularly rich in Fe lines (Fig.~\ref{fig:hotcontinuum}). In a few cases, we observe small but significant offsets from the tabulated laboratory restframe wavelength. One such noteworthy, bright line is H8 at 3890.16\,\AA, which is consistently observed about 0.2\,\AA\  bluer in the PNe than expected, most likely due to blending with \ion{He}{I} at 3889.75\,\AA.

The primary goal of our work was to measure accurate effective wavelengths of line images; line identification was secondary. There will certainly be misidentified lines in our spectral atlas, for instance a less common line from C could be mistaken for a more common Fe line, in particular if their separation is less than 0.1\,\AA; very rarely, no suitable atomic or ionic species could be identified. Likewise, no attempt was made to accurately label line blends. In case of numerous transitions of the same ionic species blended into one line, such as \ion{He}{I} at 10\,915.98\,\AA, our spectral atlas uses an approximated wavelength that is sufficiently accurate to look-up these transitions in NIST. 

Throughout this paper we use the theoretical Ritz wavelengths computed in NIST instead of their observed wavelengths. The latter are available only for a subset of the transitions. The difference between the two is negligible for our purposes: 97\% of the NIST observed and theoretical wavelengths -- up to Fe and within 0.3--2.5\,\micron\;-- agree within 0.1\,\AA, and 76\% agree within 0.01\,\AA. 

\section{\label{sc:Res}Results}
\subsection{\label{sec:dataproducts}A spectral atlas of emission lines}
The main data product of this paper is a ground-based UV-NIR (0.30--2.40\,\micron) atlas of effective, heliocentric, vacuum wavelengths of emission lines in compact PNe. The atlas includes total line fluxes and widths. It serves as a primary reference to calibrate the dispersion laws of \Euclid's low-resolution slitless spectroscopy mode covering the 0.93--1.89\,\micron\;range. All PNe have archival HST imaging and / or slitless spectroscopy. We note that the wavelength step size in the spectra is not constant, it depends on the spectrograph arm (see Sect.~\ref{sec:specjoining}).

A wide slit was used in the ground-based observations under excellent seeing conditions, (1) to capture all line flux, and (2) to propagate morphological substructures into the measured line centroids and thus effective wavelengths, as \Euclid will see them. The choice of a wide slit has little effect on the reported wavelengths. This is because we selected our PNe to be compact, largely symmetric, and mostly featureless. Their finite spatial extent thus does not bias the measured line centroid compared to a measurement obtained through a narrow slit centred on the PN. Likewise, owing to the typical rotational or point symmetry of these PNe, the slit position angle does not affect the line centroid measured in the collapsed 1D spectrum. Such effects are well within the wavelength uncertainties of our spectral atlas.

The line lists for each PN are available in FITS format at an ESA server\footnote{\texttt{\url{https://euclid.esac.esa.int/msp/refdata/nisp/PN-SPECTRAL-ATLAS-V1}}}, together with the associated full 1D spectra. The tables contain (1) the ionic species within the limitations spelled out in Sect.~\ref{sec:lineidentification}, (2) the laboratory vacuum wavelengths, (3) the observed heliocentric vacuum wavelengths, (4) line fluxes integrated over the model Gaussian(s), (5) a flag providing a goodness of fit estimate, and (6) the line FWHM. An example is shown in Table~\ref{tab:line_flux}. As a transient, STRIPE82-0130.035257 is not included in the line lists (see Sec. \ref{sec:stripe}) but the 1D spectra are available online.

\begin{figure}[t]
\centering
\includegraphics[angle=0,width=1.0\hsize]{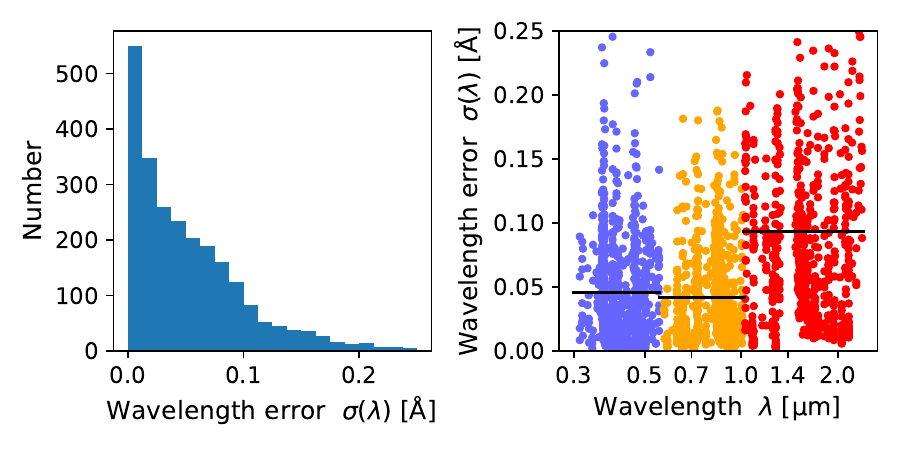}
\caption{Plots showing the statistical line centroid errors for all PNe in our sample. \textit{Left panel}: Distribution of the statistical line centroid errors. \textit{Right panel}: Line centroid error as a function of wavelength. The UVB, VIS and NIR ranges are colour coded; the black horizontal lines show the mean statistical error of 0.046, 0.042, and 0.094\,\AA, respectively. The GMOS/GNIRS data for DdDm-1 follow the same colour-coding and are not distinguished in this plot.}
\label{fig:cumulative_wave_error}
\end{figure}

\subsubsection{Notes about the 2D spectra}
Our data pack also contains 2D spectra, useful for the assessment of weak emission lines identified in the 1D spectra, and other line classification purposes. In case of X-SHOOTER, this means an identical copy of the automatically produced, pipeline-processed spectra available in the ESO archive at the time of writing. In case of GMOS/GNIRS, this means our own reduction. The 2D spectra are `observed only', that is (1) they are not flux-calibrated, (2) they are not corrected for telluric absorption, (3) the wavelengths are not corrected to a heliocentric system and thus still carry a dependence on the epoch of observation, (4) the wavelengths are not corrected for small residuals as determined by {\tt MOLECFIT} (X-SHOOTER only, see Sect.~\ref{sec:molecfitwave}), and (5) the wavelengths are in air, not vacuum (X-SHOOTER and GMOS only; GNIRS wavelengths are in vacuum).

\begin{table}
    \centering
    \begin{threeparttable}
    \caption{Systematic X-SHOOTER wavelength errors.}
    \smallskip
    \begin{tabular}{|lcc|}
    \hline
        Name & $\Delta\lambda_{\rm lab}({\rm VIS})$ & $\Delta\lambda_{\rm lab}({\rm NIR})$\\
         & [\AA] & [\AA] \\
        \hline
        PN G025.3$-$04.6 & 0.025 & 0.015\\
        PN G042.9$-$06.9 & $-$0.043 & 0.12\\
        PN G205.8$-$26.7 & 0.027 & $-$0.12\\
        PN G295.3$-$09.3 & $-$0.049 & $-$0.12\\
        PN G334.8$-$07.4 & $-$0.040 & $-$0.093\\
        He2-436 & $-$0.062 & 0.060\\
        LMC-SMP-25 & $-$0.019 & $-$0.017\\
        LMC-SMP-31 & 0.011 & 0.14\\
        LMC-SMP-53 & $-$0.095 & $-$0.045\\
        LMC-SMP-58 & $-$0.016 & 0.069\\
        LMC-Sa-104a & 0.32 & 0.53\\
        SMC-SMP-1 & 0.076 & 0.14\\
        SMC-SMP-2 & 0.18 & 0.19\\
        SMC-SMP-8 & 0.10 & 0.051\\
        SMC-SMP-13 & 0.043 & 0.23\\
        SMC-SMP-18 & 0.12 & $-$0.14\\
        SMC-SMP-20 & $-$0.083 & 0.14\\
        SMC-SMP-24 & 0.047 & 0.048\\
        SMC-SMP-25 & 0.12 & 0.51\\

        \hline
    \end{tabular}
    \footnotesize
    \tablefoot{These are systematic wavelength errors still present in the X-SHOOTER spectra, with respect to its UVB channel based on a comparison with NIST laboratory wavelengths.}
    \label{table:wavelengthoffset}
    \end{threeparttable}
\end{table}

\subsection{\label{sec:offset}Statistical and systematic wavelength errors}
The mean statistical wavelength error on the line centroids, measured over all PNe, is about 0.04\,\AA\ for UVB and VIS, and 0.08\,\AA\ for NIR (Fig.~\ref{fig:cumulative_wave_error}). These are factors 50--100 below the Euclid requirement of 5\,\AA. 

The systematic wavelength error has two components. The first component is observational, rooting in the slit centroiding errors. For X-SHOOTER, these cannot be measured as no through-slit image of a source is saved as part of the standard target acquisition procedure. We adopted the estimate of \ang{;;0.1} from the X-SHOOTER user manual, translating to 0.1\,\AA\ for UVB and VIS, and to 0.3\,\AA\ for NIR. In case of GMOS and GNIRS, the slit centroiding error was estimated from the through-slit images to be better than 0.2\,\AA\ for both instruments. 

The second component of the systematic error is instrumental, originating in residual backbone flexure in X-SHOOTER. Effectively, this modifies the slit centroiding error for each X-SHOOTER spectral arm. X-SHOOTER has a built-in flexure compensation mechanism, maintaining the alignment between its three spectrograph arms, and also correcting differential guiding effects from atmospheric refraction. Furthermore, there is a temperature dependence of the UVB arms' focal length that is actively controlled; details can be found in the user manual. 

To estimate this instrumental systematic error -- that is imperfections in the compensation mechanisms -- we first estimated the radial velocity from lines in the UVB channel (Sect.~\ref{sec:radialvelocities}; in case of DdDM-1 we used the GMOS channel). The associated Doppler shift was removed from all line wavelengths, yielding $\lambda^{\rm RV,corr}_{\rm obs}$. We then computed the difference with respect to the NIST laboratory wavelengths,
\begin{equation}
\Delta\lambda_{\rm lab} = \lambda^{\rm RV,corr}_{\rm obs}-\lambda_{\rm lab}\;.
\end{equation}

In a perfectly working instrument, $\Delta\lambda_{\rm lab}$ should be consistent with zero. However, non-zero offsets are found for both VIS and NIR with respect to UVB, summarised in Table~\ref{table:wavelengthoffset}, and visualised in Fig.~\ref{fig:PN_wave_offset}. The systematic errors are fairly random between PN. We did not correct our 1D spectra for these offsets, as their origin is not a priori clear, nor which X-SHOOTER channel is closest to the truth. Likewise, we did not correct the wavelengths in the emission-line tables that are derived from the 1D spectra, to maintain consistency with the primary 1D spectra data products. If needed, the offsets listed in Table~\ref{table:wavelengthoffset} can simply be applied to the wavelengths in the emission-line tables and to the wavelengths the 1D-spectra themselves. One may also choose to correct with respect to the VIS or the NIR channel instead of UVB. Precise wavelength ranges for the various channels are provided in Sect.~\ref{sec:specjoining}, so that these corrections can be applied accurately.

\subsection{Radial velocities\label{sec:radialvelocities}}
Using NIST wavelengths as a reference, we determined individual radial velocity estimates, typically for 10--30 emission lines in the UVB channel that appeared the most stable for X-SHOOTER (see also Sect.~\ref{sec:offset}). The weighted mean of these radial velocities then estimates the systemic radial velocity of that PN, the weights being the inverse variance of the lines' wavelength uncertainty. The lines included have ${\rm S/N}>10$; for DdDm-1 we used lines in the GMOS spectrum with ${\rm S/N}>10$. The results are given in Table~\ref{table:radialvelocity}. The Doppler shift is of course preserved in our spectral atlas of observed reference wavelengths. 

We note that the X-SHOOTER user manual states that the VIS and NIR arms have increasingly better radial velocity accuracy. However, our comparison with NIST lab wavelengths reveals the smallest wavelength residuals for UVB, and the UVB arm also contains more brighter lines for measurement. 

Comparing our radial velocities with those in the literature, we find that while some differ by only $\sim$ 1\,km s$^{-1}$ (e.g. LMC-SMP-31); others show greater differences up to 30 km s$^{-1}$ (e.g. LMC-SMP-25). The causes may vary, from dependencies on the  atomic or ionic line species, to slit centroiding errors \citep[see e.g.][]{reid2006}. We did not investigate these discrepancies further; our spectral atlas can be used for refined measurements. 

\begin{table*}[t]
    \centering
    \begin{threeparttable}
    \caption{Radial velocities for PN.}
    \smallskip
    \begin{tabular}{|lcccc|}
    \hline
        Name & \#lines & ${\rm RV}\pm\sigma_{\rm stat}\pm\sigma_{\rm sys}$ & RV literature & Reference\\
         & (RV) & [km\,s$^{-1}$] & [km\,s$^{-1}$] &\\
        \hline
        DdDm-1      & 37 & $-312.12\pm0.12\pm0.2$ & $-317\pm13$ & \cite{wright2005} \\
        PN G025.3$-$04.6 & 17 & $-91.60\pm0.11\pm0.1$ & &  \\
        PN G042.9$-$06.9   & 26 & $-67.41\pm0.11\pm0.1$ & &  \\
        PN G205.8$-$26.7  & 22 & $71.28\pm0.08\pm0.1$ & &  \\
        PN G295.3$-$09.3 & 21 & $86.74\pm0.11\pm0.1$ & &  \\
        PN G334.8$-$07.4 & 17 & $-61.36\pm0.17\pm0.1$ & &  \\
        He2-436     & 25 & $130.42\pm0.10\pm0.1$ & $131.6\pm0.1$ & \cite{richer2010} \\
        LMC-SMP-25  & 19 & $178.88\pm0.04\pm0.1$ & $208\pm4$ & \cite{reid2006} \\
        LMC-SMP-31  & 8 & $ 262.82\pm0.05\pm0.1$  & $262\pm4$  &  \cite{reid2006} \\
        LMC-SMP-53  & 15 & $274.77\pm0.05\pm0.1$ & $284\pm4$ &  \cite{reid2006} \\
        LMC-SMP-58  & 15 & $272.87\pm0.06\pm0.1$  & $295\pm4$ &  \cite{reid2006} \\
        LMC-Sa-104a  & 15 & $252.02\pm0.05\pm0.1$ &  &   \\
        SMC-SMP-1   & 11 & $144.66\pm0.04\pm0.1$ & $138.0\pm0.4$ & \cite{dopita1985} \\
        SMC-SMP-2  & 15 & $155.21\pm0.04\pm0.1$ & $157.0\pm0.4$ & \cite{dopita1985}\\
        SMC-SMP-8   & 11 & $123.59\pm0.03\pm0.1$ & & \\
        SMC-SMP-13  & 15 & $152.63\pm0.04\pm0.1$ & $153.6\pm0.4$ & \cite{dopita1985} \\
        SMC-SMP-18  & 15 & $122.42\pm0.06\pm0.1$ & $121.8\pm0.4$ & \cite{dopita1985}\\
        SMC-SMP-20  & 15 & $98.92\pm0.03\pm0.1$ & $99.3\pm0.4$ & \cite{dopita1985}\\
        SMC-SMP-24  & 15 & $137.43\pm0.05\pm0.1$ & $140.8\pm0.4$ & \cite{dopita1985}\\
        SMC-SMP-25  & 9 & $146.90\pm0.07\pm0.1$ & & \\
        \hline
    \end{tabular}
    \footnotesize
    \tablefoot{See Sect.~\ref{sec:radialvelocities} for details.}
    \label{table:radialvelocity}
    \end{threeparttable}
\end{table*}

\subsection{\label{sec:lineratios}Comparing line ratios with quantum-mechanical computations}
\cite{storey2000} have computed relativistic quantum-mechanical transition ratios for specific ionic transitions, such as [\ion{O}{III}]\,$\lambda5008/4960$, [\ion{Ne}{III}]\,$\lambda3870/3969$, and [\ion{N}{II}]\,$\lambda6585/6550$. Since both lines in each pair share the same upper level, there is virtually no dependence on temperature and density, at least not in the case of our PNe.
To compare line flux ratios (erg\,cm$^{-2}$\,s$^{-1}$) with the ratio of transition probabilities (photo-electron count rate), we must correct the flux ratio by the wavelength ratio, since a single bluer photon carries more energy than a redder photon, but both will cause just one photo-electron in the detector.

For [\ion{O}{III}] we find a weighted mean of $3.043\pm0.003$ (theoretical value: 3.013), with a standard deviation of $\sigma=0.014$ in the sample. The weights were given by the inverse flux variance of the line detections. Our measured line ratio is 1.0\% higher than expected. \cite{dimitrijevic2007} did a similar analysis for a larger quasar sample and did not find a discrepancy with respect to the same theoretical value. Since these lines are only 50\,\AA\;apart and far from X-SHOOTER's dichroic cross-over regions, problems with flux calibration over such a small wavelength range are unlikely. There could be a residual effect from detector nonlinnearity correction, as these lines are very bright and -- for some PNe -- even saturated.

The [\ion{N}{II}] and [\ion{Ne}{III}] lines have 5--50 times lower S/N than [\ion{O}{III}]. A comparison with the theoretical values from \cite{storey2000} is therefore not meaningful with our data.

\subsection{\label{sec:individualNotes}Notes about individual targets}

\subsubsection{PN G025.3--04.6}
This PN shows a double-peaked nucleus in the HST F502N image centred on the [\ion{O}{III}] line (Fig.~\ref{fig:PNmorph_linear}), which is not recognisable in the X-SHOOTER spectrum (Fig.~\ref{fig:PNlinemap}). The peaks have an angular separation of \ang{;;0.31} or 1 NISP pixel. While this is not ideal, we do not expect a problem for NISP wavelength calibration, as the X-SHOOTER observations were taken with a wide slit that propagates the morphology into the effective line wavelengths.

\subsubsection{PN G042.9--06.9}
This PN has comparatively many emission lines, and served as a starting point to construct a manually vetted list of atomic and ionic transitions from NIST and the Atomic Line Database, against which line detections in the other PNe were matched (see also Table~\ref{tab:line_flux}). The initial reference line table was then complemented by additional lines that we identified in the other PNe.

Notably, PN G042.9$-$06.9 is the only PN that shows a clear velocity splitting in some of the lower ionisation lines ([\ion{O}{I}]$\lambda$6302, [\ion{S}{II}]$\lambda\lambda$6718/33), and also in [\ion{S}{III}]$\lambda$9069. This indicates a bipolar outflow mostly along the line-of-sight; the outflow is not visible in the HST images shown in \cite{stanghellini2016}, but a ring-like morphology can be seen in our representation of the HST data (Fig.~\ref{fig:PNmorph_linear}). The line splitting is irrelevant for Euclid calibration purposes due to Euclid's $10\times$ lower spectral resolution.

\subsubsection{PN G205.8--26.7}
This is the largest PN in our sample, whose inner ring-shaped core has a diameter of \ang{;;0.8}, and is embedded in a symmetrical fainter halo of diameter \ang{;;2.5}. We included it in our sample because of its location in a particularly uncrowded area, and in case wavelength calibration with more compact PNe would be problematic due to undersampling effects. It is a backup target for NISP wavelength calibration.

\subsubsection{PN G334.8--07.4}
Compared to the other PNe, this source has a strong continuum contribution from its central star. It shows a large number of narrow absorption lines, including Ca H+K, the NaD doublet, the Ca triplet, and broad Balmer absorption lines. Using the spectral templates from \cite{kesseli2017}, we find that a K0 giant (low surface gravity) of Solar metallicity fits the spectrum best. However, the template flux is too low by 10\% or more below 4400\,\AA, and above 6700\,\AA. The emission lines are strong and therefore this PN can still serve as a wavelength calibrator. 

\subsubsection{He2-436\label{sec:he2}}
This PN is located in the Sagittarius Dwarf Galaxy. Deeper optical spectroscopy is available from \cite{otsuka2011}. He2-436 shows typical, broad Wolf-Rayet (WR) bumps from \ion{He}{II}\,$\lambda$4687, and \ion{C}{IV} at 4660\,\AA\ and 5803/14\,\AA\ \citep[see Fig.~\ref{fig:hotcontinuum}, and][]{smith1996,acker2003}. Its line fluxes are 3--5 times higher than for PNe in the Magellanic Clouds. This PN is currently our backup choice for wavelength calibration during PV phase, as its crowding index is twice as high as for SMC-SMP-20 (Sect.~\ref{sec:smc20}), our primary choice.

\subsubsection{STRIPE82-0130.035257}\label{sec:stripe}
This target was selected by us in yet unpublished S-PLUS data. The narrow-band photometric selection  \citep[see][]{gutierrezsoto2020} indicated considerable line emission, albeit not as strong as expected for typical PNe. Our X-SHOOTER observations did not reveal any emission lines at all; in addition, the continuum was considerably fainter than expected. Further analysis of the spectra revealed the typical broad Balmer absorption lines of a hydrogen WD, albeit very shallow. 

The presence of typical WD absorption lines, and the inspection of an X-SHOOTER acquisition image, confirmed that the correct target was observed. The absence of emission lines, and the weaker than expected continuum, suggest that at the time of the S-PLUS observations this source experienced an outburst with considerable nebular emission. In Gaia EDR3, this star is classified as a hot subdwarf (EDR3 \#2682273528086645632).

\subsubsection{LMC-SMP-58}
This PN is located just \ang{4.8} from the South Ecliptic Pole, close to \Euclid's continuous viewing zone that extends \ang{2.5;;} from the poles. The field, however, is heavily crowded, and thus this PN is likely not suitable to recalibrate the NISP slitless dispersion laws. It is used as a celestial absolute wavelength calibrator by JWST, though \citep{jones2023}. Similar to He2-436, the continuum shows a broad WR \ion{He}{II}\,$\lambda$4687 bump, and another bump from 4640--4660\,A that is more compatible with \ion{N}{III}. A \ion{C}{IV} bump at 5803/14\,\AA\ is absent.

\subsubsection{LMC-Sa-104a}
This PN is the only one in our sample that shows numerous high-ionisation lines from [\ion{Fe}{VII}]. We also detect [\ion{Ne}{V}] at 3347 and 3427\,\AA, and [\ion{Ne}{IV}] at 4716 and 4726\,\AA. The central star is rather unusual in the sense that it is very red, with numerous \ion{C}{I} absorption lines in the NIR.

\subsubsection{SMC-SMP-2 and SMC-SMP-25}
Besides LMC-Sa-104a, these two are the only other PNe that show strong high ionisation lines from [\ion{Ne}{V}] at 3347 and 3427\,\AA, as well as [\ion{Ne}{IV}] at 4716 and 4726\,\AA. Lines from highly ionised [\ion{Fe}{VII}] were not identified.

\subsubsection{SMC-SMP-20\label{sec:smc20}}
This PN has quite consistently the highest line fluxes among the Magellanic Cloud PNe in our sample. With $R_{\rm phot}=\ang{;;0.20}$ it is completely unresolved by NISP, it does not show any asymmetries in its HST line emission (Fig.~\ref{fig:PNmorph}), and its crowding index is one of the lowest in our sample (Table~\ref{table:observations}). In the current PV plan, we will use this PN to calibrate the NISP dispersion laws, with He2-436 (Sect.~\ref{sec:he2}) as a backup due to higher field crowding. We would only switch to He2-436 should pre-launch simulations of the PV observations reveal that the line fluxes for SMC-SMP-20 are insufficient.

\section{\label{sc:Cons}Conclusions}
The NISP instrument onboard \Euclid does not have an internal wavelength calibration lamp to update the ground-based dispersion laws \citep{maciaszek2022} its slitless spectroscopy modes after launch. To this end, \Euclid must observe compact astrophysical sources with numerous emission lines in the 0.93--1.89\,\micron\;range. In order to not bias \Euclid's cosmological measurements that are based on spectroscopic redshifts, the effective wavelengths of these emission lines must be known with an accuracy better than 5\,\AA, the overall uncertainty allowed for the NISP wavelength solution by \Euclid's galaxy clustering science. The latter requires a standard deviation of the measured redshifts around their true
values of $\sigma(z)<0.001\,(1+z)$, with a maximum allowed systematic offset of $0.2\,\sigma$ \citep[see also][]{eisenstein2005,guzzo2008}.

In this paper, we present a ground-based UV-NIR medium-resolution spectral atlas (Sect.~\ref{sec:dataproducts}) of 20 compact PNe with sub-arcsecond diameters, and typically over 100 identified emission lines. One or more of these PNe will serve as primary wavelength calibrators for \Euclid. The joint statistical and systematic error on the line images' central wavelengths is about 0.1\,\AA\ at optical, and 0.3--0.4\,\AA\ at near-infrared wavelengths for the X-SHOOTER spectra (Sect.~\ref{sec:offset}), well below the Euclid requirements. The errors on the absolute wavelengths are dominated by systematic effects that could not be reduced -- or unambiguously corrected for -- with the data at hand. The spectrum of DdDm-1, taken with GNIRS and GMOS, has larger uncertainties of 0.7\,\AA\;(GMOS) and 1.0\,\AA\;(GNIRS), partially due to the lower spectral resolution, and also because of a lower number density of usable reference arc and sky lines. For comparison, the relevant spectral resolutions of NISP, X-SHOOTER, and GNIRS/GMOS are about 450, 4000, and 2000, respectively.

Using a detailed error analysis of the uncertainty in the telluric correction for X-SHOOTER, we could also reliably detect emission lines in the deep atmospheric absorption bands (Sect.~\ref{sec:telluriccorrection}). This is essential for \Euclid, maximising line coverage across the wide spectral passbands.

The spectral atlas is useful for other purposes outside the \Euclid context. Not only can it serve as an independent check of wavelength calibration to other projects. With accurately calibrated and fluxed 1D spectra, and typically more than 100  identified lines per PN, various scientific analyses become feasible that are beyond the scope of this paper. We determined radial velocities for all PNe (Sect.~\ref{sec:radialvelocities}), and showed that our simple line-flux estimates with single Gaussians are sufficiently good to check the accuracy of relativistic quantum-mechanical computations of [\ion{O}{III}] transition probabilities (Sect.~\ref{sec:lineratios}).

As a by-product of our work, we developed a new method based on the 1st-order Gaia spectra shape coefficients (Sect.~\ref{sec:gaiaPNeSED}), to select compact line emitters in the Gaia spectroscopic data set (Sect.~\ref{sec:searchGaia2}). A cross-match analysis of the blindly selected Gaia line emitters showed that this search method is very effective. We discovered five hitherto unknown PNe candidates or stellar line emitters (listed in Table~\ref{table:pn}), albeit none of them in the Galactic halo where they would be useful for Euclid wavelength calibration. More effective filters can be built by including higher-order spectra shape coefficients.

%
%

\begin{acknowledgements}
M. Schirmer and K. Paterson thank Favio Favata, Paul McNamara and Guiseppe Racca at ESA, and Rob Ivison and Xavier Barcons at ESO for supporting and enabling our VLT/X-SHOOTER proposals; Carlo Felice Manara, Sabine Moehler, and Lodovico Coccato at ESO for deeper insight into X-SHOOTER wavelength calibration and telluric correction, Jen Andrews, Siyi Xu and German Gimeno at NOIRLAB/Gemini for their technical support with GNIRS and GMOS, the observers at Paranal and Mauna Kea for taking such excellent data; and Coryn Bailer-Jones for support with the Gaia-DR3 data.\\
The authors at MPIA acknowledge funding by the German Space Agency DLR under grant numbers 50~OR~1202, 50~QE~2003 and 50~QE~2303.\\
The plots in this publication were prepared with {\tt Matplotlib} \citep{hunter2007}. This research has made use of the HASH PN database at \url{hashpn.space}. We also used the Python package {\tt GaiaXPy}, developed and maintained by members of the Gaia Data Processing and Analysis Consortium (DPAC), and in particular, Coordination Unit 5 (CU5), and the Data Processing Centre located at the Institute of Astronomy, Cambridge, UK (DPCI).\\
This work is based on observations made with ESO Telescopes at the La Silla Paranal Observatory under programme IDs 108.MQ23.001 and 110.23Q7.001. It is also based on observations obtained under programme ID GN-2022A-FT-215 at the international Gemini Observatory, a programme of NSF’s NOIRLab, which is managed by the Association of Universities for Research in Astronomy (AURA) under a cooperative agreement with the National Science Foundation on behalf of the Gemini Observatory partnership: the National Science Foundation (United States), National Research Council (Canada), Agencia Nacional de Investigaci\'{o}n y Desarrollo (Chile), Ministerio de Ciencia, Tecnolog\'{i}a e Innovaci\'{o}n (Argentina), Minist\'{e}rio da Ci\^{e}ncia, Tecnologia, Inova\c{c}\~{o}es e Comunica\c{c}\~{o}es (Brazil), and Korea Astronomy and Space Science Institute (Republic of Korea). Observations were made with the Gemini North telescope, located within the Maunakea Science Reserve and adjacent to the summit of Maunakea. We are grateful for the privilege of observing the Universe from a place that is unique in both its astronomical quality and its cultural significance.\\
This work presents results from the European Space Agency (ESA) space mission Gaia. Gaia data are being processed by the Gaia Data Processing and Analysis Consortium (DPAC). Funding for the DPAC is provided by national institutions, in particular the institutions participating in the Gaia MultiLateral Agreement (MLA). The Gaia mission website is \url{https://www.cosmos.esa.int/gaia}. The Gaia archive website is \url{https://archives.esac.esa.int/gaia}.\\
\AckEC\\
\end{acknowledgements}

\bibliography{manuscript}

%
%

%

\begin{appendix}
  \onecolumn 

\section{Complementary information}
In this Appendix we show an example of the line atlas for one of our PN targets, HST images of our target PNe, and additional figures highlighting certain aspects of our PN spectra and the identified lines.

\begin{figure*}
\centering
\includegraphics[angle=0,width=0.90\hsize]{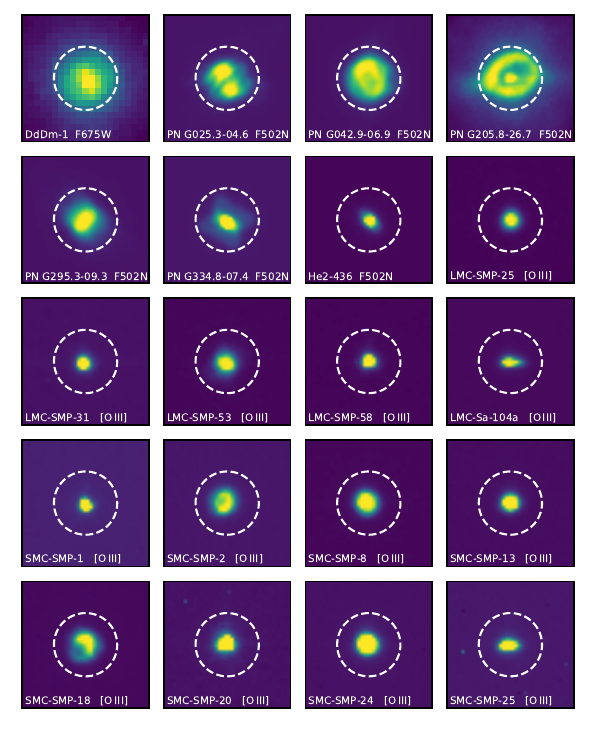}
\caption{HST images of all PNe in this paper, in linear scale (a nonlinear stretch is shown in Fig.~\ref{fig:PNmorph}). DdDm-1 was taken in the F675W broad-band filter including the H$\alpha$ line. Contrary to the other panels shown here, this image of DdDm-1 suffers from spherical aberration prior to the installation of HST's corrective optics. F502N refers to a narrow-band filter including the [\ion{O}{III}] line. These images include the emission from the central star. [\ion{O}{III}] refers to that emission line extracted from a slitless spectrum. The spectral resolution is too low to resolve the intrinsic line width, hence these can be considered 2D line images, without the central star whose continuum is dispersed along the horizontal axis. The white circle shows the NISP-S EE80 disk with a radius of \ang{;;0.5}. The field of view is $\ang{;;2}\times\ang{;;2}$.}
\label{fig:PNmorph_linear}
\end{figure*}

\begin{figure*}
\centering
\includegraphics[angle=0,width=0.90\hsize]{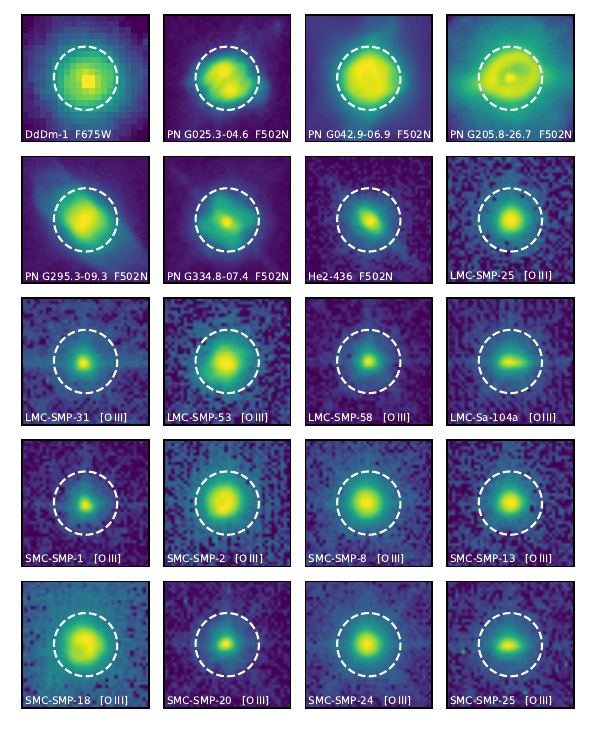}
\caption{Same as for Fig.~\ref{fig:PNmorph_linear}, but with a nonlinnear ${\rm asinh()}$ stretch to show fainter envelopes. The latter are not expected to affect the line centroid measurements. We note that the image of DdDm-1 is affected by HST's early spherical aberration.}
\label{fig:PNmorph}
\end{figure*}

\begin{figure*}[t]
\centering
\includegraphics[angle=0,width=0.90\hsize]{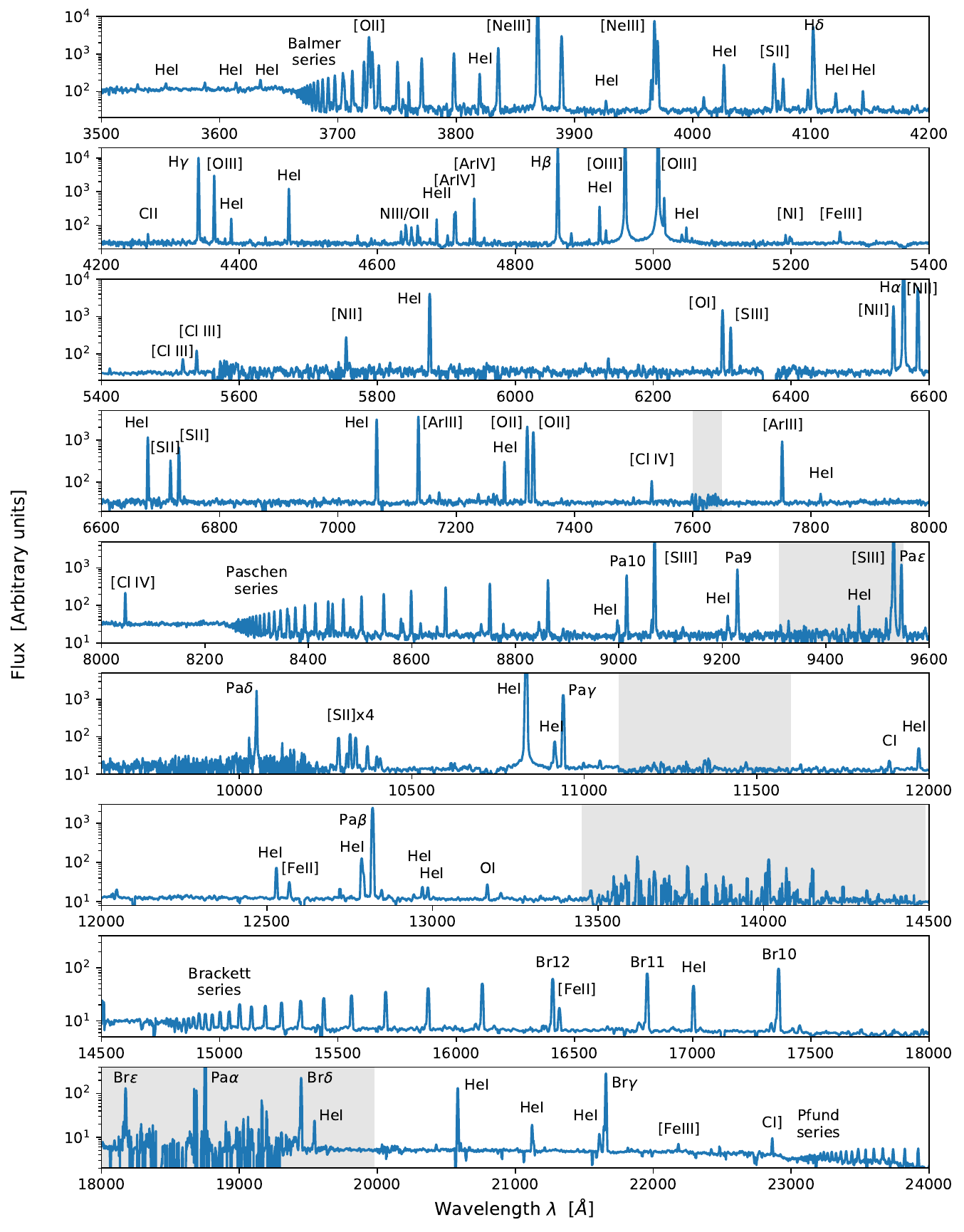}
\caption{X-SHOOTER spectrum of PN G042.9$-$06.9, rebinned by factors $5$--$20$ for illustrative purposes. The UVB range between 3000$-$3500\,\AA\ is not shown, as the only brighter line is \ion{He}{I}$\,\lambda$3189. We labelled the brightest and some of the weaker lines in the rest of the spectrum; no attempt was made to accurately label line blends. In total, we report 253 lines with measured fluxes in this spectrum (Table~\ref{tab:line_flux}), one of the richest in our sample. We note that the panels' $y$-axis ranges are different, and that the log-scale exaggerates the width of the brightest lines such as [\ion{O}{III}]$\lambda$5008 or \ion{He}{I}\,$\lambda$10830. The grey areas indicate the atmospheric $A$-band at 7600\,\AA, and other absorption bands where atmospheric transmission is reduced by at least 10\% at zenith for a precipitable water vapour of 1\,mm, as reported in the ESO VISTA \citep{sutherland2015} instrument description.}
\label{fig:PN_xshooter_spectrum}
\end{figure*}

\begin{figure*}[t]
\centering
\includegraphics[angle=0,width=1.00\hsize]{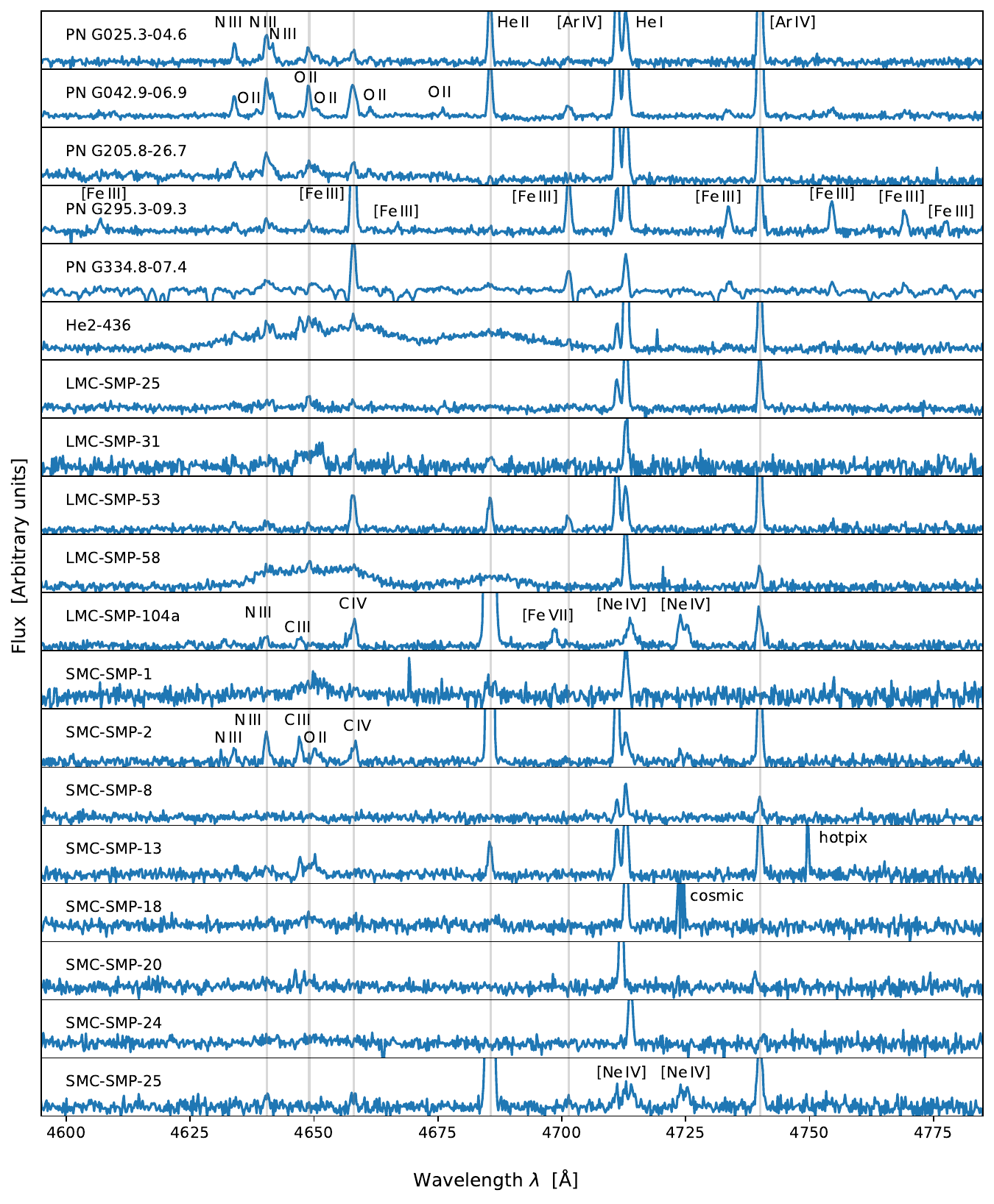}
\caption{Zoom-in onto part of the UVB spectra, corrected for radial velocity. This area highlights differences in the chemical compositions and ionisation states of the PN, as well as in the continuum of the central star. He2-436 and LMC-SMP-58 show broad bumps characteristic for WR stars. Other PNe also show features at or near 4650\,\AA, albeit mostly less pronounced.}
\label{fig:hotcontinuum}
\end{figure*}

\begin{figure*}[t]
\centering
\includegraphics[angle=0,width=0.90\hsize]{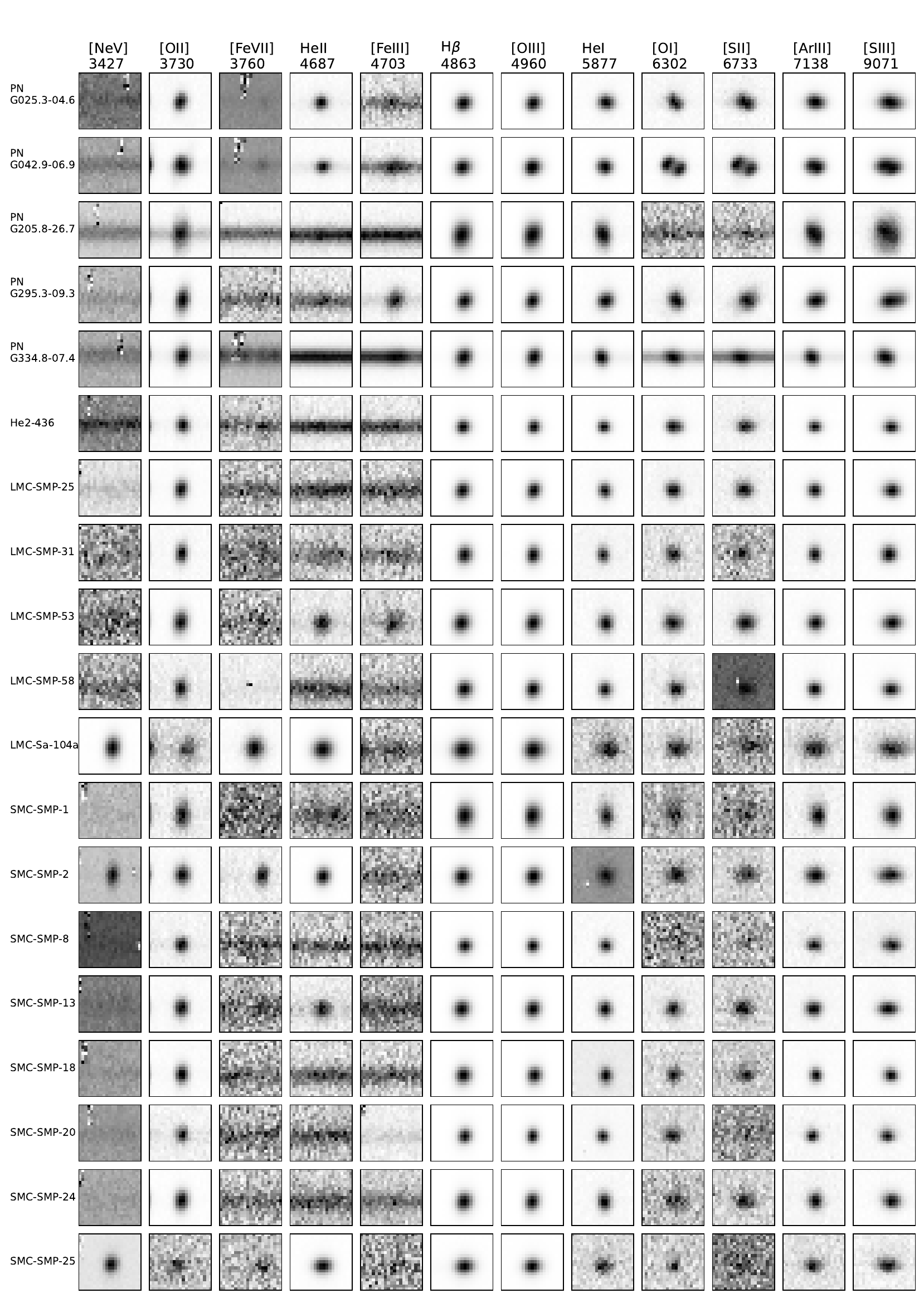}
\caption{Line images of selected ionic species in the X-SHOOTER UVB and VIS arms. The spectral width shown on the horizontal axis is 4.2\,\AA, the spatial width on the vertical axis is \ang{;;3.04}. For most sources, the lines are dynamically and spatially not or only weakly resolved. A noteworthy exception is PN G042.9$-$06.9, showing well-resolved bipolar outflow in the lower ionisation lines such as [\ion{O}{I}]$\lambda$6300 and [\ion{S}{II}]$\lambda$6731, unresolved by NISP. The line in the [\ion{Fe}{VII}]\,$\lambda$3760 panel of SMC-SMP-2 is actually nearby \ion{O}{III}\,$\lambda$3761.}
\label{fig:PNlinemap}
\end{figure*}

\input{G042.9-06.9_table.tex}

\begin{figure*}[t]
\centering
\includegraphics[angle=0,width=1\hsize]{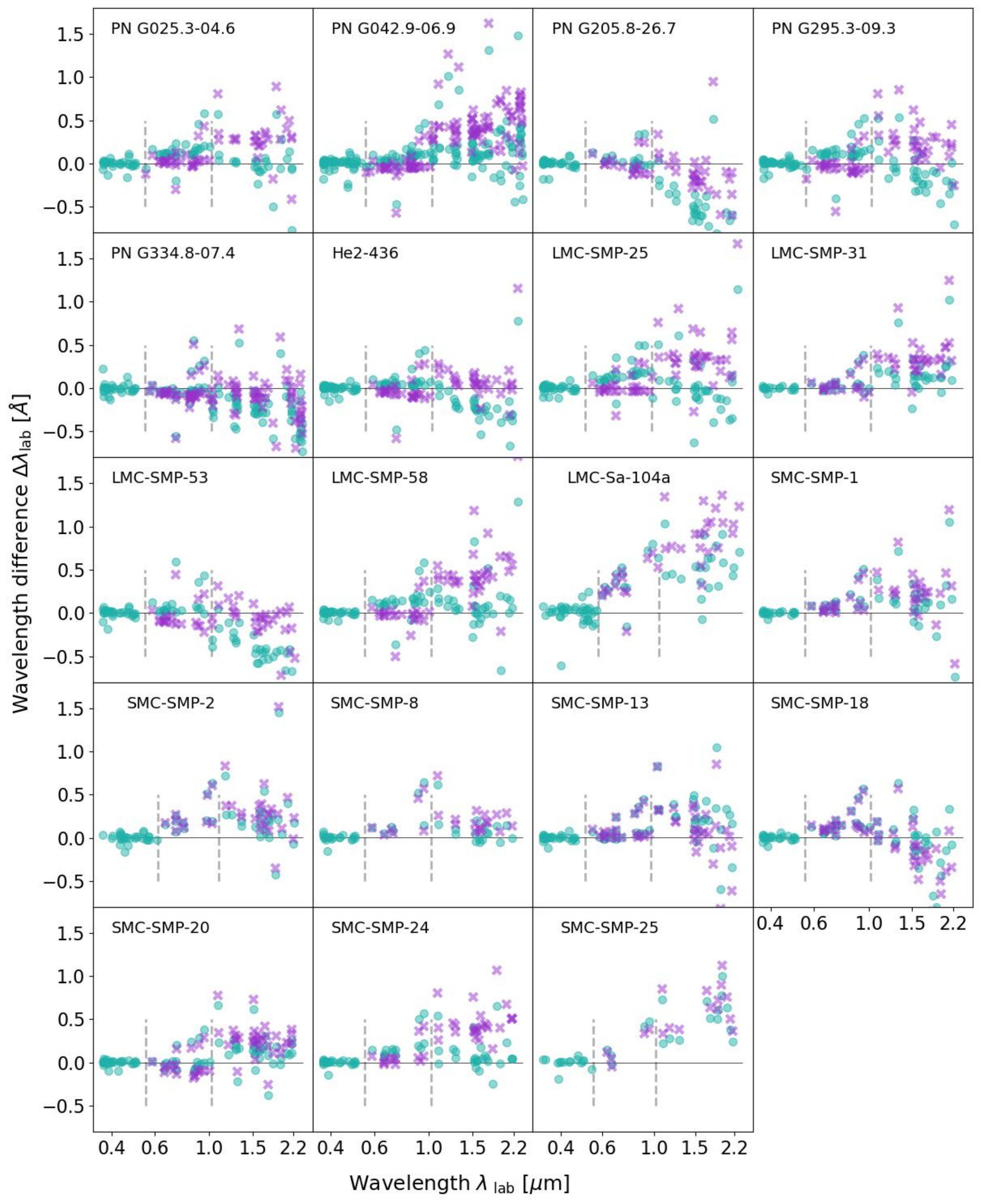}
\caption{Difference between the observed heliocentric line wavelengths, and the corresponding NIST laboratory restframe wavelengths, for lines with S/N > 3. The radial velocities were determined from the UVB spectra, which thus have a zero mean offset. The teal circles show the offsets without {\tt MOLECFIT} wavelength correction, and the purple crosses with correction. Evidently, the correction does not always work, and might even degrade the results, such as for LMC-SMP-58. Errors are not shown, as they are mostly within the size of the plotting symbols. The dashed vertical lines indicate the UVB, VIS and NIR spectral ranges (left, middle, right, respectively).}
\label{fig:PN_wave_offset}
\end{figure*}
\clearpage

\section{Emission-line selected PNe in Gaia spectroscopic data}
In this Appendix we include details of the PN-like objects found in our blind Gaia search (see Sect. \ref{sec:searchGaia2}).

\include{pn_table}

\end{appendix}

\end{document}

%% file: PN.tex
\begin{table*}[t]
    \centering
    \begin{threeparttable}
    \caption{Compact PNe presented in this paper.}
    \smallskip
    \begin{tabular}{|llcccc|}
    \hline
        Name (Alias) & RA & Dec & $R_{\rm phot}$  [$^{\prime\prime}$] & Density  & References / Comments\\
         & (J2000.0) & (J2000.0) & (HST) & [arcmin$^{-2}$] & \\
        \hline
        PN G061.9+41.3 (DdDm-1)  & 16:40:18.20 & $+$38:42:19.9 & 0.6\tablefootmark{a} & 1.3 & \cite{henry2008} \\
        PN G025.3-04.6 & 18:54:20.04 & $-$08:47:32.9 & 0.48 & 53.4 & \cite{stanghellini2016} \\
        PN G042.9-06.9   & 19:34:33.54 & $+$05:41:03.0 & 0.48 & 23.0 &  \cite{stanghellini2016} \\
        PN G205.8-26.7 & 05:03:41.85 & $-$06:10:03.0 & 1.04 & 1.8 & \cite{stanghellini2016}\\
        PN G295.3-09.3 & 11:17:43.16 & $-$70:49:32.2 & 0.48 & 12.6 &  \cite{stanghellini2016} \\
        PN G334.8-07.4 & 17:03:02.86 & $-$53:55:54.0 & 0.44 & 25.1 &  \cite{stanghellini2016} \\
        PN G004.8-22.7 (He2-436)     & 19:32:06.69 & $-$34:12:57.8 & $$>0.21$$ & 4.6 & \cite{zijlstra2006} \\
        LMC-SMP-25  & 05:06:24.00 & $-$69:03:19.2 & 0.23 & 68.4 & \cite{shaw2001} \\
        LMC-SMP-31  & 05:09:20.23 & $-$67:47:25.2 & 0.15 & 44.6 & \cite{shaw2001} \\
        LMC-SMP-53  & 05:21:32.93 & $-$67:00:05.5 & 0.40 & 19.6 & \cite{shaw2001} \\
        LMC-SMP-58\tablefootmark{b}  & 05:24:20.81 & $-$70:05:01.9 & 0.13 & 91.8 & \cite{shaw2001} \\
        LMC-Sa-104a & 04:25:32.18 & $-$66:47:16.3 & $<0.13$ & 1.9 & \cite{shaw2006}\\
        SMC-SMP-1   & 00:23:58.67 & $-$73:38:03.8 & 0.15 & 4.4 & \cite{stanghellini2003} \\
        SMC-SMP-2   & 00:32:38.81 & $-$71:41:58.7 & 0.25 & 1.8 & \cite{shaw2006} \\
        SMC-SMP-8   & 00:43:25.30 & $-$72:38:18.9 & 0.40 & 17.4 & \cite{stanghellini2003} \\
        SMC-SMP-13  & 00:49:51.71\tablefootmark{c} & $-$73:44:21.3 & 0.20 & 24.0 & \cite{stanghellini2003} \\
        SMC-SMP-18  & 00:51:57.97 & $-$73:20:31.1\tablefootmark{d} & 0.14 & 58.3 & \cite{stanghellini2003} \\
        SMC-SMP-20  & 00:56:05.39 & $-$70:19:24.7 & 0.20 & 2.2 & \cite{stanghellini2003} \\
        SMC-SMP-24  & 00:59:16.09 & $-$72:01:59.7 & 0.38 & 25.9 & \cite{stanghellini2003} \\
        SMC-SMP-25  & 00:59:40.51\tablefootmark{e} & $-$71:38:15.1\tablefootmark{e} & 0.19 & 12.8 & \cite{stanghellini2003} \\
        STRIPE82-0130 & 22:03:15.14 & $+$01:17:20.9 & -- & N/A & Transient, not a PN\\
        \hline
    \end{tabular}
    \footnotesize
    \tablefoot{
    All PNe listed here are included in our line atlas and are available as a part of our online data pack.\\
    \tablefoottext{a}{From a 1993 archival HST image, after deconvolution to correct for spherical aberration.\\}
    \tablefoottext{b}{Used as a wavelength calibration source for JWST \citep{jones2023}; most likely too crowded for Euclid calibration purposes.\\}
    \tablefoottext{c}{Incorrectly typed in \cite{stanghellini2003}.\\}
    \tablefoottext{d}{Shifted $\sim$\,\ang{;;1} south by us based on the ESO finding chart tool using the Gaia catalogue display.\\}
    \tablefoottext{e}{Wrong coordinates in \cite{stanghellini2003}, corresponding to LMC-SMP-71 in \cite{shaw2001}. We used the coordinates from \cite{meyssonnier1993}, adjusted to higher precision.\\}}
    \label{table:observations}
    \end{threeparttable}
\end{table*}

%% file: obs_table.tex
\begin{table}[t]
    \centering
    \begin{threeparttable}
    \caption{Summary of our X-SHOOTER and GNIRS/GMOS observations.}
    \smallskip
    \begin{tabular}{|lccc|}
    \hline
        Name & $t_{\rm exp}$ [s] & Airmass & Seeing  \\
         & U, V, N & (at zenith) & [$^{\prime\prime}$] \\
        \hline
        DdDm-1     & --, 480,160/320\tablefootmark{a} & 1.06 & 0.43 \\
        PN G025.3-04.6 & 240, 80, 80 & 1.05 & 0.36 \\
        PN G042.9-06.9  & 80, 40, 80 & 1.20 & 0.28 \\
        PN G205.8-26.7  & 240, 80, 240 & 1.11 & 0.41 \\
        PN G295.3-09.3  & 160, 80, 80 & 1.46 & 0.42 \\
        PN G334.8-07.4  & 40, 40, 40 & 1.30 & 0.46 \\
        He2-436    & 240, 80, 80 & 1.05 & 0.38 \\
        LMC-SMP-25   & 240, 80, 240 & 1.42 & 0.38 \\
        LMC-SMP-31  & 240, 80, 240 & 1.41 & 0.53 \\
        LMC-SMP-53 & 240, 80, 240 & 1.36 & 0.41 \\
        LMC-SMP-58 & 240, 80, 240 & 1.44 & 0.34 \\
        LMC-Sa-104a & 240, 80, 240 & 1.36 & 0.47 \\
        SMC-SMP-1 & 240, 80, 240 & 1.62 & 0.40 \\
        SMC-SMP-2  & 240, 80, 240 & 1.50 & 0.58 \\
        SMC-SMP-8 & 240, 80, 240 & 1.50 & 0.32 \\
        SMC-SMP-13 & 240, 80, 240 & 1.52 & 0.46 \\
        SMC-SMP-18 & 240, 80, 240 & 1.51 & 0.62 \\
        SMC-SMP-20 & 240, 80, 240 & 1.46 & 0.47 \\
        SMC-SMP-24 & 240, 80, 240 & 1.60 & 0.52 \\
        SMC-SMP-25 & 240, 80, 240 & 1.49 & 0.49 \\
        STRIPE82-0130\tablefootmark{b} & 240, 80, 240 & 1.41 & 0.32 \\
        \hline
    \end{tabular}
    \footnotesize
    \tablefoot{For U, V, and N we refer to the UVB, VIS and NIR channels of X-SHOOTER. In case of DdDm-1, V refers to GMOS and N to GNIRS.\\
    \tablefoottext{a}{Values for the GMOS/GNIRS setups.\\}
    \tablefoottext{b}{Transient, not a PN.\\}}
    \label{table:observations2}
    \end{threeparttable}
\end{table}

%% file: G042.9-06.9_table.tex
\begin{longtable}[c]{|lS[table-format=5.2,table-number-alignment=right]S[table-format=5.2,table-number-alignment=right]@{$\,\pm\,$}S[table-format=0.2,table-number-alignment=left]S[table-format=5.2,table-number-alignment=right]@{$\,\pm\,$}S[table-format=3.2,table-number-alignment=left]S[table-format=1.2,table-number-alignment=right]@{$\,\pm\,$}S[table-format=0.2,table-number-alignment=left]@{\,}|}
\caption{253 emission lines in the spectrum of PN G042.9$-$0.69.}\label{tab:line_flux} \\

\hline 
\multicolumn{1}{|c}{Line} & 
\multicolumn{1}{c}{$\lambda_{\rm laboratory}^{\rm vacuum}$} & 
\multicolumn{2}{c}{$\lambda_{\rm obs,\;heliocentric}^{\rm vacuum}$} &
\multicolumn{2}{c}{Flux} &
\multicolumn{2}{c|}{FWHM}\\

\multicolumn{1}{|c}{} & 
\multicolumn{1}{c}{[\AA]} &
\multicolumn{2}{c}{[\AA]} &
\multicolumn{2}{c}{[$10^{-16}$\,erg\,cm$^{-2}$\,s$^{-1}$]} &
\multicolumn{2}{c|}{[\AA]} \\ \hline 
\endfirsthead

\multicolumn{8}{c}%
{{\bfseries \tablename\ \thetable{} -- continued from previous page}} \\
\hline 
\multicolumn{1}{|c}{Line} & 
\multicolumn{1}{c}{$\lambda_{\rm laboratory}^{\rm vacuum}$} &
\multicolumn{2}{c}{$\lambda_{\rm obs,\;heliocentric}^{\rm vacuum}$} &
\multicolumn{2}{c}{Flux} &
\multicolumn{2}{c|}{FWHM}\\

\multicolumn{1}{|c}{} & 
\multicolumn{1}{c}{[\AA]} & 
\multicolumn{2}{c}{[\AA]} &
\multicolumn{2}{c}{[$10^{-16}$\,erg\,cm$^{-2}$\,s$^{-1}$]} &
\multicolumn{2}{c|}{[\AA]}\\
 \hline 

\endhead

\hline \multicolumn{8}{|r|}{{Continued on next page}} \\ \hline
\endfoot

\hline \\ \hline
\endlastfoot
\ion{O}{III} & 3133.70 & 3132.97 & 0.06 & 285.49 & 305.95 & 0.32 & 0.06\\
\ion{He}{I} & 3188.66 & 3187.97 & 0.03 & 458.00 & 276.91 & 0.35 & 0.03\\
\ion{O}{III} & 3445.04 & 3444.32 & 0.06 & 88.09 & 89.14 & 0.29 & 0.06\\
\ion{Fe}{I} & 3531.40 & 3530.68 & 0.06 & 42.35 & 55.21 & 0.26 & 0.06\\
\ion{He}{I} & 3555.56 & 3554.65 & 0.04 & 41.63 & 45.59 & 0.22 & 0.04\\
\ion{He}{I} & 3588.29 & 3587.54 & 0.05 & 60.47 & 53.55 & 0.29 & 0.05\\
\ion{He}{I} & 3614.67 & 3613.83 & 0.06 & 75.01 & 69.57 & 0.36 & 0.06\\
\ion{He}{I} & 3635.27 & 3634.46 & 0.04 & 101.43 & 73.66 & 0.34 & 0.05\\
H23 & 3674.81 & 3673.96 & 0.04 & 117.09 & 54.08 & 0.38 & 0.05\\
H22 & 3677.42 & 3676.60 & 0.02 & 130.95 & 50.02 & 0.33 & 0.03\\
H21 & 3680.41 & 3679.56 & 0.10 & 123.86 & 43.31 & 0.25 & 0.10\\
H20 & 3683.86 & 3683.00 & 0.09 & 144.03 & 42.55 & 0.25 & 0.09\\
H19 & 3687.89 & 3687.04 & 0.08 & 182.48 & 43.03 & 0.30 & 0.08\\
H18 & 3692.61 & 3691.78 & 0.07 & 195.19 & 39.01 & 0.28 & 0.07\\
H17 & 3698.21 & 3697.38 & 0.08 & 235.90 & 39.47 & 0.30 & 0.08\\
H16 & 3704.91 & 3704.09 & 0.02 & 280.83 & 41.82 & 0.37 & 0.02\\
\ion{He}{I} & 3706.06 & 3705.03 & 0.07 & 346.45 & 49.57 & 0.52 & 0.08\\
H15 & 3713.03 & 3712.19 & 0.07 & 340.07 & 37.99 & 0.32 & 0.07\\
H14 & 3723.00 & 3722.08 & 0.19 & 539.72 & 37.21 & 0.28 & 0.19\\
{[}\ion{O}{II}{]} & 3727.09 & 3726.28 & 0.03 & 3522.88 & 56.12 & 0.41 & 0.03\\
{[}\ion{O}{II}{]} & 3729.88 & 3729.02 & 0.12 & 1249.58 & 45.01 & 0.38 & 0.11\\
H13 & 3735.44 & 3734.60 & 0.22 & 419.23 & 33.43 & 0.27 & 0.22\\
H12 & 3751.2 & 3750.38 & 0.01 & 702.47 & 38.91 & 0.36 & 0.01\\
\ion{O}{III} & 3755.76 & 3754.93 & 0.06 & 34.40 & 31.76 & 0.31 & 0.07\\
H11 & 3771.71 & 3770.86 & 0.01 & 875.07 & 40.90 & 0.36 & 0.00\\
H10 & 3798.98 & 3798.12 & 0.01 & 1196.42 & 50.92 & 0.36 & 0.00\\
\ion{He}{I} & 3820.69 & 3819.84 & 0.01 & 314.54 & 47.67 & 0.35 & 0.01\\
H9 & 3836.48 & 3835.61 & 0.01 & 1689.34 & 53.31 & 0.36 & 0.00\\
{[}\ion{Ne}{III}{]} & 3869.86 & 3869.00 & 0.01 & 28670.57 & 121.81 & 0.35 & 0.00\\
H8 & 3890.16 & 3889.18 & 0.01 & 3779.44 & 49.23 & 0.41 & 0.00\\
\ion{He}{I} & 3927.66 & 3926.72 & 0.05 & 34.84 & 24.11 & 0.36 & 0.06\\
\ion{He}{I} & 3965.85 & 3964.97 & 0.01 & 178.97 & 26.25 & 0.37 & 0.01\\
{[}\ion{Ne}{III}{]} & 3968.59 & 3967.70 & 0.02 & 8984.35 & 60.45 & 0.35 & 0.02\\
H$\epsilon$ & 3971.20 & 3970.34 & 0.09 & 2487.72 & 38.04 & 0.34 & 0.09\\
\ion{He}{I} & 4010.39 & 4009.43 & 0.04 & 40.64 & 29.01 & 0.29 & 0.04\\
\ion{He}{I} & 4027.33 & 4026.43 & 0.01 & 591.02 & 35.88 & 0.37 & 0.01\\
{[}\ion{S}{II}{]} & 4069.75 & 4068.85 & 0.02 & 624.58 & 30.97 & 0.39 & 0.02\\
\ion{O}{II} & 4073.30 & 4072.39 & 0.05 & 26.55 & 23.56 & 0.28 & 0.05\\
{[}\ion{S}{II}{]} & 4077.50 & 4076.50 & 0.13 & 214.04 & 26.45 & 0.37 & 0.13\\
\ion{N}{III} & 4098.51 & 4097.55 & 0.01 & 92.18 & 20.98 & 0.34 & 0.01\\
H$\delta$ & 4102.90 & 4101.98 & 0.01 & 6533.70 & 47.89 & 0.39 & 0.00\\
\ion{He}{I} & 4121.98 & 4121.06 & 0.02 & 71.82 & 19.67 & 0.39 & 0.02\\
\ion{He}{I} & 4144.93 & 4144.00 & 0.02 & 87.23 & 18.23 & 0.35 & 0.02\\
\ion{O}{II} & 4170.40 & 4169.32 & 0.10 & 13.69 & 18.88 & 0.38 & 0.10\\
\ion{O}{II} & 4190.97 & 4190.08 & 0.07 & 11.42 & 16.25 & 0.22 & 0.07\\
\ion{C}{II} & 4268.46 & 4267.45 & 0.08 & 35.72 & 27.55 & 0.41 & 0.09\\
H$\gamma$ & 4341.69 & 4340.71 & 0.01 & 12939.00 & 57.08 & 0.42 & 0.00\\
\ion{O}{II} & 4350.65 & 4349.74 & 0.05 & 13.13 & 13.51 & 0.27 & 0.05\\
{[}\ion{O}{III}{]} & 4364.43 & 4363.45 & 0.01 & 3681.81 & 31.31 & 0.39 & 0.00\\
\ion{N}{III} & 4380.34 & 4379.24 & 0.07 & 18.25 & 16.54 & 0.47 & 0.08\\
\ion{He}{I} & 4389.16 & 4388.17 & 0.01 & 162.99 & 16.20 & 0.40 & 0.01\\
\ion{He}{I} & 4438.80 & 4437.80 & 0.07 & 20.25 & 16.33 & 0.39 & 0.08\\
\ion{He}{I} & 4472.74 & 4471.75 & 0.01 & 1494.87 & 27.31 & 0.40 & 0.00\\
\ion{Mg}{I} & 4572.4 & 4571.31 & 0.08 & 28.44 & 19.35 & 0.44 & 0.08\\
\ion{O}{II} & 4592.26 & 4591.23 & 0.10 & 20.84 & 17.51 & 0.52 & 0.12\\
\ion{O}{II} & 4597.24 & 4596.26 & 0.16 & 8.55 & 13.87 & 0.33 & 0.16\\
\ion{N}{III} & 4635.42 & 4634.38 & 0.11 & 36.40 & 11.88 & 0.29 & 0.11\\
\ion{O}{II} & 4640.15 & 4639.00 & 0.09 & 0.00 & 14.92 & 0.49 & 0.11\\
\ion{N}{III} & 4641.94 & 4640.90 & 0.02 & 65.90 & 13.88 & 0.38 & 0.03\\
\ion{N}{III} & 4643.15 & 4641.85 & 0.06 & 105.78 & 16.36 & 0.58 & 0.06\\
\ion{O}{II} & 4650.44 & 4649.37 & 0.10 & 52.15 & 11.85 & 0.30 & 0.10\\
\ion{O}{II} & 4652.14 & 4650.80 & 0.08 & 140.48 & 18.47 & 0.77 & 0.10\\
{[}\ion{Fe}{III}{]} & 4659.42 & 4658.35 & 0.07 & 109.12 & 15.52 & 0.55 & 0.07\\
\ion{O}{II} & 4662.94 & 4661.79 & 0.11 & 15.53 & 11.29 & 0.31 & 0.12\\
\ion{O}{II} & 4677.54 & 4676.48 & 0.05 & 10.31 & 10.05 & 0.21 & 0.05\\
\ion{He}{II} & 4687.04 & 4685.99 & 0.01 & 149.50 & 13.68 & 0.37 & 0.01\\
{[}FeIII{]} & 4702.87 & 4701.82 & 0.06 & 42.29 & 17.06 & 0.59 & 0.08\\
{[}\ion{Ar}{IV}{]} & 4712.58 & 4711.63 & 0.01 & 183.05 & 16.08 & 0.41 & 0.01\\
\ion{He}{I} & 4714.47 & 4713.40 & 0.05 & 242.08 & 16.33 & 0.43 & 0.05\\
\ion{C}{I} & 4735.59 & 4734.02 & 0.09 & 22.83 & 18.01 & 0.56 & 0.10\\
{[}\ion{Ar}{IV}{]} & 4741.45 & 4740.47 & 0.01 & 720.19 & 18.73 & 0.37 & 0.00\\
{[}\ion{Fe}{III}{]} & 4756.08 & 4754.99 & 0.12 & 26.30 & 20.82 & 0.64 & 0.15\\
H$\beta$ & 4862.69 & 4861.57 & 0.01 & 32714.43 & 85.93 & 0.46 & 0.00\\
{[}\ion{Fe}{III}{]} & 4882.40 & 4881.21 & 0.05 & 57.58 & 18.14 & 0.64 & 0.05\\
\ion{He}{I} & 4923.31 & 4922.18 & 0.01 & 434.05 & 14.56 & 0.43 & 0.01\\
{[}\ion{O}{III}{]} & 4932.60 & 4931.42 & 0.02 & 60.62 & 12.64 & 0.49 & 0.03\\
{[}\ion{O}{III}{]} & 4960.30 & 4959.19 & 0.01 & 0.00 & 174.34 & 0.44 & 0.00\\
{[}\ion{O}{III}{]} & 5008.24 & 5007.15 & 0.02 & 0.00 & 180.68 & 0.56 & 0.01\\
\ion{He}{I} & 5017.08 & 5015.92 & 0.01 & 787.25 & 17.30 & 0.44 & 0.01\\
\ion{Fe}{I} & 5042.48 & 5041.57 & 0.09 & 37.68 & 18.16 & 0.75 & 0.12\\
\ion{He}{I} & 5049.15 & 5048.02 & 0.03 & 67.67 & 13.76 & 0.41 & 0.03\\
? & 5057 & 5056.17 & 0.08 & 18.54 & 16.72 & 0.60 & 0.10\\
\ion{Fe}{I} & 5192.90 & 5191.87 & 0.07 & 24.26 & 13.69 & 0.37 & 0.07\\
{[}\ion{N}{I}{]} & 5199.35 & 5197.85 & 0.09 & 19.73 & 14.84 & 0.46 & 0.11\\
{[}\ion{N}{I}{]} & 5201.71 & 5200.50 & 0.23 & 9.25 & 11.02 & 0.28 & 0.25\\
{[}\ion{Fe}{III}{]} & 5271.97 & 5270.76 & 0.03 & 67.56 & 13.99 & 0.70 & 0.03\\
\ion{Ar}{II} & 5413.16 & 5411.98 & 0.05 & 25.60 & 14.10 & 0.53 & 0.06\\
{[}\ion{Cl}{III}{]} & 5519.24 & 5517.97 & 0.05 & 91.14 & 30.63 & 0.64 & 0.05\\
{[}\ion{Cl}{III}{]} & 5539.41 & 5538.04 & 0.02 & 178.94 & 25.80 & 0.53 & 0.02\\
{[}\ion{N}{II}{]} & 5756.19 & 5754.77 & 0.03 & 457.79 & 152.26 & 0.47 & 0.03\\
\ion{He}{I} & 5877.24 & 5875.92 & 0.00 & 7379.60 & 95.64 & 0.39 & 0.00\\
{[}\ion{O}{I}{]} & 6302.05 & 6300.53 & 0.01 & 2878.27 & 60.61 & 0.57 & 0.01\\
{[}\ion{S}{III}{]} & 6313.81 & 6312.32 & 0.01 & 909.98 & 49.05 & 0.45 & 0.01\\
{[}\ion{N}{II}{]} & 6549.86 & 6548.30 & 0.01 & 3668.35 & 53.70 & 0.57 & 0.01\\
H$\alpha$ & 6564.63 & 6563.06 & 0.00 & 0.00 & 253.52 & 0.46 & 0.00\\
{[}\ion{N}{II}{]} & 6585.27 & 6583.69 & 0.01 & 11251.71 & 75.46 & 0.55 & 0.01\\
\ion{He}{I} & 6680.00 & 6678.44 & 0.00 & 2093.22 & 68.47 & 0.43 & 0.00\\
{[}\ion{S}{II}{]} & 6718.30 & 6716.70 & 0.02 & 594.09 & 54.71 & 0.62 & 0.02\\
{[}\ion{S}{II}{]} & 6732.67 & 6731.06 & 0.02 & 1268.74 & 52.04 & 0.62 & 0.02\\
\ion{Fe}{I} & 6741.38 & 6739.90 & 0.11 & 26.39 & 37.94 & 0.43 & 0.12\\
\ion{He}{I} & 7067.14 & 7065.52 & 0.00 & 5707.39 & 46.78 & 0.46 & 0.00\\
{[}\ion{Ar}{III}{]} & 7137.76 & 7136.05 & 0.00 & 6675.11 & 47.77 & 0.47 & 0.00\\
{[}\ion{Fe}{II}{]} & 7157.15 & 7155.28 & 0.15 & 30.83 & 35.50 & 0.70 & 0.18\\
{[}\ion{Ar}{IV}{]} & 7172.67 & 7171.02 & 0.05 & 38.63 & 27.29 & 0.37 & 0.05\\
{[}\ion{Ar}{IV}{]} & 7239.77 & 7237.64 & 0.15 & 44.95 & 48.64 & 0.78 & 0.15\\
\ion{N}{II} & 7264.55 & 7263.14 & 0.09 & 49.30 & 42.74 & 0.66 & 0.17\\
\ion{He}{I} & 7283.36 & 7281.66 & 0.01 & 516.10 & 42.36 & 0.47 & 0.01\\
{[}\ion{O}{II}{]} & 7322.01 & 7320.18 & 0.04 & 4457.99 & 52.80 & 0.71 & 0.04\\
{[}\ion{O}{II}{]} & 7332.75 & 7330.48 & 0.08 & 3710.57 & 49.33 & 0.81 & 0.08\\
\ion{He}{I} & 7501.91 & 7499.94 & 0.13 & 25.55 & 30.45 & 0.56 & 0.15\\
{[}\ion{Cl}{IV}{]} & 7532.62 & 7530.79 & 0.01 & 131.91 & 29.88 & 0.40 & 0.01\\
{[}\ion{Ar}{III}{]} & 7753.24 & 7751.40 & 0.01 & 1724.40 & 29.53 & 0.51 & 0.00\\
\ion{He}{I} & 7818.28 & 7816.46 & 0.07 & 38.26 & 24.98 & 0.53 & 0.07\\
{[}\ion{Cl}{IV}{]} & 8047.86 & 8046.05 & 0.01 & 344.78 & 25.42 & 0.43 & 0.01\\
Pa36 & 8263.21 & 8261.30 & 0.07 & 35.98 & 27.48 & 0.59 & 0.12\\
Pa35 & 8266.56 & 8264.65 & 0.08 & 38.41 & 22.42 & 0.40 & 0.08\\
Pa34 & 8270.21 & 8268.31 & 0.07 & 36.41 & 25.57 & 0.51 & 0.08\\
Pa33 & 8274.21 & 8272.17 & 0.08 & 48.45 & 29.11 & 0.62 & 0.09\\
Pa32 & 8278.58 & 8276.71 & 0.07 & 43.19 & 28.04 & 0.49 & 0.08\\
Pa31 & 8283.40 & 8281.40 & 0.05 & 52.76 & 27.69 & 0.51 & 0.06\\
Pa30 & 8288.71 & 8286.70 & 0.05 & 62.87 & 29.58 & 0.57 & 0.06\\
Pa29 & 8294.59 & 8292.59 & 0.04 & 72.55 & 32.15 & 0.59 & 0.05\\
Pa28 & 8301.12 & 8299.26 & 0.04 & 69.98 & 30.09 & 0.51 & 0.05\\
Pa27 & 8308.40 & 8306.45 & 0.04 & 77.15 & 33.97 & 0.57 & 0.05\\
Pa26 & 8316.55 & 8314.58 & 0.04 & 94.17 & 35.41 & 0.62 & 0.05\\
Pa25 & 8325.71 & 8323.79 & 0.04 & 93.89 & 31.43 & 0.60 & 0.05\\
Pa24 & 8336.08 & 8334.14 & 0.04 & 101.95 & 30.10 & 0.56 & 0.04\\
\ion{He}{I} & 8344.65 & 8342.74 & 0.17 & 0.00 & 36.02 & 0.81 & 0.23\\
Pa23 & 8347.84 & 8345.90 & 0.03 & 119.28 & 30.56 & 0.58 & 0.03\\
Pa22 & 8361.30 & 8359.31 & 0.05 & 119.16 & 30.85 & 0.60 & 0.06\\
\ion{He}{I} & 8364.03 & 8362.02 & 0.04 & 50.17 & 29.36 & 0.55 & 0.05\\
Pa21 & 8376.78 & 8374.82 & 0.02 & 150.46 & 30.43 & 0.61 & 0.02\\
Pa20 & 8394.70 & 8392.77 & 0.02 & 176.35 & 34.20 & 0.59 & 0.02\\
Pa19 & 8415.63 & 8413.65 & 0.02 & 196.10 & 31.43 & 0.56 & 0.02\\
Pa18 & 8440.27 & 8438.30 & 0.07 & 206.21 & 28.18 & 0.51 & 0.07\\
\ion{O}{I} & 8448.62 & 8446.75 & 0.13 & 191.25 & 29.97 & 0.67 & 0.13\\
Pa17 & 8469.58 & 8467.58 & 0.01 & 270.66 & 27.22 & 0.60 & 0.01\\
Pa16 & 8504.82 & 8502.81 & 0.01 & 314.90 & 24.91 & 0.56 & 0.01\\
Pa15 & 8547.73 & 8545.70 & 0.01 & 373.59 & 23.28 & 0.57 & 0.01\\
{[}\ion{Cl}{II}{]} & 8581.05 & 8578.98 & 0.05 & 57.54 & 23.93 & 0.70 & 0.06\\
\ion{C}{III} & 8584.44 & 8582.62 & 0.06 & 27.27 & 20.53 & 0.53 & 0.07\\
Pa14 & 8600.75 & 8598.72 & 0.01 & 468.27 & 23.09 & 0.57 & 0.01\\
\ion{Fe}{I} & 8618.65 & 8616.96 & 0.06 & 43.24 & 24.35 & 0.75 & 0.07\\
Pa13 & 8667.40 & 8665.34 & 0.01 & 588.86 & 25.29 & 0.58 & 0.00\\
\ion{S}{IV} & 8735.8 & 8733.91 & 0.10 & 29.29 & 29.51 & 0.65 & 0.11\\
Pa12 & 8752.88 & 8750.82 & 0.01 & 752.56 & 33.69 & 0.59 & 0.00\\
\ion{He}{I} & 8779.13 & 8777.14 & 0.06 & 32.50 & 25.61 & 0.46 & 0.07\\
\ion{Cl}{II} & 8847.68 & 8845.69 & 0.09 & 33.27 & 30.36 & 0.57 & 0.11\\
Pa11 & 8865.22 & 8863.14 & 0.01 & 962.21 & 33.03 & 0.61 & 0.00\\
\ion{He}{I} & 8999.44 & 8997.33 & 0.06 & 45.89 & 20.89 & 0.56 & 0.07\\
Pa10 & 9017.38 & 9015.28 & 0.00 & 1280.99 & 27.47 & 0.59 & 0.00\\
\ion{He}{I} & 9065.78 & 9063.73 & 0.05 & 47.87 & 21.75 & 0.54 & 0.05\\
{[}\ion{S}{III}{]} & 9071.1 & 9069.26 & 0.01 & 15292.18 & 55.46 & 0.62 & 0.00\\
\ion{He}{I} & 9212.86 & 9210.65 & 0.07 & 75.62 & 34.28 & 0.72 & 0.07\\
\ion{He}{I} & 9215.78 & 9213.66 & 0.08 & 14.46 & 26.59 & 0.41 & 0.09\\
Pa9 & 9231.55 & 9229.37 & 0.00 & 1906.44 & 41.94 & 0.64 & 0.00\\
\ion{He}{I} & 9466.16 & 9463.94 & 0.05 & 163.77 & 65.15 & 0.63 & 0.04\\
\ion{He}{I} & 9519.18 & 9517.13 & 0.05 & 61.75 & 31.45 & 0.58 & 0.05\\
\ion{He}{I} & 9528.77 & 9526.37 & 0.10 & 0.00 & 57.67 & 0.72 & 0.10\\
{[}\ion{S}{III}{]} & 9533.2 & 9531.31 & 0.01 & 35152.42 & 111.72 & 0.67 & 0.01\\
Pa$\epsilon$ & 9548.59 & 9546.35 & 0.01 & 2689.50 & 67.26 & 0.72 & 0.01\\
\ion{He}{I} & 10030.46 & 10027.96 & 0.08 & 149.38 & 84.12 & 0.58 & 0.09\\
Pa$\delta$ & 10052.12 & 10049.76 & 0.01 & 3754.73 & 114.67 & 0.74 & 0.01\\
\ion{He}{II} & 10126.3 & 10124.04 & 0.11 & 67.15 & 114.54 & 0.38 & 0.11\\
{[}\ion{S}{II}{]} & 10289.55 & 10287.36 & 0.04 & 403.26 & 41.71 & 0.88 & 0.04\\
\ion{He}{I} & 10314.06 & 10312.04 & 0.10 & 118.44 & 41.53 & 0.90 & 0.12\\
{[}\ion{S}{II}{]} & 10323.32 & 10321.10 & 0.11 & 522.91 & 35.99 & 0.86 & 0.11\\
{[}\ion{S}{II}{]} & 10339.24 & 10337.09 & 0.17 & 390.89 & 33.13 & 0.89 & 0.17\\
{[}\ion{S}{II}{]} & 10373.34 & 10371.12 & 0.09 & 203.32 & 31.13 & 1.00 & 0.09\\
{[}\ion{N}{I}{]} & 10400.59 & 10398.53 & 0.06 & 96.19 & 22.66 & 1.07 & 0.07\\
{[}\ion{N}{I}{]} & 10410.02 & 10408.02 & 0.22 & 69.04 & 36.47 & 1.44 & 0.31\\
\ion{He}{I} & 10670.59 & 10668.28 & 0.19 & 20.82 & 27.66 & 0.44 & 0.16\\
\ion{He}{I} & 10833.25 & 10831.57 & 0.01 & 122334.08 & 80.56 & 1.06 & 0.01\\
\ion{He}{I} & 10915.98 & 10913.78 & 0.06 & 268.20 & 13.95 & 0.92 & 0.07\\
\ion{He}{I} & 10920.06 & 10918.02 & 0.10 & 25.39 & 20.63 & 0.40 & 0.08\\
Pa$\gamma$ & 10941.08 & 10938.88 & 0.01 & 6925.61 & 17.61 & 0.99 & 0.01\\
\ion{He}{I} & 10999.67 & 10997.46 & 0.08 & 31.87 & 15.06 & 0.92 & 0.09\\
\ion{He}{I} & 11048.01 & 11045.60 & 0.15 & 34.95 & 14.68 & 0.87 & 0.17\\
\ion{He}{II} & 11629.6 & 11627.35 & 0.26 & 25.15 & 20.33 & 0.95 & 0.27\\
\ion{C}{I} & 11886.23 & 11884.65 & 0.10 & 50.58 & 15.44 & 0.79 & 0.09\\
\ion{He}{I} & 11972.34 & 11969.90 & 0.02 & 193.18 & 10.99 & 1.01 & 0.02\\
\ion{He}{I} & 12530.83 & 12528.29 & 0.01 & 328.91 & 8.99 & 1.02 & 0.01\\
{[}\ion{Fe}{II}{]} & 12570.21 & 12567.46 & 0.06 & 110.97 & 11.13 & 1.51 & 0.06\\
\ion{He}{I} & 12788.41 & 12785.75 & 0.02 & 569.16 & 15.55 & 1.08 & 0.02\\
\ion{He}{I} & 12794.01 & 12791.31 & 0.02 & 156.74 & 11.92 & 1.08 & 0.02\\
Pa$\beta$ & 12821.58 & 12818.90 & 0.01 & 13418.49 & 19.89 & 1.18 & 0.01\\
\ion{He}{I} & 12849.47 & 12846.64 & 0.12 & 44.38 & 13.08 & 1.17 & 0.13\\
\ion{He}{I} & 12971.98 & 12969.21 & 0.17 & 56.82 & 8.65 & 1.02 & 0.17\\
\ion{He}{I} & 12988.43 & 12985.81 & 0.15 & 56.91 & 7.98 & 0.94 & 0.15\\
\ion{O}{I} & 13167.49 & 13165.46 & 0.04 & 90.89 & 8.32 & 1.31 & 0.04\\
{[}\ion{Fe}{II}{]} & 13209.11 & 13206.28 & 0.20 & 25.20 & 9.96 & 1.41 & 0.21\\
Br30 & 14852.22 & 14849.21 & 0.11 & 33.28 & 8.32 & 1.20 & 0.11\\
Br29 & 14871.11 & 14867.71 & 0.18 & 35.88 & 9.47 & 1.19 & 0.18\\
Br28 & 14892.09 & 14888.97 & 0.08 & 37.54 & 6.88 & 0.99 & 0.08\\
Br27 & 14915.54 & 14912.52 & 0.41 & 41.46 & 11.77 & 1.24 & 0.40\\
Br26 & 14941.83 & 14938.79 & 0.20 & 51.95 & 7.10 & 1.35 & 0.22\\
Br25 & 14971.44 & 14968.20 & 0.24 & 51.10 & 8.87 & 1.23 & 0.23\\
Br24 & 15004.98 & 15001.74 & 0.06 & 67.98 & 8.25 & 1.32 & 0.06\\
Br23 & 15043.17 & 15039.93 & 0.06 & 67.02 & 7.46 & 1.22 & 0.06\\
Br22 & 15086.92 & 15083.95 & 0.03 & 117.89 & 6.74 & 1.41 & 0.03\\
Br21 & 15137.38 & 15134.23 & 0.03 & 100.29 & 7.26 & 1.34 & 0.03\\
Br20 & 15196.01 & 15192.77 & 0.03 & 106.54 & 5.56 & 1.33 & 0.03\\
Br19 & 15264.73 & 15261.49 & 0.03 & 133.24 & 5.99 & 1.45 & 0.03\\
{[}\ion{Fe}{II}{]} & 15338.94 & 15335.15 & 0.18 & 0.00 & 5.98 & 1.57 & 0.22\\
Br18 & 15346.00 & 15342.71 & 0.03 & 143.22 & 4.42 & 1.37 & 0.03\\
Br17 & 15443.16 & 15439.86 & 0.08 & 179.33 & 5.08 & 1.40 & 0.08\\
Br16 & 15560.72 & 15557.36 & 0.01 & 206.46 & 4.33 & 1.40 & 0.01\\
Br15 & 15704.97 & 15701.65 & 0.02 & 250.77 & 5.56 & 1.37 & 0.02\\
Br14 & 15884.90 & 15881.49 & 0.01 & 304.45 & 5.42 & 1.42 & 0.01\\
Br13 & 16113.73 & 16110.31 & 0.01 & 379.71 & 4.76 & 1.44 & 0.01\\
Br12 & 16411.69 & 16408.19 & 0.04 & 481.97 & 4.17 & 1.45 & 0.04\\
{[}\ion{Fe}{II}{]} & 16440.02 & 16436.37 & 1.41 & 59.40 & 3.26 & 1.16 & 1.42\\
{[}\ion{Fe}{II}{]} & 16773.34 & 16769.96 & 0.23 & 13.15 & 5.42 & 1.72 & 0.27\\
\ion{Fe}{I} & 16800.96 & 16797.55 & 0.13 & 15.64 & 5.70 & 1.36 & 0.13\\
Br11 & 16811.13 & 16807.68 & 0.01 & 636.64 & 5.27 & 1.47 & 0.01\\
\ion{He}{I} & 17007.04 & 17003.43 & 0.01 & 344.14 & 5.06 & 1.47 & 0.01\\
\ion{He}{I} & 17334.41 & 17330.85 & 0.13 & 16.56 & 3.23 & 1.22 & 0.14\\
\ion{He}{I} & 17356.45 & 17352.78 & 0.05 & 29.85 & 3.41 & 1.48 & 0.07\\
Br10 & 17366.89 & 17363.21 & 0.02 & 801.73 & 3.86 & 1.58 & 0.02\\
\ion{He}{I} & 17427.12 & 17423.40 & 0.07 & 7.52 & 2.34 & 0.88 & 0.08\\
\ion{He}{I} & 17454.40 & 17451.89 & 0.31 & 19.98 & 4.79 & 2.67 & 0.34\\
Br$\epsilon$ & 18179.10 & 18174.96 & 0.04 & 1119.34 & 12.55 & 1.79 & 0.03\\
Pa$\alpha$ & 18756.10 & 18752.36 & 0.08 & 28725.58 & 207.17 & 1.64 & 0.05\\
\ion{He}{I} & 19411.45 & 19407.08 & 0.48 & 26.36 & 6.71 & 0.96 & 0.54\\
\ion{He}{I} & 19438.89 & 19435.04 & 0.11 & 48.31 & 12.44 & 1.98 & 0.12\\
Br$\delta$ & 19450.89 & 19447.02 & 0.07 & 1629.82 & 26.12 & 1.83 & 0.06\\
\ion{He}{I} & 19548.46 & 19543.99 & 0.04 & 159.23 & 9.57 & 1.23 & 0.03\\
\ion{He}{I} & 20430.51 & 20426.02 & 0.13 & 15.03 & 2.75 & 1.76 & 0.15\\
\ion{He}{I} & 20586.90 & 20582.70 & 0.03 & 1086.11 & 4.53 & 1.73 & 0.03\\
{[}\ion{N}{I}{]} & 20606.9 & 20602.89 & 0.14 & 23.03 & 3.39 & 1.67 & 0.16\\
\ion{He}{I} & 21125.84 & 21121.46 & 0.06 & 124.94 & 2.91 & 1.78 & 0.07\\
\ion{He}{I} & 21137.80 & 21133.47 & 0.17 & 25.72 & 12.47 & 1.87 & 0.19\\
\ion{He}{I} & 21613.69 & 21609.15 & 0.11 & 58.44 & 2.78 & 1.90 & 0.11\\
\ion{He}{I} & 21622.91 & 21618.11 & 0.88 & 12.06 & 1.85 & 1.23 & 0.93\\
\ion{He}{I} & 21647.4 & 21642.85 & 0.31 & 104.00 & 2.24 & 1.69 & 0.33\\
Br$\gamma$ & 21661.21 & 21656.60 & 0.02 & 2566.73 & 3.17 & 1.95 & 0.02\\
{[}\ion{Fe}{III}{]} & 22187.3 & 22182.00 & 0.16 & 22.52 & 3.07 & 2.49 & 0.17\\
\ion{C}{I}{]} & 22866.0 & 22862.47 & 0.05 & 50.92 & 2.62 & 1.84 & 0.06\\
Pf37 & 23218.02 & 23212.68 & 0.42 & 12.14 & 3.06 & 2.84 & 0.46\\
Pf36 & 23242.39 & 23237.16 & 0.36 & 12.54 & 2.39 & 1.83 & 0.38\\
Pf35 & 23268.90 & 23264.08 & 0.48 & 9.54 & 2.22 & 1.80 & 0.50\\
Pf34 & 23297.87 & 23293.18 & 0.25 & 12.39 & 2.22 & 1.68 & 0.25\\
Pf33 & 23329.59 & 23324.78 & 0.28 & 13.82 & 2.54 & 2.16 & 0.29\\
Pf32 & 23364.44 & 23359.66 & 0.28 & 18.84 & 2.78 & 2.60 & 0.29\\
Pf31 & 23402.83 & 23398.12 & 0.21 & 21.39 & 3.14 & 2.75 & 0.22\\
Pf30 & 23445.28 & 23440.41 & 0.18 & 21.68 & 2.80 & 2.69 & 0.19\\
Pf29 & 23492.37 & 23485.96 & 0.25 & 32.22 & 3.56 & 3.99 & 0.27\\
Pf28 & 23544.81 & 23540.00 & 0.20 & 24.04 & 2.87 & 2.15 & 0.21\\
Pf27 & 23603.47 & 23598.75 & 0.14 & 25.52 & 2.62 & 2.37 & 0.15\\
Pf26 & 23669.37 & 23664.39 & 0.14 & 26.50 & 2.76 & 2.35 & 0.15\\
Pf25 & 23743.77 & 23738.69 & 0.11 & 26.37 & 2.59 & 2.13 & 0.11\\
Pf24 & 23828.23 & 23823.10 & 0.09 & 31.74 & 2.37 & 2.30 & 0.09\\
Pf23 
\footnote{Emission lines that could be detected automatically and matched with transitions in NIST and the Atomic Line List; the latter was also used to verify the forbidden or semi-forbidden states. The line fluxes were integrated  within $\pm3$ times the Gaussian width of the line centre. A flag indicating the goodness of the fit (0: good; 1: saturated line; 2: bad fit) is provided for each flux measurement. The FWHM were obtained from the best-fit Gaussian line profile. The tables for the other PNe are available in the online material. The data with full numeric precision and for all PNe are available as FITS tables online.}
    & 23924.68 & 23919.03 & 0.31 & 33.94 & 8.26 & 2.28 & 0.21\\
\end{longtable}

%% file: pn_table.tex
\begin{longtable}{|lcccc|}
    \caption{PN-like objects with ${\rm RP} < -0.4$ found in Gaia DR3.} \label{table:pn} \\
\hline
\multicolumn{1}{|c}{Name} & 
\multicolumn{1}{c}{Radius} &
\multicolumn{1}{c}{Density} & 
\multicolumn{1}{c}{Type} &
\multicolumn{1}{c|}{Reference for radius estimate}\\
\multicolumn{1}{|c}{} & 
\multicolumn{1}{c}{[$^{\prime\prime}$]} &
\multicolumn{1}{c}{[arcmin$^{-2}$]} & 
\multicolumn{1}{c}{} &
\multicolumn{1}{c|}{}\\
\hline 
\endfirsthead

\multicolumn{5}{c}%
{{\bfseries \tablename\ \thetable{} -- continued from previous page}} \\
\hline 
\multicolumn{1}{|c}{Name} & 
\multicolumn{1}{c}{Radius} &
\multicolumn{1}{c}{Density} & 
\multicolumn{1}{c}{Type} &
\multicolumn{1}{c|}{Comments}\\
\multicolumn{1}{|c}{} & 
\multicolumn{1}{c}{[$^{\prime\prime}$]} &
\multicolumn{1}{c}{[arcmin$^{-2}$]} & 
\multicolumn{1}{c}{} &
\multicolumn{1}{c|}{}\\
\hline 

\endhead

\hline \multicolumn{5}{|r|}{{Continued on next page}} \\ \hline
\endfoot

\hline \hline
\endlastfoot

\hline
        LMC-SMP-81 & 0.15 & 4.8 & PN & \cite{stanghellini2003}\\
        LMC-SMP-79 & 0.22 & 6.7 & PN & \cite{stanghellini2003}\\
        SMC-SMP-27 & 0.23 & 8.3 & PN & \cite{stanghellini2003}\\
        SMC-SMP-15 & 0.17 & 14.6 & PN & \cite{stanghellini2003}\\
        SMC-SMP-5 & 0.31 & 15.9 & PN & \cite{stanghellini2003}\\
        SMC-SMP-16 & 0.18 & 32.1 & PN & \cite{stanghellini2003}\\
        LMC-SMP-71 & 0.31 & 47.4 & PN? & \cite{stanghellini2003}\\
        LMC-SMP-34 & 0.32 & 68.1 & PN & \cite{stanghellini2003}\\
        LMC-SMP-80 & 0.21 & 71.3 & PN & \cite{stanghellini2003}\\
        LMC-SMP-48 & 0.2 & 74.5 & PN & \cite{stanghellini2003}\\
        \hline
        PN G315.4+09.4 & 2.5 & 11.5 & PN & \cite{stanghellini2010}\\
        LMC-SMP-67 & 5.4 & 17.5 & PN & \cite{reid2006}\\
        LMC-SMP-76 & 4.7 & 18.1 & PN & \cite{reid2006}\\
        LMC-SMP-37 & 4.9 & 20.0 & PN & \cite{reid2006}\\
        LMC-SMP-55 & 5.7 & 20.0 & PN & \cite{reid2006}\\
        He 2-25 & 2.2 & 23.2 & PN & \cite{stanghellini2010}\\
        PN G319.0$-$04.1 & 6.0 & 23.3 & PN & \cite{ledu2022}\\
        LMC-SMP-23 & 5.7 & 26.7 & PN & \cite{reid2006}\\
        LMC-SMP-75 & 5.0 & 29.3 & PN & \cite{reid2006}\\
        LMC-SMP-78 & 5.7 & 36.9 & PN & \cite{reid2006}\\
        LMC-SMP-36 & 3.8 & 39.1 & PN & \cite{reid2006}\\
        LMC-SMP-56 & 5.0 & 42.7 & PN & \cite{reid2006}\\
        LMC-SMP-63 & 5.0 & 46.5 & PN & \cite{reid2006}\\
        LMC-SMP-38 & 6.6 & 46.8 & PN & \cite{reid2006}\\
        LMC-SMP-77 & 6.0 & 47.1 & PN & \cite{reid2006}\\
        LMC-SMP-73 & 5.8 & 52.5 & PN? & \cite{reid2006}\\
        LMC-SMP-26 & 5.7 & 71.0 & PN & \cite{reid2006}\\
        LMC-SMP-29 & 6.0 & 76.1 & PN & \cite{reid2006}\\
        PN G356.9$-$05.8 & 3.4 & 87.2 & PN & \cite{stanghellini2010}\\
        LMC-SMP-64 & 6.0 & 94.2 & PN & \cite{reid2006}\\
        MGPN LMC 39 & 4.4 & 122.5 & PN & \cite{reid2006}\\
        LMC-SMP-47 & 5.0 & 149.9 & PN & \cite{reid2006}\\
        \hline
        EC 05147$-$5727 & -- & 0.6 & Cataclysmic variable & \\
        J132856.71+310846.0 & -- & 0.6 & Cataclysmic variable & \\
        {[}M94b{]} 48 & -- & 1.6 & PN candidate & \\
        {[}M94b{]} 54 & -- & 1.6 & PN candidate & \\
        LMC-SMP-3 & -- & 1.9 & PN & \\
        SMC-SMP-4 & -- & 2.5 & PN & \\
        SMC-SMP-28 & -- & 2.5 & PN & \\
        WDJ214206.81+030322.26 & -- & 2.5 & WD & \\
        LMC-SMP-61 & -- & 3.8 & PN & \\
        {[}M94b{]} 49 & -- & 3.8 & PN & \\
        J174506.57$-$020844.1 & -- & 4.5 & QSO & \\
        He 2-375 & -- & 4.5 & PN & \\
        SMC-SMP-17 & -- & 4.5 & PN & \\
        LMC-SMP-94 & -- & 4.8 & PN & \\
        MGPN LMC 45 & -- & 4.8 & PN & \\
        LMC-SMP-2 & -- & 5.7 & PN & \\
        LM 1-60 & -- & 6.0 & PN candidate & \\
        PN G053.6$-$12.3 & -- & 6.4 & PN candidate & \\
        2MASS 382453513 & -- & 6.7 & AGN candidate & \\
        Gaia DR3 \#6128295187479810560 & -- & 7.0 & stellar & Previously unknown PN?\\
        LMC-SMP-85 & -- & 7.6 & PN & \\
        LMC-SMP-1 & -- & 8.3 & PN & \\
        LMC-SMP-5 & -- & 8.3 & PN & \\
        LMC-SMP-84 & -- & 14.3 & PN & \\
        V* V339 Del & -- & 14.3 & Classical Nova & \\
        He 2-90 & -- & 15.0 & PN & \\
        {[}MA93{]} 1721 & -- & 15.6 & Em* & \\
        V* V2659 Cyg & -- & 16.2 & Classical Nova & \\
        SMC-SMP-6 & -- & 16.9 & PN & \\
        Gaia DR3 \#4663598364758836992 & -- & 16.9 & stellar & Previously unknown PN?\\
        WDJ172916.00$-$525338.87 & -- & 17.8 & WD & \\
        J052744.6$-$665300 & -- & 19.7 & stellar & \\
        Gaia DR3 \#4660554298082054400 & -- & 20.0 & stellar & Previously unknown PN?\\
        LMC-SM-62 & -- & 20.0 & PN & \\
        V* V1016 Cyg & -- & 20.4 & PN?/symbiotic & \\
        He 2-437 & -- & 20.7 & PN & \\
        LMC-SMP-8 & -- & 21.3 & PN & \\
        V* V2944 Oph & -- & 21.3 & Classical nova & \\
        Gaia DR3 \#4660478947196927232 & -- & 22.0 & stellar & Previously unknown PN?\\
        Lin 49 & -- & 25.1 & PN & \\
        PN MaC 1-15 & -- & 26.1 & PN candidate & \\
        J054519$-$711603 & -- & 26.7 & PN & \\
        LMC-SMP-74 & -- & 28.0 & PN & \\
        PN G342.8+04.1 & -- & 29.6 & PN & \\
        J053048.9$-$671010 & -- & 31.5 & stellar & \\
        J185938.6$-$091102 & -- & 32.8 & stellar & \\
        LMC-SMP-89 & -- & 37.2 & PN & \\
        LM 1-14 & -- & 39.8 & PN candidate & \\
        PN G007.5+04.3 & -- & 43.6 & PN & \\
        LMC-SMP-51 & -- & 50.0 & PN & \\
        Gaia DR3 \#5836588052566682880 & -- & 51.9 & stellar & Previously unknown PN?\\
        SMC-SMP-21 & -- & 73.5 & PN & \\
        G354.98$-$02.87 & -- & 85.6 & Symbiotic & \\
        V* V5667 Sgr 
        \footnote{PN-like objects found using our search for compact, zero-redshifted emission-line sources (orange lines in the bottom panels of Fig.~\ref{fig:gaia_spectra}). The first section of this table shows objects with radii $\leq$ \ang{;;0.5}, sorted by stellar field density (local number density of Gaia sources with Gmag$<$19\,mag). The second section shows objects with radii $>$ \ang{;;0.5}. The third section shows objects without size estimate in HASH.}
        & -- & 117.5 & Dwarf nova & \\

\end{longtable}

%% file: main.bbl
\begin{thebibliography}{107}
\expandafter\ifx\csname natexlab\endcsname\relax\def\natexlab#1{#1}\fi

\bibitem[{{Acker} \& {Neiner}(2003)}]{acker2003}
{Acker}, A. \& {Neiner}, C. 2003, \aap, 403, 659

\bibitem[{{Amendola} \& {Tsujikawa}(2010)}]{amendola2010}
{Amendola}, L. \& {Tsujikawa}, S. 2010, {Dark Energy: Theory and Observations}
  (Cambridge University Press, New York)

\bibitem[{{Astropy Collaboration} {et~al.}(2018){Astropy Collaboration},
  {Price-Whelan}, {Sip{\H{o}}cz}, {G{\"u}nther}, {Lim}, {Crawford}, {Conseil},
  {Shupe}, {Craig}, {Dencheva}, {Ginsburg}, {Vand erPlas}, {Bradley},
  {P{\'e}rez-Su{\'a}rez}, {de Val-Borro}, {Aldcroft}, {Cruz}, {Robitaille},
  {Tollerud}, {Ardelean}, {Babej}, {Bach}, {Bachetti}, {Bakanov}, {Bamford},
  {Barentsen}, {Barmby}, {Baumbach}, {Berry}, {Biscani}, {Boquien}, {Bostroem},
  {Bouma}, {Brammer}, {Bray}, {Breytenbach}, {Buddelmeijer}, {Burke},
  {Calderone}, {Cano Rodr{\'\i}guez}, {Cara}, {Cardoso}, {Cheedella}, {Copin},
  {Corrales}, {Crichton}, {D'Avella}, {Deil}, {Depagne}, {Dietrich}, {Donath},
  {Droettboom}, {Earl}, {Erben}, {Fabbro}, {Ferreira}, {Finethy}, {Fox},
  {Garrison}, {Gibbons}, {Goldstein}, {Gommers}, {Greco}, {Greenfield},
  {Groener}, {Grollier}, {Hagen}, {Hirst}, {Homeier}, {Horton}, {Hosseinzadeh},
  {Hu}, {Hunkeler}, {Ivezi{\'c}}, {Jain}, {Jenness}, {Kanarek}, {Kendrew},
  {Kern}, {Kerzendorf}, {Khvalko}, {King}, {Kirkby}, {Kulkarni}, {Kumar},
  {Lee}, {Lenz}, {Littlefair}, {Ma}, {Macleod}, {Mastropietro}, {McCully},
  {Montagnac}, {Morris}, {Mueller}, {Mumford}, {Muna}, {Murphy}, {Nelson},
  {Nguyen}, {Ninan}, {N{\"o}the}, {Ogaz}, {Oh}, {Parejko}, {Parley}, {Pascual},
  {Patil}, {Patil}, {Plunkett}, {Prochaska}, {Rastogi}, {Reddy Janga},
  {Sabater}, {Sakurikar}, {Seifert}, {Sherbert}, {Sherwood-Taylor}, {Shih},
  {Sick}, {Silbiger}, {Singanamalla}, {Singer}, {Sladen}, {Sooley},
  {Sornarajah}, {Streicher}, {Teuben}, {Thomas}, {Tremblay}, {Turner},
  {Terr{\'o}n}, {van Kerkwijk}, {de la Vega}, {Watkins}, {Weaver}, {Whitmore},
  {Woillez}, {Zabalza}, \& {Astropy Contributors}}]{astropy2018}
{Astropy Collaboration}, {Price-Whelan}, A.~M., {Sip{\H{o}}cz}, B.~M., {et~al.}
  2018, \aj, 156, 123

\bibitem[{{Astropy Collaboration} {et~al.}(2013){Astropy Collaboration},
  {Robitaille}, {Tollerud}, {Greenfield}, {Droettboom}, {Bray}, {Aldcroft},
  {Davis}, {Ginsburg}, {Price-Whelan}, {Kerzendorf}, {Conley}, {Crighton},
  {Barbary}, {Muna}, {Ferguson}, {Grollier}, {Parikh}, {Nair}, {Unther},
  {Deil}, {Woillez}, {Conseil}, {Kramer}, {Turner}, {Singer}, {Fox}, {Weaver},
  {Zabalza}, {Edwards}, {Azalee Bostroem}, {Burke}, {Casey}, {Crawford},
  {Dencheva}, {Ely}, {Jenness}, {Labrie}, {Lim}, {Pierfederici}, {Pontzen},
  {Ptak}, {Refsdal}, {Servillat}, \& {Streicher}}]{astropy2013}
{Astropy Collaboration}, {Robitaille}, T.~P., {Tollerud}, E.~J., {et~al.} 2013,
  \aap, 558, A33

\bibitem[{Birch \& Downs(1993)}]{birch1993}
Birch, K.~P. \& Downs, M.~J. 1993, Metrologia, 30, 155

\bibitem[{Birch \& Downs(1994)}]{birch1994}
Birch, K.~P. \& Downs, M.~J. 1994, Metrologia, 31, 315

\bibitem[{{Blanton} \& {Roweis}(2007)}]{blanton2007}
{Blanton}, M.~R. \& {Roweis}, S. 2007, \aj, 133, 734

\bibitem[{Bougoin \& Lavenac(2011)}]{bougoin2011}
Bougoin, M. \& Lavenac, J. 2011, in Optical Manufacturing and Testing IX, ed.
  J.~H. Burge, O.~W. Fähnle, \& R.~Williamson, Vol. 8126, International
  Society for Optics and Photonics (SPIE), 248--257

\bibitem[{Bougoin {et~al.}(2017)Bougoin, Lavenac, Pamplona, Martin, Gimenez,
  Castel, \& Maciaszek}]{bougoin2017}
Bougoin, M., Lavenac, J., Pamplona, T., {et~al.} 2017, in International
  Conference on Space Optics -- ICSO 2016, ed. B.~Cugny, N.~Karafolas, \&
  Z.~Sodnik, Vol. 10562, International Society for Optics and Photonics (SPIE),
  1329--1337

\bibitem[{{Bougoin} {et~al.}(2019){Bougoin}, {Mallet}, {Lavenac},
  {Gerbert-Gaillard}, {Ballhause}, \& {Chaumeil}}]{bougoin2019}
{Bougoin}, M., {Mallet}, F., {Lavenac}, J., {et~al.} 2019, in Society of
  Photo-Optical Instrumentation Engineers (SPIE) Conference Series, Vol. 11180,
  International Conference on Space Optics -- ICSO 2018, 111801P

\bibitem[{{Castorina} {et~al.}(2019){Castorina}, {Hand}, {Seljak}, {Beutler},
  {Chuang}, {Zhao}, {Gil-Mar{\'\i}n}, {Percival}, {Ross}, {Choi}, {Dawson}, {de
  la Macorra}, {Rossi}, {Ruggeri}, {Schneider}, \& {Zhao}}]{castorina2019}
{Castorina}, E., {Hand}, N., {Seljak}, U., {et~al.} 2019, \jcap, 2019, 010

\bibitem[{{Chambers} {et~al.}(2016){Chambers}, {Magnier}, {Metcalfe},
  {Flewelling}, {Huber}, {Waters}, {Denneau}, {Draper}, {Farrow}, {Finkbeiner},
  {Holmberg}, {Koppenhoefer}, {Price}, {Rest}, {Saglia}, {Schlafly}, {Smartt},
  {Sweeney}, {Wainscoat}, {Burgett}, {Chastel}, {Grav}, {Heasley}, {Hodapp},
  {Jedicke}, {Kaiser}, {Kudritzki}, {Luppino}, {Lupton}, {Monet}, {Morgan},
  {Onaka}, {Shiao}, {Stubbs}, {Tonry}, {White}, {Ba{\~n}ados}, {Bell},
  {Bender}, {Bernard}, {Boegner}, {Boffi}, {Botticella}, {Calamida},
  {Casertano}, {Chen}, {Chen}, {Cole}, {Deacon}, {Frenk}, {Fitzsimmons},
  {Gezari}, {Gibbs}, {Goessl}, {Goggia}, {Gourgue}, {Goldman}, {Grant},
  {Grebel}, {Hambly}, {Hasinger}, {Heavens}, {Heckman}, {Henderson}, {Henning},
  {Holman}, {Hopp}, {Ip}, {Isani}, {Jackson}, {Keyes}, {Koekemoer}, {Kotak},
  {Le}, {Liska}, {Long}, {Lucey}, {Liu}, {Martin}, {Masci}, {McLean}, {Mindel},
  {Misra}, {Morganson}, {Murphy}, {Obaika}, {Narayan}, {Nieto-Santisteban},
  {Norberg}, {Peacock}, {Pier}, {Postman}, {Primak}, {Rae}, {Rai}, {Riess},
  {Riffeser}, {Rix}, {R{\"o}ser}, {Russel}, {Rutz}, {Schilbach}, {Schultz},
  {Scolnic}, {Strolger}, {Szalay}, {Seitz}, {Small}, {Smith}, {Soderblom},
  {Taylor}, {Thomson}, {Taylor}, {Thakar}, {Thiel}, {Thilker}, {Unger},
  {Urata}, {Valenti}, {Wagner}, {Walder}, {Walter}, {Watters}, {Werner},
  {Wood-Vasey}, \& {Wyse}}]{chambers2016}
{Chambers}, K.~C., {Magnier}, E.~A., {Metcalfe}, N., {et~al.} 2016,
  arXiv:1612.05560

\bibitem[{{Chornay} \& {Walton}(2020)}]{chornay2020}
{Chornay}, N. \& {Walton}, N.~A. 2020, \aap, 638, A103

\bibitem[{{Chornay} \& {Walton}(2021)}]{chornay2021a}
{Chornay}, N. \& {Walton}, N.~A. 2021, \aap, 656, A110

\bibitem[{{Chornay} {et~al.}(2021){Chornay}, {Walton}, {Jones}, {Boffin},
  {Rejkuba}, \& {Wesson}}]{chornay2021c}
{Chornay}, N., {Walton}, N.~A., {Jones}, D., {et~al.} 2021, \aap, 648, A95

\bibitem[{{Clairmont} {et~al.}(2022){Clairmont}, {Steffen}, \&
  {Koning}}]{clairmont2022}
{Clairmont}, R., {Steffen}, W., \& {Koning}, N. 2022, \mnras, 516, 2711

\bibitem[{{Cropper} {et~al.}(2012){Cropper}, {Cole}, {James}, {Mellier},
  {Martignac}, {Di Giorgio}, {Paltani}, {Genolet}, {Fourmond}, {Cara},
  {Amiaux}, {Guttridge}, {Walton}, {Thomas}, {Rees}, {Pool}, {Endicott},
  {Holland}, {Gow}, {Murray}, {Duvet}, {Augueres}, {Laureijs}, {Gondoin},
  {Kitching}, {Massey}, \& {Hoekstra}}]{cropper2012}
{Cropper}, M., {Cole}, R., {James}, A., {et~al.} 2012, in Society of
  Photo-Optical Instrumentation Engineers (SPIE) Conference Series, Vol. 8442,
  Space Telescopes and Instrumentation 2012: Optical, Infrared, and Millimeter
  Wave, ed. M.~C. {Clampin}, G.~G. {Fazio}, H.~A. {MacEwen}, \& J.~{Oschmann},
  Jacobus~M., 84420V

\bibitem[{{Cutri} {et~al.}(2003){Cutri}, {Skrutskie}, {van Dyk}, {Beichman},
  {Carpenter}, {Chester}, {Cambresy}, {Evans}, {Fowler}, {Gizis}, {Howard},
  {Huchra}, {Jarrett}, {Kopan}, {Kirkpatrick}, {Light}, {Marsh}, {McCallon},
  {Schneider}, {Stiening}, {Sykes}, {Weinberg}, {Wheaton}, {Wheelock}, \&
  {Zacarias}}]{cutri2003}
{Cutri}, R.~M., {Skrutskie}, M.~F., {van Dyk}, S., {et~al.} 2003, VizieR Online
  Data Catalog, II/246

\bibitem[{{De Angeli} {et~al.}(2022){De Angeli}, {Weiler}, {Montegriffo},
  {Evans}, {Riello}, {Andrae}, {Carrasco}, {Busso}, {Burgess}, {Cacciari},
  {Davidson}, {Harrison}, {Hodgkin}, {Jordi}, {Osborne}, {Pancino},
  {Altavilla}, {Barstow}, {Bailer-Jones}, {Bellazzini}, {Brown}, {Castellani},
  {Cowell}, {Delchambre}, {De Luise}, {Diener}, {Fabricius}, {Fouesneau},
  {Fremat}, {Gilmore}, {Giuffrida}, {Hambly}, {Hidalgo}, {Holland},
  {Kostrzewa-Rutkowska}, {van Leeuwen}, {Lobel}, {Marinoni}, {Miller},
  {Pagani}, {Palaversa}, {Piersimoni}, {Pulone}, {Ragaini}, {Rainer},
  {Richards}, {Rixon}, {Ruz-Mieres}, {Sanna}, {Sarro}, {Rowell}, {Sordo},
  {Walton}, \& {Yoldas}}]{deangeli2022}
{De Angeli}, F., {Weiler}, M., {Montegriffo}, P., {et~al.} 2022,
  arXiv:2206.06143

\bibitem[{{Delgado-Inglada} {et~al.}(2020){Delgado-Inglada},
  {Garc{\'\i}a-Rojas}, {Stasi{\'n}ska}, \&
  {Rechy-Garc{\'\i}a}}]{delgadoinglada2020}
{Delgado-Inglada}, G., {Garc{\'\i}a-Rojas}, J., {Stasi{\'n}ska}, G., \&
  {Rechy-Garc{\'\i}a}, J.~S. 2020, \mnras, 498, 5367

\bibitem[{{Dimitrijevi{\'c}} {et~al.}(2007){Dimitrijevi{\'c}},
  {Kova{\v{c}}evi{\'c}}, {Popovi{\'c}}, {Da{\v{c}}i{\'c}}, \&
  {Ili{\'c}}}]{dimitrijevic2007}
{Dimitrijevi{\'c}}, M.~S., {Kova{\v{c}}evi{\'c}}, J., {Popovi{\'c}},
  L.~{\v{C}}., {Da{\v{c}}i{\'c}}, M., \& {Ili{\'c}}, D. 2007, in American
  Institute of Physics Conference Series, Vol. 895, Fifty Years of Romanian
  Astrophysics, ed. C.~{Dumitrache}, N.~A. {Popescu}, M.~D. {Suran}, \&
  V.~{Mioc}, 313--316

\bibitem[{{Dopita} {et~al.}(1985){Dopita}, {Ford}, {Lawrence}, \&
  {Webster}}]{dopita1985}
{Dopita}, M.~A., {Ford}, H.~C., {Lawrence}, C.~J., \& {Webster}, B.~L. 1985,
  \apj, 296, 390

\bibitem[{{Eisenstein} {et~al.}(2005){Eisenstein}, {Zehavi}, {Hogg},
  {Scoccimarro}, {Blanton}, {Nichol}, {Scranton}, {Seo}, {Tegmark}, {Zheng},
  {Anderson}, {Annis}, {Bahcall}, {Brinkmann}, {Burles}, {Castander},
  {Connolly}, {Csabai}, {Doi}, {Fukugita}, {Frieman}, {Glazebrook}, {Gunn},
  {Hendry}, {Hennessy}, {Ivezi{\'c}}, {Kent}, {Knapp}, {Lin}, {Loh}, {Lupton},
  {Margon}, {McKay}, {Meiksin}, {Munn}, {Pope}, {Richmond}, {Schlegel},
  {Schneider}, {Shimasaku}, {Stoughton}, {Strauss}, {SubbaRao}, {Szalay},
  {Szapudi}, {Tucker}, {Yanny}, \& {York}}]{eisenstein2005}
{Eisenstein}, D.~J., {Zehavi}, I., {Hogg}, D.~W., {et~al.} 2005, \apj, 633, 560

\bibitem[{{Elias} {et~al.}(2006{\natexlab{a}}){Elias}, {Joyce}, {Liang},
  {Muller}, {Hileman}, \& {George}}]{elias2006b}
{Elias}, J.~H., {Joyce}, R.~R., {Liang}, M., {et~al.} 2006{\natexlab{a}}, in
  Society of Photo-Optical Instrumentation Engineers (SPIE) Conference Series,
  Vol. 6269, Society of Photo-Optical Instrumentation Engineers (SPIE)
  Conference Series, ed. I.~S. {McLean} \& M.~{Iye}, 62694C

\bibitem[{{Elias} {et~al.}(2006{\natexlab{b}}){Elias}, {Rodgers}, {Joyce},
  {Lazo}, {Doppmann}, {Winge}, \& {Rodr{\'\i}guez-Ardila}}]{elias2006a}
{Elias}, J.~H., {Rodgers}, B., {Joyce}, R.~R., {et~al.} 2006{\natexlab{b}}, in
  Society of Photo-Optical Instrumentation Engineers (SPIE) Conference Series,
  Vol. 6269, Society of Photo-Optical Instrumentation Engineers (SPIE)
  Conference Series, ed. I.~S. {McLean} \& M.~{Iye}, 626914

\bibitem[{{Euclid Collaboration: Blanchard} {et~al.}(2020){Euclid
  Collaboration: Blanchard}, {Camera}, {Carbone}, {Cardone}, {Casas}, {Clesse},
  {Ili{\'c}}, {Kilbinger}, {Kitching}, {Kunz}, {Lacasa}, {Linder}, {Majerotto},
  {Markovi{\v{c}}}, {Martinelli}, {Pettorino}, {Pourtsidou}, {Sakr},
  {S{\'a}nchez}, {Sapone}, {Tutusaus}, {Yahia-Cherif}, {Yankelevich},
  {Andreon}, {Aussel}, {Balaguera-Antol{\'\i}nez}, {Baldi}, {Bardelli},
  {Bender}, {Biviano}, {Bonino}, {Boucaud}, {Bozzo}, {Branchini}, {Brau-Nogue},
  {Brescia}, {Brinchmann}, {Burigana}, {Cabanac}, {Capobianco}, {Cappi},
  {Carretero}, {Carvalho}, {Casas}, {Castander}, {Castellano}, {Cavuoti},
  {Cimatti}, {Cledassou}, {Colodro-Conde}, {Congedo}, {Conselice}, {Conversi},
  {Copin}, {Corcione}, {Coupon}, {Courtois}, {Cropper}, {Da Silva}, {de la
  Torre}, {Di Ferdinando}, {Dubath}, {Ducret}, {Duncan}, {Dupac}, {Dusini},
  {Fabbian}, {Fabricius}, {Farrens}, {Fosalba}, {Fotopoulou}, {Fourmanoit},
  {Frailis}, {Franceschi}, {Franzetti}, {Fumana}, {Galeotta}, {Gillard},
  {Gillis}, {Giocoli}, {G{\'o}mez-Alvarez}, {Graci{\'a}-Carpio}, {Grupp},
  {Guzzo}, {Hoekstra}, {Hormuth}, {Israel}, {Jahnke}, {Keihanen}, {Kermiche},
  {Kirkpatrick}, {Kohley}, {Kubik}, {Kurki-Suonio}, {Ligori}, {Lilje}, {Lloro},
  {Maino}, {Maiorano}, {Marggraf}, {Martinet}, {Marulli}, {Massey},
  {Medinaceli}, {Mei}, {Mellier}, {Metcalf}, {Metge}, {Meylan}, {Moresco},
  {Moscardini}, {Munari}, {Nichol}, {Niemi}, {Nucita}, {Padilla}, {Paltani},
  {Pasian}, {Percival}, {Pires}, {Polenta}, {Poncet}, {Pozzetti}, {Racca},
  {Raison}, {Renzi}, {Rhodes}, {Romelli}, {Roncarelli}, {Rossetti}, {Saglia},
  {Schneider}, {Scottez}, {Secroun}, {Sirri}, {Stanco}, {Starck}, {Sureau},
  {Tallada-Cresp{\'\i}}, {Tavagnacco}, {Taylor}, {Tenti}, {Tereno},
  {Toledo-Moreo}, {Torradeflot}, {Valenziano}, {Vassallo}, {Verdoes Kleijn},
  {Viel}, {Wang}, {Zacchei}, {Zoubian}, \& {Zucca}}]{blanchard2020}
{Euclid Collaboration: Blanchard}, A., {Camera}, S., {Carbone}, C., {et~al.}
  2020, \aap, 642, A191

\bibitem[{{Euclid Collaboration: Gabarra} {et~al.}(2023){Euclid Collaboration:
  Gabarra}, {Mancini}, {Rodriguez Munoz}, {Rodighiero}, {Sirignano},
  {Scodeggio}, {Talia}, {Dusini}, {Gillard}, {Granett}, {Maiorano}, {Moresco},
  {Paganin}, {Palazzi}, {Pozzetti}, {Renzi}, {Rossetti}, {Vergani}, {Allevato},
  {Bisigello}, {Castignani}, {De Caro}, {Fumana}, {Ganga}, {Garilli},
  {Hirschmann}, {La Franca}, {Laigle}, {Passalacqua}, {Schirmer}, {Stanco},
  {Troja}, {Yung}, {Zamorani}, {Zoubian}, {Aghanim}, {Amara}, {Auricchio},
  {Baldi}, {Bender}, {Bodendorf}, {Bonino}, {Branchini}, {Brescia},
  {Brinchmann}, {Camera}, {Capobianco}, {Carbone}, {Carretero}, {Castander},
  {Castellano}, {Cavuoti}, {Cledassou}, {Congedo}, {Conselice}, {Conversi},
  {Copin}, {Corcione}, {Costille}, {Courbin}, {Da Silva}, {Degaudenzi},
  {Dinis}, {Dubath}, {Dupac}, {Ealet}, {Farrens}, {Ferriol}, {Frailis},
  {Franceschi}, {Franzetti}, {Galeotta}, {Gillis}, {Giocoli}, {Grazian},
  {Grupp}, {Guzzo}, {Holmes}, {Hornstrup}, {Hudelot}, {Jahnke}, {K{\"u}mmel},
  {Kermiche}, {Kiessling}, {Kilbinger}, {Kitching}, {Kohley}, {Kubik}, {Kunz},
  {Kurki-Suonio}, {Ligori}, {Lilje}, {Lloro}, {Mansutti}, {Marggraf},
  {Markovic}, {Marulli}, {Massey}, {Maurogordato}, {Mei}, {Meneghetti},
  {Meylan}, {Moscardini}, {Munari}, {Nichol}, {Niemi}, {Nightingale},
  {Padilla}, {Paltani}, {Pasian}, {Pedersen}, {Percival}, {Pettorino},
  {Polenta}, {Poncet}, {Raison}, {Rhodes}, {Riccio}, {Romelli}, {Roncarelli},
  {Saglia}, {Sapone}, {Schneider}, {Secroun}, {Seidel}, {Serrano}, {Sirri},
  {Surace}, {Tallada-Cresp{\'\i}}, {Tavagnacco}, {Taylor}, {Tereno},
  {Toledo-Moreo}, {Torradeflot}, {Trifoglio}, {Tutusaus}, {Valentijn},
  {Valenziano}, {Vassallo}, {Wang}, {Weller}, {Zacchei}, {Andreon}, {Aussel},
  {Bardelli}, {Bolzonella}, {Boucaud}, {Bozzo}, {Colodro-Conde}, {Di
  Ferdinando}, {Farina}, {Graci{\'a}-Carpio}, {Keih{\"a}nen}, {Lindholm},
  {Maino}, {Mauri}, {Mellier}, {Neissner}, {Scottez}, {Tenti}, {Zucca},
  {Akrami}, {Baccigalupi}, {Ballardini}, {Bernardeau}, {Biviano}, {Borlaff},
  {Borsato}, {Burigana}, {Cabanac}, {Cappi}, {Carvalho}, {Casas}, {Castro},
  {Chambers}, {Cooray}, {Coupon}, {Courtois}, {Davini}, {de la Torre}, {De
  Lucia}, {Desprez}, {Dole}, {Escartin}, {Escoffier}, {Ferrero}, {Finelli},
  {Fotopoulou}, {Garcia-Bellido}, {George}, {Giacomini}, {Gozaliasl},
  {Hildebrandt}, {Hook}, {Ilbert}, {Jimenez Mu{\~n}oz}, {Kajava}, {Kansal},
  {Kirkpatrick}, {Legrand}, {Loureiro}, {Macias-Perez}, {Magliocchetti},
  {Mainetti}, {Maoli}, {Marcin}, {Martinelli}, {Martinet}, {Martins},
  {Matthew}, {Maurin}, {Metcalf}, {Morgante}, {Nadathur}, {Nucita}, {Patrizii},
  {Popa}, {Porciani}, {Potter}, {P{\"o}ntinen}, {S{\'a}nchez}, {Sakr},
  {Schneider}, {Sefusatti}, {Sereno}, {Shulevski}, {Spurio Mancini}, {Stadel},
  {Steinwagner}, {Teyssier}, {Valiviita}, {Veropalumbo}, {Viel}, \&
  {Zinchenko}}]{gabarra2023}
{Euclid Collaboration: Gabarra}, L., {Mancini}, C., {Rodriguez Munoz}, L.,
  {et~al.} 2023, arXiv:2302.09372

\bibitem[{{Euclid Collaboration: Scaramella} {et~al.}(2022){Euclid
  Collaboration: Scaramella}, {Amiaux}, {Mellier}, {Burigana}, {Carvalho},
  {Cuillandre}, {Da Silva}, {Derosa}, {Dinis}, {Maiorano}, {Maris}, {Tereno},
  {Laureijs}, {Boenke}, {Buenadicha}, {Dupac}, {Gaspar Venancio},
  {G{\'o}mez-{\'A}lvarez}, {Hoar}, {Lorenzo Alvarez}, {Racca},
  {Saavedra-Criado}, {Schwartz}, {Vavrek}, {Schirmer}, {Aussel}, {Azzollini},
  {Cardone}, {Cropper}, {Ealet}, {Garilli}, {Gillard}, {Granett}, {Guzzo},
  {Hoekstra}, {Jahnke}, {Kitching}, {Maciaszek}, {Meneghetti}, {Miller},
  {Nakajima}, {Niemi}, {Pasian}, {Percival}, {Pottinger}, {Sauvage},
  {Scodeggio}, {Wachter}, {Zacchei}, {Aghanim}, {Amara}, {Auphan}, {Auricchio},
  {Awan}, {Balestra}, {Bender}, {Bodendorf}, {Bonino}, {Branchini},
  {Brau-Nogue}, {Brescia}, {Candini}, {Capobianco}, {Carbone}, {Carlberg},
  {Carretero}, {Casas}, {Castander}, {Castellano}, {Cavuoti}, {Cimatti},
  {Cledassou}, {Congedo}, {Conselice}, {Conversi}, {Copin}, {Corcione},
  {Costille}, {Courbin}, {Degaudenzi}, {Douspis}, {Dubath}, {Duncan}, {Dusini},
  {Farrens}, {Ferriol}, {Fosalba}, {Fourmanoit}, {Frailis}, {Franceschi},
  {Franzetti}, {Fumana}, {Gillis}, {Giocoli}, {Grazian}, {Grupp}, {Haugan},
  {Holmes}, {Hormuth}, {Hudelot}, {Kermiche}, {Kiessling}, {Kilbinger},
  {Kohley}, {Kubik}, {K{\"u}mmel}, {Kunz}, {Kurki-Suonio}, {Lahav}, {Ligori},
  {Lilje}, {Lloro}, {Mansutti}, {Marggraf}, {Markovic}, {Marulli}, {Massey},
  {Maurogordato}, {Melchior}, {Merlin}, {Meylan}, {Mohr}, {Moresco}, {Morin},
  {Moscardini}, {Munari}, {Nichol}, {Padilla}, {Paltani}, {Peacock},
  {Pedersen}, {Pettorino}, {Pires}, {Poncet}, {Popa}, {Pozzetti}, {Raison},
  {Rebolo}, {Rhodes}, {Rix}, {Roncarelli}, {Rossetti}, {Saglia}, {Schneider},
  {Schrabback}, {Secroun}, {Seidel}, {Serrano}, {Sirignano}, {Sirri},
  {Skottfelt}, {Stanco}, {Starck}, {Tallada-Cresp{\'\i}}, {Tavagnacco},
  {Taylor}, {Teplitz}, {Toledo-Moreo}, {Torradeflot}, {Trifoglio}, {Valentijn},
  {Valenziano}, {Verdoes Kleijn}, {Wang}, {Welikala}, {Weller}, {Wetzstein},
  {Zamorani}, {Zoubian}, {Andreon}, {Baldi}, {Bardelli}, {Boucaud}, {Camera},
  {Di Ferdinando}, {Fabbian}, {Farinelli}, {Galeotta}, {Graci{\'a}-Carpio},
  {Maino}, {Medinaceli}, {Mei}, {Neissner}, {Polenta}, {Renzi}, {Romelli},
  {Rosset}, {Sureau}, {Tenti}, {Vassallo}, {Zucca}, {Baccigalupi},
  {Balaguera-Antol{\'\i}nez}, {Battaglia}, {Biviano}, {Borgani}, {Bozzo},
  {Cabanac}, {Cappi}, {Casas}, {Castignani}, {Colodro-Conde}, {Coupon},
  {Courtois}, {Cuby}, {de la Torre}, {Desai}, {Dole}, {Fabricius}, {Farina},
  {Ferreira}, {Finelli}, {Flose-Reimberg}, {Fotopoulou}, {Ganga}, {Gozaliasl},
  {Hook}, {Keihanen}, {Kirkpatrick}, {Liebing}, {Lindholm}, {Mainetti},
  {Martinelli}, {Martinet}, {Maturi}, {McCracken}, {Metcalf}, {Morgante},
  {Nightingale}, {Nucita}, {Patrizii}, {Potter}, {Riccio}, {S{\'a}nchez},
  {Sapone}, {Schewtschenko}, {Schultheis}, {Scottez}, {Teyssier}, {Tutusaus},
  {Valiviita}, {Viel}, {Vriend}, \& {Whittaker}}]{scaramella2022}
{Euclid Collaboration: Scaramella}, R., {Amiaux}, J., {Mellier}, Y., {et~al.}
  2022, \aap, 662, A112

\bibitem[{{Evans} {et~al.}(2018){Evans}, {Riello}, {De Angeli}, {Carrasco},
  {Montegriffo}, {Fabricius}, {Jordi}, {Palaversa}, {Diener}, {Busso},
  {Cacciari}, {van Leeuwen}, {Burgess}, {Davidson}, {Harrison}, {Hodgkin},
  {Pancino}, {Richards}, {Altavilla}, {Balaguer-N{\'u}{\~n}ez}, {Barstow},
  {Bellazzini}, {Brown}, {Castellani}, {Cocozza}, {De Luise}, {Delgado},
  {Ducourant}, {Galleti}, {Gilmore}, {Giuffrida}, {Holl}, {Kewley}, {Koposov},
  {Marinoni}, {Marrese}, {Osborne}, {Piersimoni}, {Portell}, {Pulone},
  {Ragaini}, {Sanna}, {Terrett}, {Walton}, {Wevers}, \&
  {Wyrzykowski}}]{evans2018}
{Evans}, D.~W., {Riello}, M., {De Angeli}, F., {et~al.} 2018, \aap, 616, A4

\bibitem[{{Fouesneau} {et~al.}(2022){Fouesneau}, {Fr{\'e}mat}, {Andrae},
  {Korn}, {Soubiran}, {Kordopatis}, {Vallenari}, {Heiter}, {Creevey}, {Sarro},
  {de Laverny}, {Lanzafame}, {Lobel}, {Sordo}, {Rybizki}, {Slezak},
  {{\'A}lvarez}, {Drimmel}, {Garabato}, {Delchambre}, {Bailer-Jones},
  {Hatzidimitriou}, {Lorca}, {Le Fustec}, {Pailler}, {Mary}, {Robin},
  {Utrilla}, {Abreu Aramburu}, {Bakker}, {Bellas-Velidis}, {Bijaoui}, {Blomme},
  {Bouret}, {Brouillet}, {Brugaletta}, {Burlacu}, {Carballo}, {Casamiquela},
  {Chaoul}, {Chiavassa}, {Contursi}, {Cooper}, {Dafonte}, {Demouchy},
  {Dharmawardena}, {Garc{\'\i}a-Lario}, {Garc{\'\i}a-Torres}, {Gomez},
  {Gonz{\'a}lez-Santamar{\'\i}a}, {Jean-Antoine Piccolo}, {Kontizas},
  {Lebreton}, {Licata}, {Lindstr{\o}m}, {Livanou}, {Magdaleno Romeo},
  {Manteiga}, {Marocco}, {Martayan}, {Marshall}, {Nicolas}, {Ordenovic},
  {Palicio}, {Pallas-Quintela}, {Pichon}, {Poggio}, {Recio-Blanco}, {Riclet},
  {Santove{\~n}a}, {Schultheis}, {Segol}, {Silvelo}, {Smart}, {S{\"u}veges},
  {Th{\'e}venin}, {Torralba Elipe}, {Ulla}, {van Dillen}, {Zhao}, \&
  {Zorec}}]{fouesneau2022}
{Fouesneau}, M., {Fr{\'e}mat}, Y., {Andrae}, R., {et~al.} 2022,
  arXiv:2206.05992

\bibitem[{{Gaia Collaboration: Prusti} {et~al.}(2016){Gaia Collaboration:
  Prusti}, {de Bruijne}, {Brown}, {Vallenari}, {Babusiaux}, {Bailer-Jones},
  {Bastian}, {Biermann}, {Evans}, \& et~al.}]{gaia2016}
{Gaia Collaboration: Prusti}, T., {de Bruijne}, J.~H.~J., {Brown}, A.~G.~A.,
  {et~al.} 2016, \aap, 595, A1

\bibitem[{{Gaia Collaboration: Vallenari} {et~al.}(2022){Gaia Collaboration:
  Vallenari}, {Brown}, {Prusti}, {de Bruijne}, {Arenou}, {Babusiaux},
  {Biermann}, {Creevey}, {Ducourant}, {Evans}, {Eyer}, {Guerra}, {Hutton},
  {Jordi}, {Klioner}, {Lammers}, {Lindegren}, {Luri}, {Mignard}, {Panem},
  {Pourbaix}, {Randich}, {Sartoretti}, {Soubiran}, {Tanga}, {Walton},
  {Bailer-Jones}, {Bastian}, {Drimmel}, {Jansen}, {Katz}, {Lattanzi}, {van
  Leeuwen}, {Bakker}, {Cacciari}, \& et~al.}]{gaiaDR3}
{Gaia Collaboration: Vallenari}, A., {Brown}, A.~G.~A., {Prusti}, T., {et~al.}
  2022, arXiv:2208.00211

\bibitem[{{Garc{\'\i}a-D{\'\i}az} {et~al.}(2012){Garc{\'\i}a-D{\'\i}az},
  {L{\'o}pez}, {Steffen}, \& {Richer}}]{garcia2012}
{Garc{\'\i}a-D{\'\i}az}, M.~T., {L{\'o}pez}, J.~A., {Steffen}, W., \& {Richer},
  M.~G. 2012, \apj, 761, 172

\bibitem[{{Gesicki} \& {Zijlstra}(2000)}]{gesicki2000}
{Gesicki}, K. \& {Zijlstra}, A.~A. 2000, \aap, 358, 1058

\bibitem[{Gessner {et~al.}(2001)Gessner, Smith, Kolm, Endemann, \&
  Gare}]{gessner2001}
Gessner, R., Smith, D.~J., Kolm, M., Endemann, M.~J., \& Gare, P. 2001, in
  Sensors, Systems, and Next-Generation Satellites IV, ed. H.~Fujisada, J.~B.
  Lurie, A.~Ropertz, \& K.~Weber, Vol. 4169, International Society for Optics
  and Photonics (SPIE), 133--143

\bibitem[{{Glazebrook} \& {Bland-Hawthorn}(2001)}]{glazebrook2001}
{Glazebrook}, K. \& {Bland-Hawthorn}, J. 2001, \pasp, 113, 197

\bibitem[{{Goldoni} {et~al.}(2012){Goldoni}, {Flores}, {Fran{\c{c}}ois},
  {Royer}, \& {Haigron}}]{goldoni2012}
{Goldoni}, P., {Flores}, H., {Fran{\c{c}}ois}, P., {Royer}, F., \& {Haigron},
  R. 2012, in Astronomical Society of the Pacific Conference Series, Vol. 461,
  Astronomical Data Analysis Software and Systems XXI, ed. P.~{Ballester},
  D.~{Egret}, \& N.~P.~F. {Lorente}, 741

\bibitem[{{Gonz{\'a}lez-Santamar{\'\i}a}
  {et~al.}(2021){Gonz{\'a}lez-Santamar{\'\i}a}, {Manteiga}, {Manchado}, {Ulla},
  {Dafonte}, \& {L{\'o}pez Varela}}]{gonzalez2021}
{Gonz{\'a}lez-Santamar{\'\i}a}, I., {Manteiga}, M., {Manchado}, A., {et~al.}
  2021, \aap, 656, A51

\bibitem[{{Groh} {et~al.}(2007){Groh}, {Damineli}, \& {Jablonski}}]{groh2007}
{Groh}, J.~H., {Damineli}, A., \& {Jablonski}, F. 2007, \aap, 465, 993

\bibitem[{{Grupp} {et~al.}(2019){Grupp}, {Kaminski}, {Bodendorf}, {Geis},
  {Penka}, \& {Bender}}]{grupp2019}
{Grupp}, F., {Kaminski}, J., {Bodendorf}, C., {et~al.} 2019, in Society of
  Photo-Optical Instrumentation Engineers (SPIE) Conference Series, Vol. 11116,
  Astronomical Optics: Design, Manufacture, and Test of Space and Ground
  Systems II, 1111618

\bibitem[{Grupp {et~al.}(2012)Grupp, Prieto, Geis, Bode, Katterloher, Grange,
  Junk, \& Bender}]{grupp2012}
Grupp, F., Prieto, E., Geis, N., {et~al.} 2012, in Space Telescopes and
  Instrumentation 2012: Optical, Infrared, and Millimeter Wave, ed. M.~C.
  Clampin, G.~G. Fazio, H.~A. MacEwen, \& J.~M.~O. Jr., Vol. 8442,
  International Society for Optics and Photonics (SPIE), 358--368

\bibitem[{{Guti{\'e}rrez-Soto} {et~al.}(2020){Guti{\'e}rrez-Soto},
  {Gon{\c{c}}alves}, {Akras}, {Cortesi}, {L{\'o}pez-Sanjuan}, {Guerrero},
  {Daflon}, {Borges Fernandes}, {Mendes de Oliveira}, {Ederoclite},
  {Sodr{\'e}}, {Pereira}, {Kanaan}, {Werle}, {V{\'a}zquez Rami{\'o}},
  {Alcaniz}, {Angulo}, {Cenarro}, {Crist{\'o}bal-Hornillos}, {Dupke},
  {Hern{\'a}ndez-Monteagudo}, {Mar{\'\i}n-Franch}, {Moles}, {Varela},
  {Ribeiro}, {Schoenell}, {Alvarez-Candal}, {Galbany}, {Jim{\'e}nez-Esteban},
  {Logro{\~n}o-Garc{\'\i}a}, \& {Sobral}}]{gutierrezsoto2020}
{Guti{\'e}rrez-Soto}, L.~A., {Gon{\c{c}}alves}, D.~R., {Akras}, S., {et~al.}
  2020, \aap, 633, A123

\bibitem[{{Guzzo} {et~al.}(2008){Guzzo}, {Pierleoni}, {Meneux}, {Branchini},
  {Le F{\`e}vre}, {Marinoni}, {Garilli}, {Blaizot}, {De Lucia}, {Pollo},
  {McCracken}, {Bottini}, {Le Brun}, {Maccagni}, {Picat}, {Scaramella},
  {Scodeggio}, {Tresse}, {Vettolani}, {Zanichelli}, {Adami}, {Arnouts},
  {Bardelli}, {Bolzonella}, {Bongiorno}, {Cappi}, {Charlot}, {Ciliegi},
  {Contini}, {Cucciati}, {de la Torre}, {Dolag}, {Foucaud}, {Franzetti},
  {Gavignaud}, {Ilbert}, {Iovino}, {Lamareille}, {Marano}, {Mazure}, {Memeo},
  {Merighi}, {Moscardini}, {Paltani}, {Pell{\`o}}, {Perez-Montero}, {Pozzetti},
  {Radovich}, {Vergani}, {Zamorani}, \& {Zucca}}]{guzzo2008}
{Guzzo}, L., {Pierleoni}, M., {Meneux}, B., {et~al.} 2008, \nat, 451, 541

\bibitem[{{Henry} {et~al.}(2008){Henry}, {Kwitter}, {Dufour}, \&
  {Skinner}}]{henry2008}
{Henry}, R.~B.~C., {Kwitter}, K.~B., {Dufour}, R.~J., \& {Skinner}, J.~N. 2008,
  \apj, 680, 1162

\bibitem[{{Hook} {et~al.}(2004){Hook}, {J{\o}rgensen}, {Allington-Smith},
  {Davies}, {Metcalfe}, {Murowinski}, \& {Crampton}}]{hook2004}
{Hook}, I.~M., {J{\o}rgensen}, I., {Allington-Smith}, J.~R., {et~al.} 2004,
  \pasp, 116, 425

\bibitem[{{Howell}(1984)}]{howell1984}
{Howell}, K.~C. 1984, Celestial Mechanics, 32, 53

\bibitem[{Hunter(2007)}]{hunter2007}
Hunter, J.~D. 2007, Computing in Science \& Engineering, 9, 90

\bibitem[{{Jacob} {et~al.}(2013){Jacob}, {Sch{\"o}nberner}, \&
  {Steffen}}]{jacob2013}
{Jacob}, R., {Sch{\"o}nberner}, D., \& {Steffen}, M. 2013, \aap, 558, A78

\bibitem[{{Jones} {et~al.}(2023){Jones}, {{\'A}lvarez-M{\'a}rquez}, {Sloan},
  {Kavanagh}, {Argyriou}, {Labiano}, {Law}, {Patapis}, {Mueller}, {Larson},
  {Bright}, {Klaassen}, {Fox}, {Gasman}, {Geers}, {Glauser}, {Guillard},
  {Nayak}, {Noriega-Crespo}, {Ressler}, {Sargent}, {Temim}, {Vandenbussche}, \&
  {Garc{\'\i}a Mar{\'\i}n}}]{jones2023}
{Jones}, O.~C., {{\'A}lvarez-M{\'a}rquez}, J., {Sloan}, G.~C., {et~al.} 2023,
  arXiv:2301.13233

\bibitem[{{Kausch} {et~al.}(2015){Kausch}, {Noll}, {Smette}, {Kimeswenger},
  {Barden}, {Szyszka}, {Jones}, {Sana}, {Horst}, \& {Kerber}}]{kausch2015}
{Kausch}, W., {Noll}, S., {Smette}, A., {et~al.} 2015, \aap, 576, A78

\bibitem[{{Kesseli} {et~al.}(2017){Kesseli}, {West}, {Veyette}, {Harrison},
  {Feldman}, \& {Bochanski}}]{kesseli2017}
{Kesseli}, A.~Y., {West}, A.~A., {Veyette}, M., {et~al.} 2017, \apjs, 230, 16

\bibitem[{Kramida {et~al.}(2021)Kramida, Ralchenko, Reader, \&
  Team}]{kramida2021}
Kramida, A., Ralchenko, Y., Reader, J., \& Team, N.~A. 2021, online; National
  Institute of Standards and Technology, Gaithersburg, MD.

\bibitem[{Kwitter \& Henry(2022)}]{kwitter2022}
Kwitter, K.~B. \& Henry, R. B.~C. 2022, Publications of the Astronomical
  Society of the Pacific, 134, 022001

\bibitem[{{Labiano} {et~al.}(2021){Labiano}, {Argyriou},
  {{\'A}lvarez-M{\'a}rquez}, {Glasse}, {Glauser}, {Patapis}, {Law}, {Brandl},
  {Justtanont}, {Lahuis}, {Mart{\'\i}nez-Galarza}, {Mueller}, {Noriega-Crespo},
  {Royer}, {Shaughnessy}, \& {Vandenbussche}}]{labiano2021}
{Labiano}, A., {Argyriou}, I., {{\'A}lvarez-M{\'a}rquez}, J., {et~al.} 2021,
  \aap, 656, A57

\bibitem[{{Latour} {et~al.}(2015){Latour}, {Fontaine}, {Green}, \&
  {Brassard}}]{Latour2015}
{Latour}, M., {Fontaine}, G., {Green}, E.~M., \& {Brassard}, P. 2015, \aap,
  579, A39

\bibitem[{{Laureijs} {et~al.}(2011){Laureijs}, {Amiaux}, {Arduini},
  {Augu{\`e}res}, {Brinchmann}, {Cole}, {Cropper}, {Dabin}, {Duvet}, {Ealet},
  \& et~al.}]{laureijs2011}
{Laureijs}, R., {Amiaux}, J., {Arduini}, S., {et~al.} 2011, arXiv:1110.3193

\bibitem[{{Le D{\^u}} {et~al.}(2022){Le D{\^u}}, {Mulato}, {Parker}, {Petit},
  {Ritter}, {Drechsler}, {Strottner}, {Patchick}, {Prestgard}, {Garde},
  {Outters}, \& {Raffaelli}}]{ledu2022}
{Le D{\^u}}, P., {Mulato}, L., {Parker}, Q.~A., {et~al.} 2022, \aap, 666, A152

\bibitem[{{L{\'o}pez} {et~al.}(2016){L{\'o}pez}, {Richer}, {Pereyra}, \&
  {Garc{\'\i}a-D{\'\i}az}}]{lopez2016}
{L{\'o}pez}, J.~A., {Richer}, M.~G., {Pereyra}, M., \& {Garc{\'\i}a-D{\'\i}az},
  M.~T. 2016, in Journal of Physics Conference Series, Vol. 728, Journal of
  Physics Conference Series, 032002

\bibitem[{{Maciaszek} {et~al.}(2022){Maciaszek}, {Ealet}, {Gillard}, {Jahnke},
  {Barbier}, {Prieto}, {Bon}, {Bonnefoi}, {Caillat}, {Carle}, {Costille},
  {Ducret}, {Fabron}, {Foulon}, {Gimenez}, {Grassi}, {Jaquet}, {Le Mignant},
  {Martin}, {Pamplona}, {Sanchez}, {Cl{\'e}mens}, {Caillat}, {Niclas},
  {Secroun}, {Kubik}, {Ferriol}, {Berthe}, {Barri{\`e}re}, {Fontignie},
  {Valenziano}, {Auricchio}, {Battaglia}, {De Rosa}, {Farinelli}, {Franceschi},
  {Medinaceli}, {Morgante}, {Sortino}, {Trifoglio}, {Corcione}, {Capobianco},
  {Ligori}, {Dusini}, {Borsato}, {Dal Corso}, {Laudisio}, {Sirignano},
  {Stanco}, {Ventura}, {Patrizii}, {Chiarusi}, {Fornari}, {Giacomini},
  {Margiotta}, {Mauri}, {Pasqualini}, {Sirri}, {Spurio}, {Tenti}, {Travaglini},
  {Bonoli}, {Bortoletto}, {Balestra}, {Dalessandro}, {Grupp}, {Penka},
  {Steinwagner}, {Hormuth}, {Schirmer}, {Seidel}, {Padilla}, {Casas}, {Lloro},
  {Toledo-Moreo}, {Gomez}, {Colodro-Conde}, {Liz{\'a}n}, {Diaz}, {Lilje},
  {Andersen}, {Andersen}, {S{\o}rensen}, {Hornstrup}, {Jessen}, {Thizy},
  {Holmes}, {Pniel}, {Jhabvala}, {Pravdo}, {Seiffert}, {Waczynski}, {Laureij},
  {Racca}, {Salvignol}, {Boenke}, {Strada}, \& {Mellier}}]{maciaszek2022}
{Maciaszek}, T., {Ealet}, A., {Gillard}, W., {et~al.} 2022, in Society of
  Photo-Optical Instrumentation Engineers (SPIE) Conference Series, Vol. 12180,
  Space Telescopes and Instrumentation 2022: Optical, Infrared, and Millimeter
  Wave, ed. L.~E. {Coyle}, S.~{Matsuura}, \& M.~D. {Perrin}, 121801K

\bibitem[{{Maciaszek} {et~al.}(2016){Maciaszek}, {Ealet}, {Jahnke}, {Prieto},
  {Barbier}, {Mellier}, {Beaumont}, {Bon}, {Bonnefoi}, {Carle}, {Caillat},
  {Costille}, {Dormoy}, {Ducret}, {Fabron}, {Febvre}, {Foulon}, {Garcia},
  {Gimenez}, {Grassi}, {Laurent}, {Le Mignant}, {Martin}, {Rossin}, {Pamplona},
  {Sanchez}, {Vives}, {Cl{\'e}mens}, {Gillard}, {Niclas}, {Secroun}, {Serra},
  {Kubik}, {Ferriol}, {Amiaux}, {Barri{\`e}re}, {Berthe}, {Rosset},
  {Macias-Perez}, {Auricchio}, {De Rosa}, {Franceschi}, {Guizzo}, {Morgante},
  {Sortino}, {Trifoglio}, {Valenziano}, {Patrizii}, {Chiarusi}, {Fornari},
  {Giacomini}, {Margiotta}, {Mauri}, {Pasqualini}, {Sirri}, {Spurio}, {Tenti},
  {Travaglini}, {Dusini}, {Dal Corso}, {Laudisio}, {Sirignano}, {Stanco},
  {Ventura}, {Borsato}, {Bonoli}, {Bortoletto}, {Balestra}, {D'Alessandro},
  {Medinaceli}, {Farinelli}, {Corcione}, {Ligori}, {Grupp}, {Wimmer},
  {Hormuth}, {Seidel}, {Wachter}, {Padilla}, {Lamensans}, {Casas}, {Lloro},
  {Toledo-Moreo}, {Gomez}, {Colodro-Conde}, {Liz{\'a}n}, {Diaz}, {Lilje},
  {Toulouse-Aastrup}, {Andersen}, {S{\o}rensen}, {Jakobsen}, {Hornstrup},
  {Jessen}, {Thizy}, {Holmes}, {Israelsson}, {Seiffert}, {Waczynski},
  {Laureijs}, {Racca}, {Salvignol}, {Boenke}, \& {Strada}}]{maciaszek2016}
{Maciaszek}, T., {Ealet}, A., {Jahnke}, K., {et~al.} 2016, in Society of
  Photo-Optical Instrumentation Engineers (SPIE) Conference Series, Vol. 9904,
  Space Telescopes and Instrumentation 2016: Optical, Infrared, and Millimeter
  Wave, ed. H.~A. {MacEwen}, G.~G. {Fazio}, M.~{Lystrup}, N.~{Batalha},
  N.~{Siegler}, \& E.~C. {Tong}, 99040T

\bibitem[{{Magrini} {et~al.}(2003){Magrini}, {Corradi}, {Greimel}, {Leisy},
  {Lennon}, {Mampaso}, {Perinotto}, {Pollacco}, {Walsh}, {Walton}, \&
  {Zijlstra}}]{magrini2003}
{Magrini}, L., {Corradi}, R.~L.~M., {Greimel}, R., {et~al.} 2003, \aap, 407, 51

\bibitem[{{Marigo} {et~al.}(2001){Marigo}, {Girardi}, {Groenewegen}, \&
  {Weiss}}]{marigo2001}
{Marigo}, P., {Girardi}, L., {Groenewegen}, M.~A.~T., \& {Weiss}, A. 2001,
  \aap, 378, 958

\bibitem[{{Martins} \& {Viegas}(2002)}]{martins2002}
{Martins}, L.~P. \& {Viegas}, S.~M. 2002, \aap, 387, 1074

\bibitem[{{Meyssonnier} \& {Azzopardi}(1993)}]{meyssonnier1993}
{Meyssonnier}, N. \& {Azzopardi}, M. 1993, \aaps, 102, 451

\bibitem[{{Modigliani} {et~al.}(2010){Modigliani}, {Goldoni}, {Royer},
  {Haigron}, {Guglielmi}, {Fran{\c{c}}ois}, {Horrobin}, {Bristow}, {Vernet},
  {Moehler}, {Kerber}, {Ballester}, {Mason}, \& {Christensen}}]{modigliani2010}
{Modigliani}, A., {Goldoni}, P., {Royer}, F., {et~al.} 2010, in Society of
  Photo-Optical Instrumentation Engineers (SPIE) Conference Series, Vol. 7737,
  Observatory Operations: Strategies, Processes, and Systems III, ed. D.~R.
  {Silva}, A.~B. {Peck}, \& B.~T. {Soifer}, 773728

\bibitem[{{Montegriffo} {et~al.}(2022){Montegriffo}, {De Angeli}, {Andrae},
  {Riello}, {Pancino}, {Sanna}, {Bellazzini}, {Evans}, {Carrasco}, {Sordo},
  {Busso}, {Cacciari}, {Jordi}, {van Leeuwen}, {Vallenari}, {Altavilla},
  {Barstow}, {Brown}, {Burgess}, {Castellani}, {Cowell}, {Davidson}, {De
  Luise}, {Delchambre}, {Diener}, {Fabricius}, {Fremat}, {Fouesneau},
  {Gilmore}, {Giuffrida}, {Hambly}, {Harrison}, {Hidalgo}, {Hodgkin},
  {Holland}, {Marinoni}, {Osborne}, {Pagani}, {Palaversa}, {Piersimoni},
  {Pulone}, {Ragaini}, {Rainer}, {Richards}, {Rowell}, {Ruz-Mieres}, {Sarro},
  {Walton}, \& {Yoldas}}]{montegriffo2022}
{Montegriffo}, P., {De Angeli}, F., {Andrae}, R., {et~al.} 2022,
  arXiv:2206.06205

\bibitem[{{Mora} {et~al.}(2016){Mora}, {Biermann}, {Bombrun}, {Boyadjian},
  {Chassat}, {Corberand}, {Davidson}, {Doyle}, {Escolar}, {Gielesen},
  {Guilpain}, {Hernandez}, {Kirschner}, {Klioner}, {Koeck}, {Laine},
  {Lindegren}, {Serpell}, {Tatry}, \& {Thoral}}]{mora2016}
{Mora}, A., {Biermann}, M., {Bombrun}, A., {et~al.} 2016, in Society of
  Photo-Optical Instrumentation Engineers (SPIE) Conference Series, Vol. 9904,
  Space Telescopes and Instrumentation 2016: Optical, Infrared, and Millimeter
  Wave, ed. H.~A. {MacEwen}, G.~G. {Fazio}, M.~{Lystrup}, N.~{Batalha},
  N.~{Siegler}, \& E.~C. {Tong}, 99042D

\bibitem[{{Ochsenbein} {et~al.}(2000){Ochsenbein}, {Bauer}, \&
  {Marcout}}]{ochsenbein2000}
{Ochsenbein}, F., {Bauer}, P., \& {Marcout}, J. 2000, \aaps, 143, 23

\bibitem[{{Otsuka} {et~al.}(2009){Otsuka}, {Hyung}, {Lee}, {Izumiura}, \&
  {Tajitsu}}]{otsuka2009}
{Otsuka}, M., {Hyung}, S., {Lee}, S.-J., {Izumiura}, H., \& {Tajitsu}, A. 2009,
  \apj, 705, 509

\bibitem[{{Otsuka} {et~al.}(2011){Otsuka}, {Meixner}, {Riebel}, {Hyung},
  {Tajitsu}, \& {Izumiura}}]{otsuka2011}
{Otsuka}, M., {Meixner}, M., {Riebel}, D., {et~al.} 2011, \apj, 729, 39

\bibitem[{Pamplona {et~al.}(2016)Pamplona, Gimenez, Febvre, Ceria, Martin,
  Prieto, Maciaszek, Foulon, Ducret, Bougoin, \& Castel}]{pamplona2016}
Pamplona, T., Gimenez, J.-L., Febvre, A., {et~al.} 2016, in Advances in Optical
  and Mechanical Technologies for Telescopes and Instrumentation II, ed.
  R.~Navarro \& J.~H. Burge, Vol. 9912, International Society for Optics and
  Photonics (SPIE), 197--209

\bibitem[{{Parker}(2022)}]{parker2022}
{Parker}, Q.~A. 2022, Frontiers in Astronomy and Space Sciences, 9, 895287

\bibitem[{{Parker} {et~al.}(2016){Parker}, {Boji{\v{c}}i{\'c}}, \&
  {Frew}}]{parker2016}
{Parker}, Q.~A., {Boji{\v{c}}i{\'c}}, I.~S., \& {Frew}, D.~J. 2016, in Journal
  of Physics Conference Series, Vol. 728, Journal of Physics Conference Series,
  032008

\bibitem[{{Pe{\~n}a} {et~al.}(2007){Pe{\~n}a}, {Stasi{\'n}ska}, \&
  {Richer}}]{pena2007}
{Pe{\~n}a}, M., {Stasi{\'n}ska}, G., \& {Richer}, M.~G. 2007, \aap, 476, 745

\bibitem[{{Piskunov} {et~al.}(1995){Piskunov}, {Kupka}, {Ryabchikova}, {Weiss},
  \& {Jeffery}}]{piskunov1995}
{Piskunov}, N.~E., {Kupka}, F., {Ryabchikova}, T.~A., {Weiss}, W.~W., \&
  {Jeffery}, C.~S. 1995, \aaps, 112, 525

\bibitem[{{Porter} \& {Rivinius}(2003)}]{porter2003}
{Porter}, J.~M. \& {Rivinius}, T. 2003, \pasp, 115, 1153

\bibitem[{{Prieto} {et~al.}(2012){Prieto}, {Amiaux}, {Augu{\`e}res},
  {Barri{\`e}re}, {Bonoli}, {Bortoletto}, {Cerna}, {Corcione}, {Duvet},
  {Ealet}, {Garilli}, {Gondoin}, {Grupp}, {Jahnke}, {Laureijs}, {Ligori}, {Le
  F{\`e}vre}, {Maciaszek}, {Madrid}, {Martignac}, {Martin}, {Morgante},
  {Mellier}, {Pamplona}, {Holmes}, {Grange}, {Riva}, {Rossin}, {Seidel},
  {Smadja}, {Toledo-Moreo}, {Trifoglio}, {Valenziano}, \& {Zerbi}}]{prieto2012}
{Prieto}, E., {Amiaux}, J., {Augu{\`e}res}, J.-L., {et~al.} 2012, in Society of
  Photo-Optical Instrumentation Engineers (SPIE) Conference Series, Vol. 8442,
  Space Telescopes and Instrumentation 2012: Optical, Infrared, and Millimeter
  Wave, ed. M.~C. {Clampin}, G.~G. {Fazio}, H.~A. {MacEwen}, \& J.~{Oschmann},
  Jacobus~M., 84420W

\bibitem[{{Racca} {et~al.}(2016){Racca}, {Laureijs}, {Stagnaro}, {Salvignol},
  {Lorenzo Alvarez}, {Saavedra Criado}, {Gaspar Venancio}, {Short}, {Strada},
  {B{\"o}nke}, {Colombo}, {Calvi}, {Maiorano}, {Piersanti}, {Prezelus},
  {Rosato}, {Pinel}, {Rozemeijer}, {Lesna}, {Musi}, {Sias}, {Anselmi},
  {Cazaubiel}, {Vaillon}, {Mellier}, {Amiaux}, {Berth{\'e}}, {Sauvage},
  {Azzollini}, {Cropper}, {Pottinger}, {Jahnke}, {Ealet}, {Maciaszek},
  {Pasian}, {Zacchei}, {Scaramella}, {Hoar}, {Kohley}, {Vavrek}, {Rudolph}, \&
  {Schmidt}}]{racca2016}
{Racca}, G.~D., {Laureijs}, R., {Stagnaro}, L., {et~al.} 2016, in Society of
  Photo-Optical Instrumentation Engineers (SPIE) Conference Series, Vol. 9904,
  Space Telescopes and Instrumentation 2016: Optical, Infrared, and Millimeter
  Wave, ed. H.~A. {MacEwen}, G.~G. {Fazio}, M.~{Lystrup}, N.~{Batalha},
  N.~{Siegler}, \& E.~C. {Tong}, 99040O

\bibitem[{{Reid} \& {Parker}(2006)}]{reid2006}
{Reid}, W.~A. \& {Parker}, Q.~A. 2006, \mnras, 373, 521

\bibitem[{{Richer} {et~al.}(2010){Richer}, {L{\'o}pez}, {D{\'\i}az-M{\'e}ndez},
  {Riesgo}, {B{\'a}ez}, {Garc{\'\i}a-D{\'\i}az}, {Meaburn}, {Clark},
  {Calder{\'o}n Olvera}, {L{\'o}pez Soto}, \& {Toledano Rebolo}}]{richer2010}
{Richer}, M.~G., {L{\'o}pez}, J.~A., {D{\'\i}az-M{\'e}ndez}, E., {et~al.} 2010,
  \rmxaa, 46, 191

\bibitem[{{Richer} \& {McCall}(2007)}]{richer2007}
{Richer}, M.~G. \& {McCall}, M.~L. 2007, \apj, 658, 328

\bibitem[{{Riello} {et~al.}(2021){Riello}, {De Angeli}, {Evans}, {Montegriffo},
  {Carrasco}, {Busso}, {Palaversa}, {Burgess}, {Diener}, {Davidson}, {Rowell},
  {Fabricius}, {Jordi}, {Bellazzini}, {Pancino}, {Harrison}, {Cacciari}, {van
  Leeuwen}, {Hambly}, {Hodgkin}, {Osborne}, {Altavilla}, {Barstow}, {Brown},
  {Castellani}, {Cowell}, {De Luise}, {Gilmore}, {Giuffrida}, {Hidalgo},
  {Holland}, {Marinoni}, {Pagani}, {Piersimoni}, {Pulone}, {Ragaini}, {Rainer},
  {Richards}, {Sanna}, {Walton}, {Weiler}, \& {Yoldas}}]{riello2021}
{Riello}, M., {De Angeli}, F., {Evans}, D.~W., {et~al.} 2021, \aap, 649, A3

\bibitem[{{Rousselot} {et~al.}(2000){Rousselot}, {Lidman}, {Cuby}, {Moreels},
  \& {Monnet}}]{rousselot2000}
{Rousselot}, P., {Lidman}, C., {Cuby}, J.~G., {Moreels}, G., \& {Monnet}, G.
  2000, \aap, 354, 1134

\bibitem[{{Ryabchikova} {et~al.}(2015){Ryabchikova}, {Piskunov}, {Kurucz},
  {Stempels}, {Heiter}, {Pakhomov}, \& {Barklem}}]{ryabchikova2015}
{Ryabchikova}, T., {Piskunov}, N., {Kurucz}, R.~L., {et~al.} 2015, \physscr,
  90, 054005

\bibitem[{{Schirmer}(2016)}]{schirmer2016}
{Schirmer}, M. 2016, \pasp, 128, 114001

\bibitem[{Schönberner(2016)}]{schoenberner2016}
Schönberner, D. 2016, Journal of Physics: Conference Series, 728, 032001

\bibitem[{{Shaw} {et~al.}(2001){Shaw}, {Stanghellini}, {Mutchler}, {Balick}, \&
  {Blades}}]{shaw2001}
{Shaw}, R.~A., {Stanghellini}, L., {Mutchler}, M., {Balick}, B., \& {Blades},
  J.~C. 2001, \apj, 548, 727

\bibitem[{{Shaw} {et~al.}(2006){Shaw}, {Stanghellini}, {Villaver}, \&
  {Mutchler}}]{shaw2006}
{Shaw}, R.~A., {Stanghellini}, L., {Villaver}, E., \& {Mutchler}, M. 2006,
  \apjs, 167, 201

\bibitem[{{Smette} {et~al.}(2015){Smette}, {Sana}, {Noll}, {Horst}, {Kausch},
  {Kimeswenger}, {Barden}, {Szyszka}, {Jones}, {Gallenne}, {Vinther},
  {Ballester}, \& {Taylor}}]{smette2015}
{Smette}, A., {Sana}, H., {Noll}, S., {et~al.} 2015, \aap, 576, A77

\bibitem[{{Smith} {et~al.}(1996){Smith}, {Shara}, \& {Moffat}}]{smith1996}
{Smith}, L.~F., {Shara}, M.~M., \& {Moffat}, A. F.~J. 1996, \mnras, 281, 163

\bibitem[{{Soubiran} {et~al.}(2013){Soubiran}, {Jasniewicz}, {Chemin}, {Crifo},
  {Udry}, {Hestroffer}, \& {Katz}}]{soubiran2013}
{Soubiran}, C., {Jasniewicz}, G., {Chemin}, L., {et~al.} 2013, \aap, 552, A64

\bibitem[{{Stanghellini} {et~al.}(2020){Stanghellini}, {Bucciarelli},
  {Lattanzi}, \& {Morbidelli}}]{stanghellini2020}
{Stanghellini}, L., {Bucciarelli}, B., {Lattanzi}, M.~G., \& {Morbidelli}, R.
  2020, \apj, 889, 21

\bibitem[{{Stanghellini} \& {Haywood}(2010)}]{stanghellini2010}
{Stanghellini}, L. \& {Haywood}, M. 2010, \apj, 714, 1096

\bibitem[{{Stanghellini} {et~al.}(2003){Stanghellini}, {Shaw}, {Balick},
  {Mutchler}, {Blades}, \& {Villaver}}]{stanghellini2003}
{Stanghellini}, L., {Shaw}, R.~A., {Balick}, B., {et~al.} 2003, \apj, 596, 997

\bibitem[{{Stanghellini} {et~al.}(2002){Stanghellini}, {Shaw}, {Mutchler},
  {Palen}, {Balick}, \& {Blades}}]{stanghellini2002}
{Stanghellini}, L., {Shaw}, R.~A., {Mutchler}, M., {et~al.} 2002, \apj, 575,
  178

\bibitem[{{Stanghellini} {et~al.}(2016){Stanghellini}, {Shaw}, \&
  {Villaver}}]{stanghellini2016}
{Stanghellini}, L., {Shaw}, R.~A., \& {Villaver}, E. 2016, \apj, 830, 33

\bibitem[{{Steffen} {et~al.}(2009){Steffen}, {Esp{\'\i}ndola}, {Mart{\'\i}nez},
  \& {Koning}}]{steffen2009}
{Steffen}, W., {Esp{\'\i}ndola}, M., {Mart{\'\i}nez}, S., \& {Koning}, N. 2009,
  \rmxaa, 45, 143

\bibitem[{{Storey} \& {Zeippen}(2000)}]{storey2000}
{Storey}, P.~J. \& {Zeippen}, C.~J. 2000, \mnras, 312, 813

\bibitem[{{Sutherland} {et~al.}(2015){Sutherland}, {Emerson}, {Dalton},
  {Atad-Ettedgui}, {Beard}, {Bennett}, {Bezawada}, {Born}, {Caldwell}, {Clark},
  {Craig}, {Henry}, {Jeffers}, {Little}, {McPherson}, {Murray}, {Stewart},
  {Stobie}, {Terrett}, {Ward}, {Whalley}, \& {Woodhouse}}]{sutherland2015}
{Sutherland}, W., {Emerson}, J., {Dalton}, G., {et~al.} 2015, \aap, 575, A25

\bibitem[{{Vernet} {et~al.}(2011){Vernet}, {Dekker}, {D'Odorico}, {Kaper},
  {Kjaergaard}, {Hammer}, {Randich}, {Zerbi}, {Groot}, {Hjorth}, {Guinouard},
  {Navarro}, {Adolfse}, {Albers}, {Amans}, {Andersen}, {Andersen}, {Binetruy},
  {Bristow}, {Castillo}, {Chemla}, {Christensen}, {Conconi}, {Conzelmann},
  {Dam}, {de Caprio}, {de Ugarte Postigo}, {Delabre}, {di Marcantonio},
  {Downing}, {Elswijk}, {Finger}, {Fischer}, {Flores}, {Fran{\c{c}}ois},
  {Goldoni}, {Guglielmi}, {Haigron}, {Hanenburg}, {Hendriks}, {Horrobin},
  {Horville}, {Jessen}, {Kerber}, {Kern}, {Kiekebusch}, {Kleszcz}, {Klougart},
  {Kragt}, {Larsen}, {Lizon}, {Lucuix}, {Mainieri}, {Manuputy}, {Martayan},
  {Mason}, {Mazzoleni}, {Michaelsen}, {Modigliani}, {Moehler}, {M{\o}ller},
  {Norup S{\o}rensen}, {N{\o}rregaard}, {P{\'e}roux}, {Patat}, {Pena}, {Pragt},
  {Reinero}, {Rigal}, {Riva}, {Roelfsema}, {Royer}, {Sacco}, {Santin},
  {Schoenmaker}, {Spano}, {Sweers}, {Ter Horst}, {Tintori}, {Tromp}, {van
  Dael}, {van der Vliet}, {Venema}, {Vidali}, {Vinther}, {Vola}, {Winters},
  {Wistisen}, {Wulterkens}, \& {Zacchei}}]{vernet2011}
{Vernet}, J., {Dekker}, H., {D'Odorico}, S., {et~al.} 2011, \aap, 536, A105

\bibitem[{Wang {et~al.}(2004)Wang, Stiller, von Clarmann, Garcia-Comas,
  Lopez-Puertas, Kiefer, Hoepfner, Glatthor, Funke, Gil-Lopez, Grabowski,
  Kellmann, Linden, Tsidu, Milz, Steck, Fischer, III, Remsberg, Mertens, \&
  Mlynczak}]{wang2004}
Wang, D.-Y., Stiller, G.~P., von Clarmann, T., {et~al.} 2004, in Remote Sensing
  of Clouds and the Atmosphere VIII, ed. K.~P. Schaefer, A.~Comeron, M.~R.
  Carleer, \& R.~H. Picard, Vol. 5235, International Society for Optics and
  Photonics (SPIE), 196--207

\bibitem[{Wang(2010)}]{wang2010}
Wang, Y. 2010, Dark Energy (John Wiley \& Sons, Ltd)

\bibitem[{{Weinberg} {et~al.}(2013){Weinberg}, {Mortonson}, {Eisenstein},
  {Hirata}, {Riess}, \& {Rozo}}]{weinberg2013}
{Weinberg}, D.~H., {Mortonson}, M.~J., {Eisenstein}, D.~J., {et~al.} 2013,
  \physrep, 530, 87

\bibitem[{{Wright} {et~al.}(2005){Wright}, {Corradi}, \&
  {Perinotto}}]{wright2005}
{Wright}, S.~A., {Corradi}, R.~L.~M., \& {Perinotto}, M. 2005, \aap, 436, 967

\bibitem[{Yuan \& Liu(2013)}]{yuan2013}
Yuan, H.~B. \& Liu, X.~W. 2013, Monthly Notices of the Royal Astronomical
  Society, 436, 718

\bibitem[{{Zhang} {et~al.}(2004){Zhang}, {Liu}, {Wesson}, {Storey}, {Liu}, \&
  {Danziger}}]{zhang2004}
{Zhang}, Y., {Liu}, X.~W., {Wesson}, R., {et~al.} 2004, \mnras, 351, 935

\bibitem[{{Zijlstra} {et~al.}(2006){Zijlstra}, {Gesicki}, {Walsh},
  {P{\'e}quignot}, {van Hoof}, \& {Minniti}}]{zijlstra2006}
{Zijlstra}, A.~A., {Gesicki}, K., {Walsh}, J.~R., {et~al.} 2006, \mnras, 369,
  875

\end{thebibliography}
